\documentclass[11pt,a4paper]{article}

\usepackage{grffile}  
\usepackage{doi}
\usepackage{hyperref}  
\usepackage{a4wide}

\usepackage{natbib}
\usepackage{amsmath}
\usepackage{amssymb}
\usepackage{graphicx}

\usepackage{amsfonts,amssymb}
\usepackage{natbib}

\title{Variability modes in core flows inverted from geomagnetic field models}

\author
{M. A. Pais$^{1,2}$, A. L. Morozova$^1$, N. Schaeffer$^{3,4}$ \\
  $^1$ CGUC, Geophysical and Astronomical Observatory, University of Coimbra, Portugal\\
  $^2$ Department of Physics, University of Coimbra, P-3004-516 Coimbra, Portugal \\
  $^3$ Univ. Grenoble Alpes, ISTerre, F-38041 Grenoble, France \\
  $^4$ CNRS, ISTerre, F-38041 Grenoble, France
}

\date{\today \\ 
Published in \emph{Geophys. J. Int.} (2015) \textbf{200}, 402--420, 
doi: \href{http://dx.doi.org/10.1093/gji/ggu403}{10.1093/gji/ggu403}}

\begin{document}
\label{firstpage}
\maketitle

\begin{abstract}

The flow of liquid metal inside the Earth's core produces the geomagnetic field and its time variations. Understanding the variability of those deep currents is crucial to improve the forecast of geomagnetic field variations, which affect human spacial and aeronautic activities. Moreover, it may provide relevant information on the core dynamics. The main goal of this study is to extract and characterize the leading variability modes of core flows over centennial periods, and to assess their statistical robustness.
To this end, we use flows that we invert from two geomagnetic field models ({\it gufm1} and {\it COV-OBS}), and apply Principal Component Analysis and Singular Value Decomposition of coupled fields.
The quasi geostrophic (QG) flows inverted from both geomagnetic field models show similar features. However, {\it COV-OBS} has a less energetic mean and larger time variability. 
The statistical significance of flow components is tested from analyses performed on subareas of the whole domain.
Bootstrapping methods are also used to extract significant flow features required by both {\it gufm1} and {\it COV-OBS}.

Three main empirical circulation modes emerge, simultaneously constrained by both geomagnetic field models and expected to be robust against the particular {\it a priori} used to build them (large scale QG dynamics). Mode 1 exhibits three large vortices at medium/high latitudes, with opposite circulation under the Atlantic and the Pacific hemispheres.
Mode 2 interestingly accounts for most of the variations of the Earth's core angular momentum. In this mode, the regions close to the tangent cylinder and to the equator are correlated, and oscillate with a period between 80 and 90 years.
Each of these two modes is energetic enough to alter the mean flow, sometimes reinforcing the eccentric gyre, and other times breaking it up into smaller circulations. The three main circulation modes added together to the mean flow account for about 70\% of the flows variability, 90\% of the root mean square total velocities, and 95\% of the secular variation induced by the total flows.

Direct physical interpretation of the computed modes is not straightforward. Nonetheless, similarities found between the two first modes and time/spatial features identified in different studies of core dynamics, suggest that our approach can help to pinpoint the relevant physical processes inside the core on centennial timescales.

\end{abstract}


\section{Introduction}
\label{sec1}

Time variation of the main geomagnetic field on timescales smaller than about a couple of centuries (secular variation, SV) have been scrutinized for deeper understanding of the Earth's core dynamics. New insights have been gained both from numerical simulation of the Navier-Stokes, energy and induction equations inside the core \citep[see e.g.][]{jones01} and inversion of magnetic field models fitting observatory and satellite observations \citep[e.g.][]{holme07}. Although the parameter regime characterizing the Earth's short timescale dynamics is not yet attainable in three-dimensional numerical simulations, different schemes have been used to extrapolate results. These include derivation of scaling laws and/or the use of simplifying approximations that allow the use of more Earth-like parameters. 

The Quasi-Geostrophic (QG) approximation is such an example, since it reduces the system dimensionality from three to two (3D to 2D) making it possible to investigate a range of Ekman (ratio of viscous to Coriolis forces) and Lundquist (ratio of Alfv\'en wave to diffusion timescales) numbers that approach the Earth's core conditions. Convective quasi-geostrophic rolls are known since the work by \cite{roberts68} and \cite{busse70} and recent studies show that for timescales characteristic of SV and large scale flows, elongated structures along the $z$-rotation axis will not be destroyed by Lorentz (magnetic) forces \citep{jault08, gillet11}. 
One must however keep in mind the limitations of the QG approximation, mainly that it does not allow for anti-symmetric flows (the ones that cross the equator) which could in reality be produced by any non-symmetric forcing.

From the point of view of flows computed from inversion of geomagnetic field models, the QG approximation is one example of constraint used to mitigate the intrinsic non-uniqueness characterizing the inverse problem.
Other constraints relying on different dynamical assumptions or empirical reasons have also been used (e.g. purely toroidal flow, steady flow, tangential geostrophy, steady flow in a drifting reference frame, helical flow, and tangential magnetostrophy). An account on these previously used constraints can be found in \cite{holme07} and \cite{finlay10}.
In the following, we will restrict to QG flows due to their importance in SV dynamics \citep{gillet11}.
In addition, although the flow is inverted on the core surface, it has a unique prolongation inside the core.
QG flows inverted from different geomagnetic field models have been previously computed that explain the observed SV \citep{pais08, gillet09}. In these studies, a large scale spatial structure was identified and described as a large eccentric anticyclonic gyre.
It interestingly gathers into one single structure the well-known westward drift under the low latitude Atlantic region, and the high latitude jet under the Bering Sea and close to the tangent cylinder (TC, cylinder coaxial with the Earth's rotation axis and tangent to the inner core surface) \citep[see e.g. Figure 3 in][]{holme07}.
This structure seems to be more prominent in inversions derived from recent satellite geomagnetic field models, suggesting some time variability on relatively short timescales characteristic of SV.
Other structures are also retrieved, a large scale cyclonic vortex beneath the Pacific Hemisphere and smaller scale vortices dominantly cyclonic under the Pacific Hemisphere and anticyclonic under the Atlantic Hemisphere.
\cite{gillet09} pointed out that the smaller scale vortices may not in fact be well resolved, since they strongly depend on the induction effects that involve the small scale main field (above harmonic degree 13-14), which are difficult to infer reliably from surface and satellite observations. In this respect, \cite{finlay12} stress the fact that the spectral slope above degree 12 depends significantly on the modeling regularizations in space.
But there has not yet been a dedicated study to establish the significance of the flow structures.
In this paper, we study the time variability of the inverted flow, and we assess their significance with statistical tools largely used in climate studies  and oceanography.

Recently, a new geomagnetic field model has been computed, {\it COV-OBS}, which covers the period of historical geomagnetic field data provided by the international network of observatories (1840-1990), but also includes the recent period of satellite missions (1990-2010) with a dense coverage of high quality data \citep{gillet13}. It avoids employing the regularized least squares inversion approach used to compute the {\it gufm1} model \citep{jackson00}, where roughness in the spatial and in the temporal domains is minimized when fitting the data to a truncated spherical harmonic expansion and using low-order splines in the time domain.
Claiming that such regularizations are mainly {\it ad hoc} and do completely constrain the model above some harmonic degree (smaller at older epochs, larger at recent ones), \cite{gillet13} preferred to use {\it a priori} information on the statistical properties of the geomagnetic field, which they consider to be relatively well known, and follow a stochastic modeling approach
(although their approach is not completely free from some imposed temporal constraints, since the {\it COV-OBS} model is expanded on a spline basis with a 2-year knot spacing).
The Gauss coefficients of the geomagnetic field are treated as resulting from a stationary process, their statistical properties being condensed into a large {\it prior} covariance matrix with variances and time correlations for all  coefficients.
Although the two models {\it gufm1} and {\it COV-OBS} fit the shared data at the same level for the common period (1840-1990), the different intrinsic characteristics are expected to distinguish flows inverted from one and the other. 

In this study, we use QG flow models inverted from both {\it gufm1} and from {\it COV-OBS}.
The computation of these flows, which is outlined in section 2, is explained and discussed in the appendix \ref{secA.1}.
Our main purpose is to test the possibility to describe the global circulation of the core flow in terms of a small number of patterns of variability, much like the climate variability can be described as a combination of patterns like the ENSO (El Ni\~no-Southern Oscillation), PNA (Pacific - North America teleconnection pattern), NAO (North Atlantic Oscillation), etc \citep[e.g.][]{barry01}.
We apply Principal Component Analysis (PCA) tools, of standard use in the geophysical sciences of meteorology and oceanography, where we find the main references \citep[e.g][]{preisendorfer88,vonstorch95}. These are non-parametric methods that do not assume any particular statistical distribution for the analyzed data or any particular dynamical equation for the flow. In geomagnetic studies, PCA has been previously applied to space-time data grids in studies of geoelectromagnetic induction \citep{fujii02,balasis06}, and using ionospheric data \citep{matsuo02,zhang13}. The PCA approach has several limitations, as explained by \cite{richman86}. These may be due to domain shape dependence through the space orthogonality condition imposed,  to sensitivity to the extent of spatial and time domains considered and to incomplete separation of consecutive modes. Also, there may be no correspondence between PCA-computed and physical modes. Nevertheless, PCA variability modes generally allow for a complexity reduction and, in certain cases, can provide information on the underlying dynamics \citep[e.g][]{north84}. They have accordingly been extensively used in data analysis. Besides PCA, singular Value Decomposition (SVD) of coupled fields is also applied in this study to identify pairs of coupled spatial patterns explaining the covariance between {\it gufm1} and {\it COV-OBS} \citep[see e.g.][]{bjornsson97}.
At the same time, we propose to carry out significance tests that can decide which are the modes carrying relevant information on the system variability. 
The obtained results receive statistical support from analyses performed on subareas of the whole domain. Bootstrapping resampling methods are used to assess the statistical robustness of particular flow features.
In the end, a physical interpretation of the most important/significant modes is sought.

The flow data to be analyzed is described in section 2. The basis of the PCA method and the more technical aspects that are used in this study are introduced in section 3. Results are discussed in section 4. They include results obtained when analyzing flows inverted from {\it gufm1} and from {\it COV-OBS} separately, and results obtained by extracting correlated information contained in both of them. Section 5 contains a discussion on the identified structures and their possible dynamical interpretation.
Finally, the last section summarizes the main conclusions in this study.

\section{Inverted QG flow models}

Two core surface flow models were computed, using regularized inversion, from the time-varying geomagnetic field models for the observatory era {\it gufm1} (1840 -- 1990) \citep{jackson00} and COV-OBS (1840 -- 2010) \citep{gillet13}. These two geomagnetic field models employ cubic B-splines to model the time variation, with 2.5-year and 2-year knot spacing, respectively (but data are supplied twice the knot spacing). Our flow inversions generate a flow snapshot for each year in the geomagnetic field model period, and the whole set of flows is denoted {\it flow}$^{gufm1}$ and {\it flow}$^{COV-OBS}$, respectively. The whole inversion procedure is explained in detail in the appendix \ref{secA.1}.

In the inversion, the flows were assumed to be columnar in the whole volume and incompressible. This implies that they can be completely retrieved from a single scalar pseudo streamfunction $\xi$, symmetric with respect to the equatorial plane, which in turn can be inverted from geomagnetic field models. At the core surface, the relation between the core surface flow $\mathbf{u}$ and $\xi $ at a certain epoch is \citep{schaeffer05, pais08, amit13}:
\begin{equation}
\mathbf{u} = \dfrac{1}{\cos \theta} \nabla_H \wedge \xi (\theta, \phi) \, \hat{r} + \dfrac{\sin \theta}{R_c \cos^2 \theta} \xi^{NZ} (\theta, \phi) \, \hat{\phi}  \label{eq1}
\end{equation}
where $(r, \theta, \phi)$ are spherical coordinates, $R_c$ is the core radius, and $\xi^{NZ}$ is the non-zonal ($\phi$-dependent) part of $\xi$. The first term in (\ref{eq1}) has exactly the same expression as tangential geostrophic (TG) flows computed at the CMB \citep[e.g.]{chulliat00}, where $\xi$ would represent the geostrophic pressure to within a constant factor, $p_{geo} = -2 \rho \Omega \xi$ (with $\rho$ the core fluid density and $\Omega$ the Earth's rotation rate).
For columnar flows touching the CMB, if they conserve mass they no longer follow isocontours of $\xi$ because of the second term in (\ref{eq1}). This term does not contribute to the zonal flow, is everywhere latitudinal and dependent on co-latitude.
It is required for fluid mass conservation, leading to equal contributions to upwellings and downwelling from the two terms \citep[see e.g.][]{amit13}.
As also discussed in \cite{amit13}, the $\xi$ field is uniquely determined from inversion of geomagnetic field models.
Close to the equator, $\xi^{NZ}(\theta, \phi) \sim \cos^2 \theta$ in order for the flow to remain finite.
\cite{canet14} use other pseudo streamfunction $\Psi(\theta,\phi)$ to study hydromagnetic quasi-geostrophic modes in planetary interiors, the two scalar functions being related through $\Psi^{NZ}= - R_c \cos \theta\,  \xi^{NZ}$ and $\partial_\theta \Psi^Z = -R_c \cos \theta \, \partial_\theta \, \xi^Z$ (where the $Z$ superscript denotes the zonal component).

The assumption of equatorially symmetric flow due to columnar convection does not hold inside the tangent cylinder.
Different regimes can be envisioned there, depending on the dominant force balance believed to exist in that region.
In particular, this can lead either to independent columnar convection in the North and South hemispheres \citep{pais08} or thermal wind convection \citep[e.g][]{aurnou03} that can change the sign of circulation close to the inner core and the outer core boundaries.

Because of this uncertainty, and for convenience, we use the same assumptions inside and outside the TC for computing $\mathbf{u}$.
Hence, to guarantee that flows computed for this small region (that usually tend to have high root mean square velocity $u_{rms}$) will not affect the PCA modes, the $20^{\circ}$ caps around the North and South CMB poles were excluded from the analysis as explained in the following section.


\section{Principal Component Analysis of QG flows }
\label{secPCA}

We follow the implementation of the method as explained in \cite{bjornsson97, hannachi07}. The field $\xi$, closely related to a geostrophic pressure especially at high latitudes, can be used  to characterize the spatial flow features and its time variability over some time-period, with a similar role to sea-level pressure in studies of variability of atmospheric currents \citep[e.g.][]{hannachi07}. In short, we will analyze the $\xi$ field into a sum of orthogonal functions in space, multiplied by coefficients which are uncorrelated functions of time, by applying a linear transformation to the dataset that concentrates as much of the variance as possible into a small number of terms in the expansion.

\subsection{Empirical Orthogonal Functions and Principal Components}
\label{secSEOF}

The spatio-temporal information retrieved from inverted flow models is gathered into matrices:
\begin{equation}
\mathbf{X} = \left[\begin{array}{cccc}
X_{11}&X_{12} &\cdots& X_{1N_p}\\ X_{21}&X_{22} &\cdots& X_{2N_p} \\ \vdots &\vdots &\vdots & \vdots\\ X_{N_e1}&X_{N_e2} &\cdots& X_{N_eN_p} \\ \end{array}\right] \label{eq2}
\end{equation}
with
\begin{equation}
X_{ij} \equiv  X(t_i, \vec{r}_j)  =  \sqrt{\sin\theta} \left( \xi(t_i, \theta_j, \phi_j) - \frac{1}{N_e}\sum_{i=1}^{N_e} \xi(t_i, \theta_j, \phi_j) \right) \label{eq2b} \quad.
\end{equation}
In a few words, $X_{ij}$ is the (scalar) value of the $\xi$ pseudo-streamfunction at the core surface point $ \vec{r}_j = \vec{r}(R_c, \theta_j, \phi_j)$ at time $t_i$, once the time-average has been subtracted. It is also weighted by a latitude-dependent factor $\sqrt{\sin \theta}$  to compensate for the fact that in a regular grid on a spherical surface the number of points to cover a given area increases with latitude.
The index $i$ takes values from 1 to the total number of epochs, $N_e$, and the index $j$ takes values from 1 to the total number of grid points, $N_p$.
For {\it flow}$^{gufm1}$ and {\it flow}$^{COV-OBS}$, a set of $5\,^{\circ} \times 5\,^{\circ}$ latitude/longitude grids of the scalar function $\xi$ were computed, one for each year, covering the CMB region under study defined by $20^\circ < \theta < 160^\circ$ for all values of longitude.
We exclude the polar caps because, as referred above, the computed QG flows are not reliable inside the tangent cylinder (TC).

The Empirical Orthogonal Function/Principal Component analysis (EOF/PCA) relies on the description of the space-time data in terms of decorrelated modes.
They are the eigenvectors of the covariance (or variance-covariance) matrix $\mathbf{C}_X=\mathbf{X}^T \mathbf{X}$: each element $(\mathbf{C}_X)_{ij}=X_{li} X_{lj}=\sum_{l=1}^{N_e} X(t_l, \vec{r}_i) X(t_l, \vec{r}_j)$ is the discrete form of the (temporal) covariance between the data values at points $\vec{r}_i$ and $\vec{r}_j$. In search for decorrelated modes, we need to transform $\mathbf{C}_X$ in a diagonal matrix, for which the standard procedure is to transform the data matrix $\mathbf{X}$ into
\begin{equation}
\mathbf{Y} =   \mathbf{X} \mathbf{P} \label{eq3}
\end{equation}
where $\mathbf{P}$ is the orthogonal matrix of eigenvectors of $\mathbf{C}_X$, each eigenvector $\mathbf{p}_i$ forming a column. The covariance matrix of $\mathbf{Y}$ is then
\begin{equation}
\mathbf{C}_Y =  \mathbf{P}^T \mathbf{C}_X \mathbf{P} = \mathbf{\Lambda} \label{eq4}
\end{equation}
and is diagonal with non-zero elements $\lambda_i$, ordered from the highest to the lowest.

The transformed data-matrix $\mathbf{Y} $ is now expressed in terms of decorrelated modes, as was intended. From (\ref{eq3}) the reconstructed data matrix yields
\begin{equation}
\mathbf{X} =   \mathbf{Y} \mathbf{P}^T \label{eq5} \qquad,
\end{equation}
which can be written
\begin{equation}
\mathbf{X}= \mathbf{y}_1 \mathbf{p}_1^T + \mathbf{y}_2 \mathbf{p}_2^T + \cdots + \mathbf{y}^{\,}_{N_p} \mathbf{p}_{N_p}^T = \sum_{k=1}^{N_p} \mathbf{y}^{\,}_k \mathbf{p}_k^T \label{eq6} \quad,
\end{equation}
i.e., as a linear combination of all eigenvectors of $\mathbf{C}_X$, the coefficients of the expansion being the different columns of matrix $\mathbf{Y}$.

Here, and according to \cite{bjornsson97}, the following notation is used:
\begin{itemize}
\item each column $\mathbf{p}^{\,}_k$ of $\mathbf{P}$ is an Empirical Orthogonal Function (EOF) and represents a certain spatial function;
\item each column $\mathbf{y}^{\,}_k$ of $\mathbf{Y}$ is a Principal Component (PC) and represents a time series.
\end{itemize}

This decomposition presents the further advantage of data reduction, since there is no practical need to keep the whole set of $N_p$ $EOF$s (the same number as grid points) in the analysis.
The different eigenvalues $\lambda_i$ of $\mathbf{C}_X$, determine which modes are to be kept in the expansion and in practice only the first few ones are important, those that explain the highest percentage of data variance as given by
\begin{equation}
f_i = \lambda_i \, \left( \sum\limits_{k=1}^{N_p} \lambda_k \right)^{-1} \label{eqSC} \quad .
\end{equation}

In terms of describing the data covariance matrix, upon which the whole framework is based, the EOFs provide a simplified and (hopefully) compressed way to decompose this matrix. From (\ref{eq4}) it follows
\begin{equation}
\mathbf{C}_X =  \displaystyle\sum\limits_{k=1}^{N_p} \lambda_k \mathbf{p}^{\,}_k \mathbf{p}^T_k  \label{eq7_1} \quad,
\end{equation}
with only the first modes required.

We note that some studies apply PCA using correlation instead of covariance matrices, by normalizing each time series to unit standard deviation \citep[see e.g.][]{bretherton92}. Here, such a choice could  enhance small, non-meaningful signal and was not adopted.

\subsection{Singular Value Decomposition of coupled fields}
\label{secSVD}

Singular Value Decomposition (SVD) of coupled fields has been widely used in meteorology and  oceanography, to study two combined data fields of different physical quantities such as e.g. Sea Level Pressure (SLP) and Sea Suface Temperature (SST).  In this study, it is used as a framework to isolate important coupled modes of variability between the two time series of core flows, {\it flow}$^{gufm1}$ and {\it flow}$^{COV-OBS}$, representing the same physical quantity but derived from geomagnetic field models computed using different {\it a priori} on the data and the model itself. We argue that by applying SVD to the correlation  between the two flows  the most robust features can be recovered, those that do not depend specifically on the {\it a priori} used in one or the other geomagnetic field models. 

We follow the implementation of the method as explained in \cite{bretherton92}, \cite{bjornsson97} and \cite{hannachi07}. Starting from two data matrices $\mathbf{X}_1$ and $\mathbf{X}_2$ for {\it flow}$^{gufm1}$ and {\it flow}$^{COV-OBS}$ respectively, built as explained above, the temporal cross-variance matrix is constructed, $\mathbf{C}_{X_1 X_2} = \mathbf{X}^T_1 \mathbf{X}_2$. As will be made clear in the following, it makes no difference for the spatial patterns, the temporal functions or the percentage of cross-covariance explained, that this covariance matrix or its transpose, $\mathbf{C}_{X_2 X_1}$, be considered. The SVD general matrix operation is applied to $\mathbf{C}_{X_1 X_2}$:
\begin{equation}
\mathbf{C}_{X_1 X_2} = \mathbf{U} \Lambda \mathbf{V}^T \qquad,  \label{eq7_2}
\end{equation}
where $ \mathbf{U}$ and $ \mathbf{V}$ are both orthogonal matrices, the former having in columns the eigenvectors of $\mathbf{C}^{\,}_{X_1 X_2} \mathbf{C}^T_{X_1 X_2}$ and the later having the eigenvectors of $\mathbf{C}^T_{X_1 X_2} \mathbf{C}^{\,}_{X_1 X_2}$. As to $ \Lambda$, it is a diagonal matrix with singular values $\lambda_i$ in its diagonal, ordered from the highest to the lowest. These $\lambda_i$ are the (positive) square roots of the eigenvalues of $\mathbf{C}^T_{X_1 X_2} \mathbf{C}^{\,}_{X_1 X_2}$, the same as for $\mathbf{C}^{\,}_{X_1 X_2} \mathbf{C}^T_{X_1 X_2} $.

Linear transformation of data matrices $\mathbf{X}_1$ and $\mathbf{X}_2$ with the orthogonal matrices $\mathbf{U}$ and $\mathbf{V}$ respectively, makes the new cross-correlation matrix diagonal, $\mathbf{C}_{Y_1 Y_2} = \Lambda$, where $\mathbf{Y}_1=\mathbf{X}_1 \mathbf{U}$ and $\mathbf{Y}_2=\mathbf{X}_2 \mathbf{V}$. The original data matrices are reconstructed as
\begin{eqnarray}
\mathbf{X}_1&=&\mathbf{Y}_1 \mathbf{U}^T \nonumber  \\
\mathbf{X}_2&=&\mathbf{Y}_2 \mathbf{V}^T \label{eq8}    \quad,
\end{eqnarray}
analogously to (\ref{eq5}). It is nonetheless worth noting that $\mathbf{U}$ and $\mathbf{V}$ integrate information on both {\it flow}$^{gufm1}$ and {\it flow}$^{COV-OBS}$, contrary to the transformation matrix $\mathbf{P}$ in section \ref{secSEOF}. As a result, spatial patterns obtained from $\mathbf{U}$ for instance are different from EOFs obtained by applying  PCA to $\mathbf{X}_1$. Also, the expansion coefficients time series in (\ref{eq8}), namely $\mathbf{Y}_1$ and $\mathbf{Y}_2$, are different from the PCs referred in (\ref{secSEOF}). 

In analogy with (\ref{eq7_1}), the column vectors $\mathbf{u}_i$ and $\mathbf{v}_i$ of the transformation matrices $\mathbf{U}$ and $\mathbf{V}$, respectively, can be used to decompose the matrix $\mathbf{C}_{X_1 X_2}$ :
\begin{equation}
\mathbf{C}_{X_1 X_2}=  \displaystyle\sum\limits_{k=1}^{N_p} \lambda_k \mathbf{u}^{\,}_k \mathbf{v}^T_k  \label{eq9} \quad.
\end{equation}
This expression resorts the fact that each pair $k$ of coupled modes describes a fraction of the cross-covariance between the two fields, the fraction being given by $f_k$ (see \ref{eqSC}).

\subsection{Homogeneous and heterogeneous correlation maps}
\label{sec-homo-hetero}

Besides direct representation of PCA or SVD modes in the form of maps of the EOFs and time plots of corresponding PCs, correlation maps are a means to highlight in the analyzed spatial region, those regions that are evolving in time in a correlated manner.
In each point of the space grid, a correlation coefficient is computed between a PC time function and the `observed' time-evolving streamfunction in that same point.
For homogeneous correlation maps, the concept is very similar to the one leading to different EOFs and correlating the different PCs with the original data gives rise to spatial structures that are the same as the corresponding EOFs.
Nonetheless, it involves looking for strong correlation instead of strong covariance and, as such, it is normalised to local variance and the values in each grid point are between $-1$ and $+1$. In this way, regions of low amplitude variations that show up as faint regions in the charts of EOFs, will show up as strong as regions of high amplitude variations, if they belong to the same EOF structure.
Heterogeneous correlation maps are meaningful in SVD analysis to show how PCs extracted from a certain flow model are correlated with the other flow model.
If the identified SVD modes are to be given some physical meaning, it is required that a certain PC correlated with the two flow models gives similar structures.

From studies of climatic patterns analyses, we borrow standard tests used to assess the robustness of variability modes. The relevant quantities are introduced in appendix \ref{sec_signif}.

\section{Results of statistical analyses}

\begin{figure}
\centering
            \includegraphics[width=0.45\textwidth]{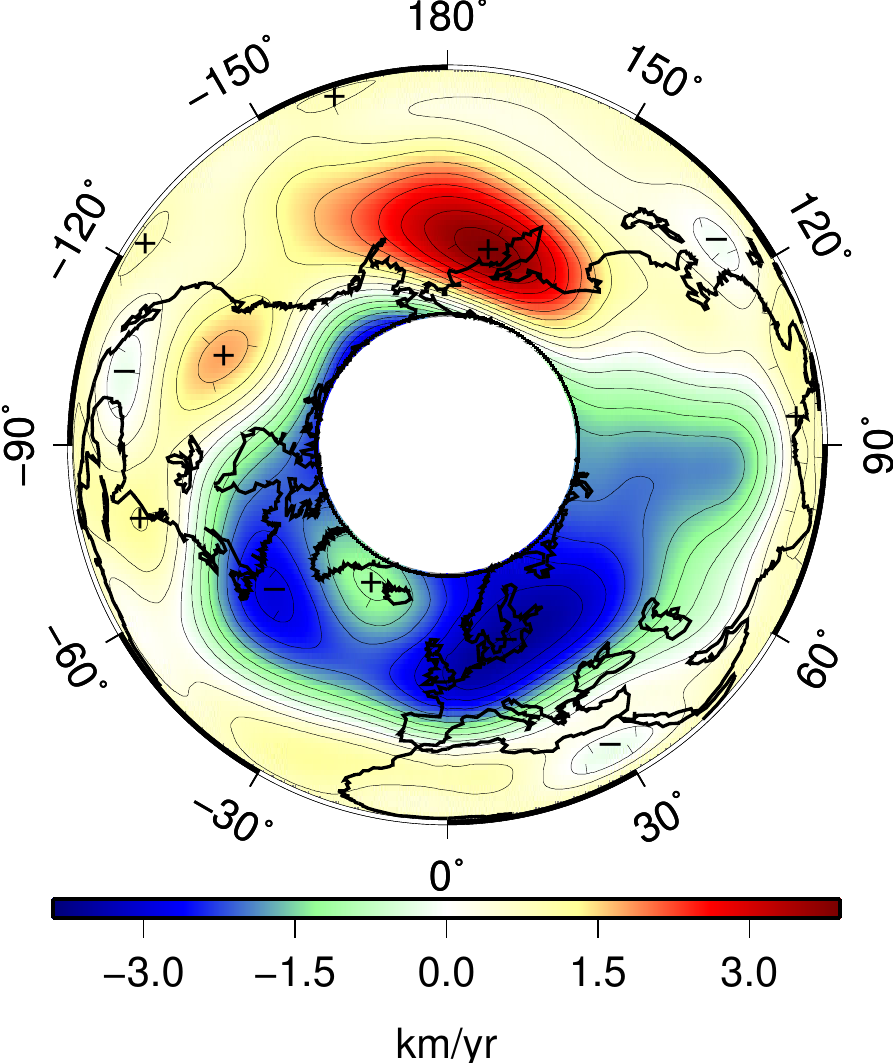}
            \includegraphics[width=0.45\textwidth]{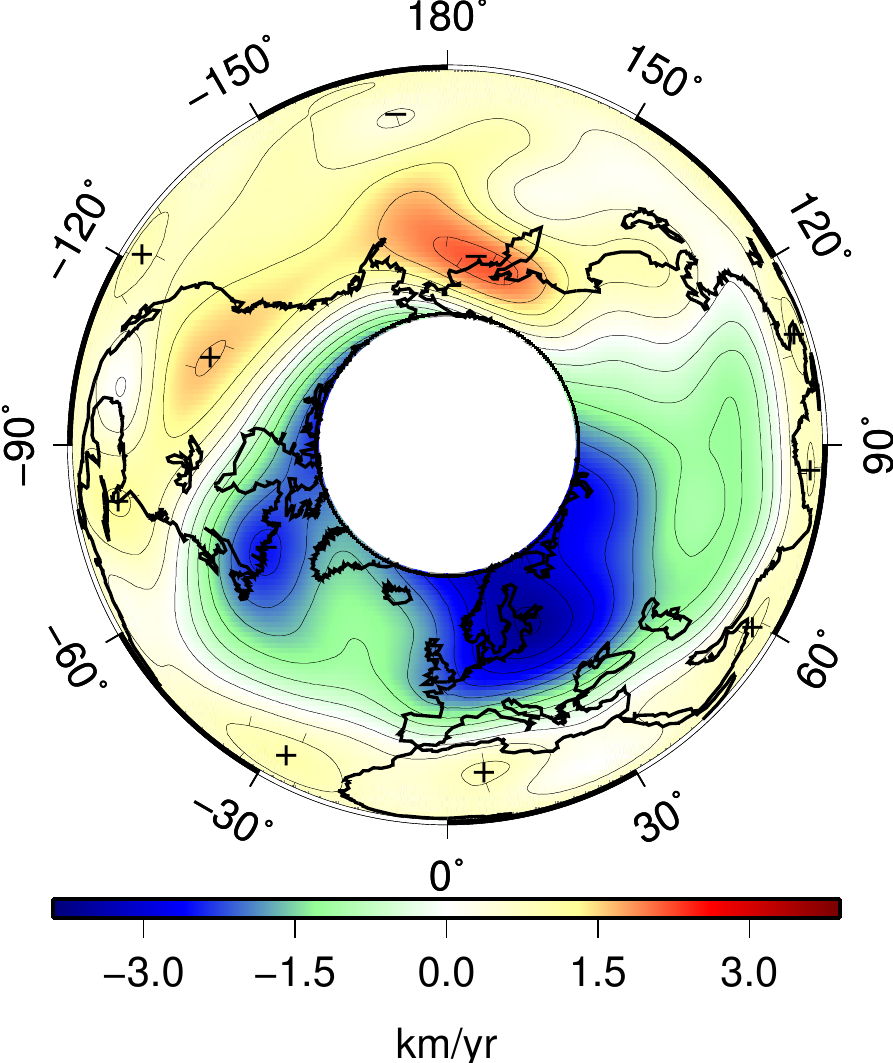} \\
            \includegraphics[width=0.45\textwidth]{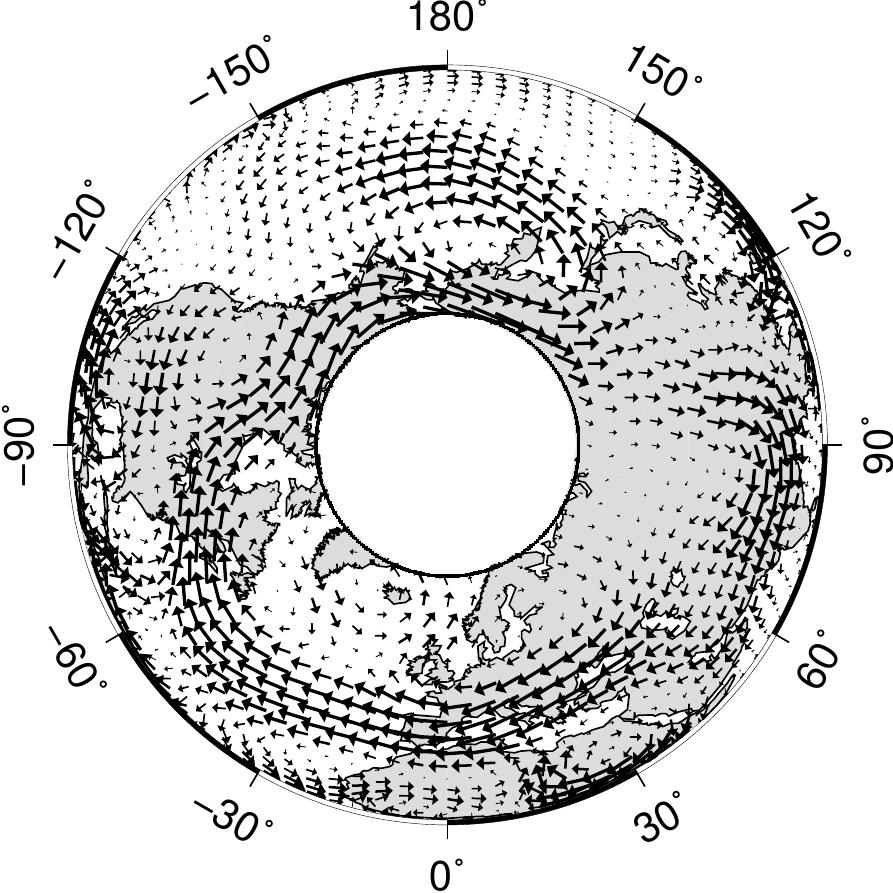}
            \includegraphics[width=0.45\textwidth]{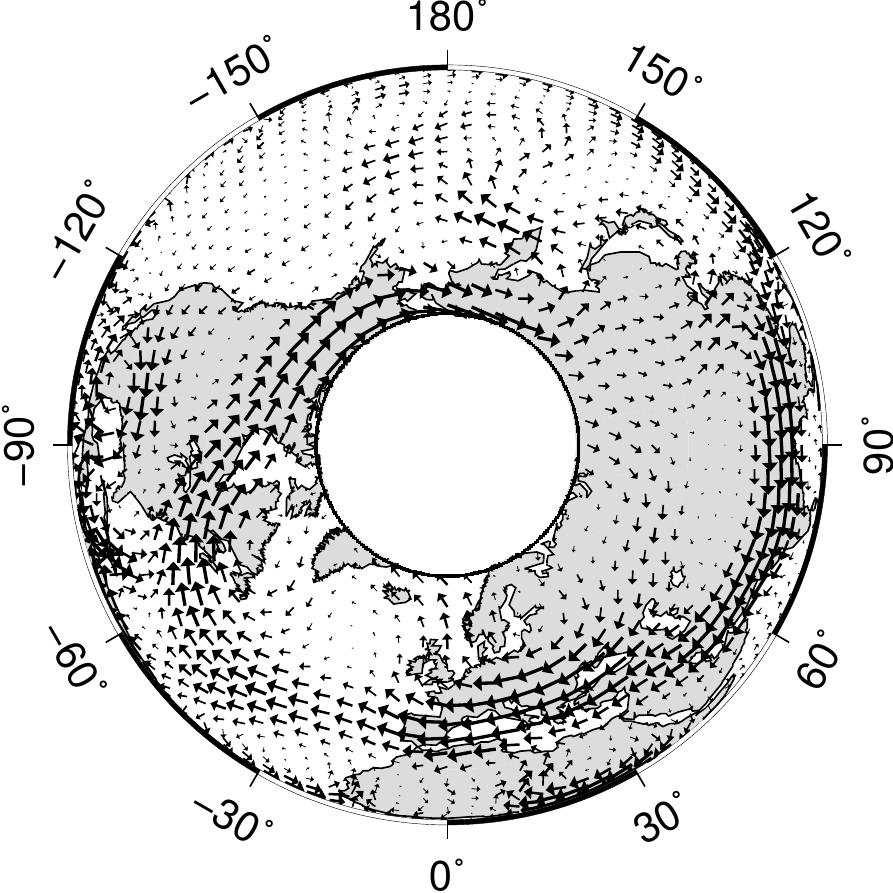}
 \caption{The mean QG-flow on the equatorial plane over the time period 1840 - 1990, as seen from the North pole. For {\it gufm1} (left) and {\it COV-OBS} (right). At the top, the pseudo streamfunction $\xi / R_c$ in units of km/yr; at the bottom, arrows visualize the flow. }
\label{fig1}
 \end{figure}

The mean flows shown in Fig. \ref{fig1} are computed by averaging the flow over the corresponding time-period: 1840 - 1990 for {\it flow}$^{gufm1}$ and 1840 - 2010 for {\it flow}$^{COV-OBS}$. There are no significant differences if we consider the shorter 1840 - 1990 period for {\it COV-OBS}. The root mean square (rms) velocity for the surface mean flow, $\overline{u}_{rms}$, is stronger for {\it flow}$^{gufm1}$ ($12$ km/yr) than for {\it flow}$^{COV-OBS}$ ($10$ km/yr), basically due to a stronger anticyclone under the Pacific Hemisphere and a stronger radial jet under the Eastern Asian continent, from high to low latitudes. The two different kind of charts in Fig. \ref{fig1} illustrate the fact that the flow very closely follows contours of the $\xi$ streamfunction  at medium-high latitudes. The rule to keep in mind when interpreting $\xi$-contour charts, from eq. \ref{eq1}, is that the fluid circulates anti-clockwise around centers of positive $\xi$ and clockwise around centers of negative $\xi$.

The EOF/PCA tools can sort out the time variability associated to the main structures in the mean flow, and other structures which average out during the inspected period. In the following, we will be showing results for the analysis carried out over the computed flows, using tools described in section \ref{secPCA} and appendix \ref{sec_signif}.

\subsection{PCA applied to {\it gufm1} and {\it COV-OBS}, separately}
\label{secPCA-raw}

\begin{table}

                \begin{tabular}{l r r r r r}
                  \hline \hline
                 \,          & \multicolumn{5}{c}{Fraction of variance, $f_i$ [\%] ($\delta f_i$ [\%])}      \\
                  \cline{2-6}
                  & \textit{mode-1}  & \textit{mode-2}  & \textit{mode-3} & \textit{mode-4} & \textit{mode-5}  \\
                  \cline{1-6}       
                  {\it COV-OBS} (1840 - 2010)            & 34.1 (4.1)          & 16.6 (2.0)         & 11.0 (1.3)                & 10.3 (1.2)           & 7.5 (0.9)    \\
                  
                   {\it COV-OBS} (1840 - 1990)             & 34.4 (4.5)            & 18.3 (2.4)          & 12.7 (1.7)                & 8.6 (1.1)            & 7.7 (1.0)      \\
                 
                  {\it gufm1} (1840 - 1990)               & 33.3 (4.3)             & 22.9 (3.0)          & 16.2 (2.1)                & 7.9 (1.0)            & 6.6 (0.9)    \\
                \hline \hline
                \end{tabular}  
                \caption{Standard PCA applied to ${\it flow}^{gufm1}$ and ${\it flow}^{COV-OBS}$ separately. Percentage of variance explained by each mode ($100 f_i$, using eq. \ref{eqSC}). In parenthesis, the $\delta f_i$ error due to $\delta \lambda_i$ as given by eq. \ref {north_error}.}   
                \label{tab1}

\end{table}

\begin{figure}
\centering
      \begin{minipage}[c]{0.45\textwidth}
            \includegraphics[scale=0.6]{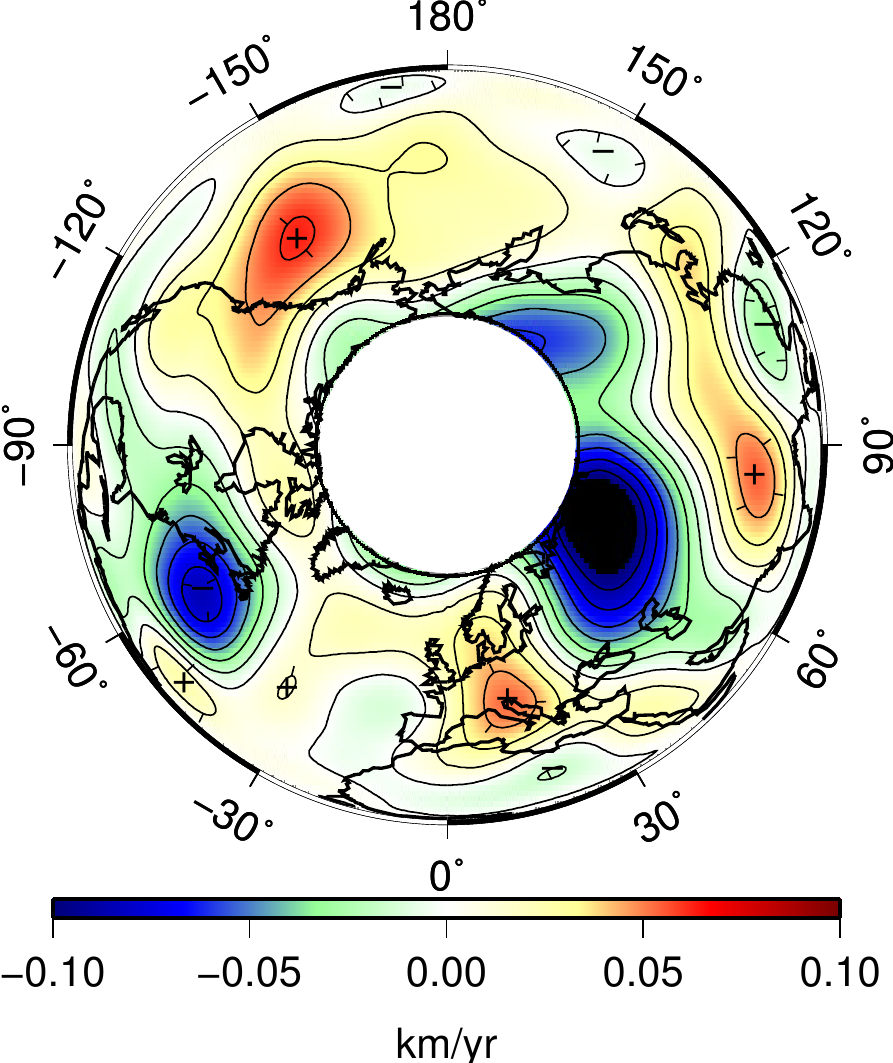}
      \end{minipage} 
      \begin{minipage}[c]{0.45\textwidth}
            \includegraphics[scale=0.6]{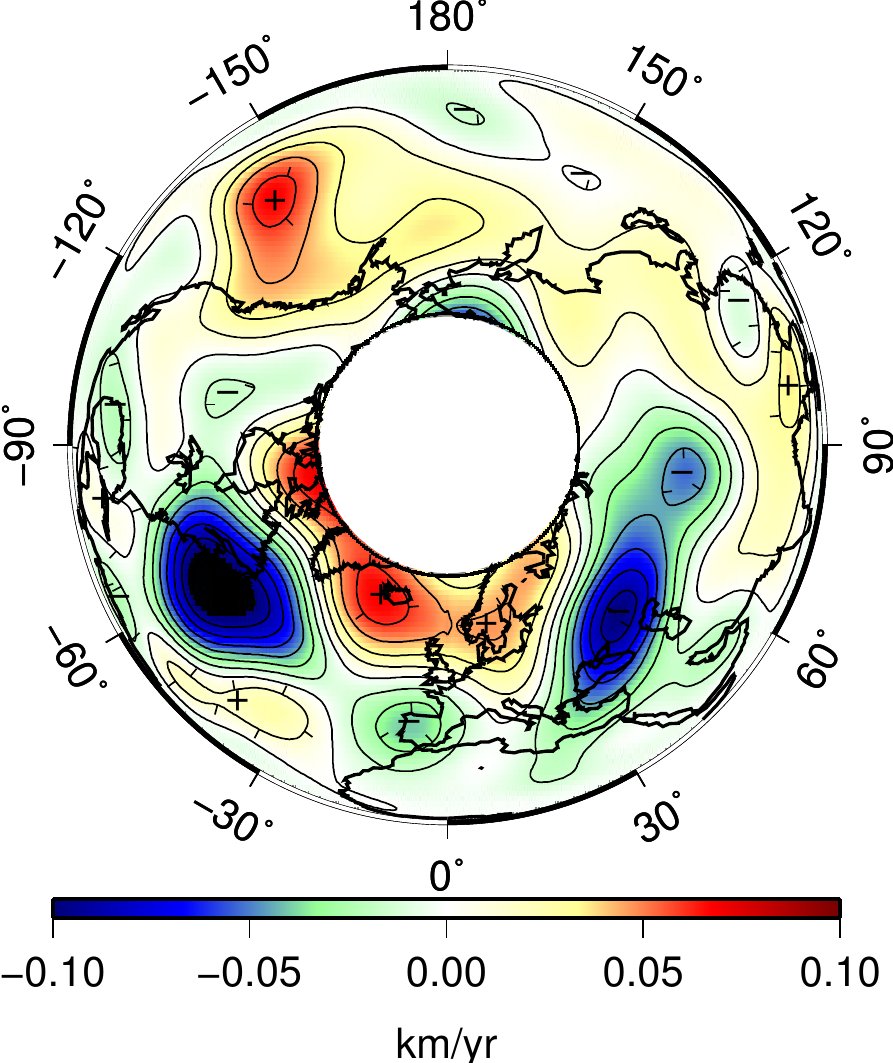}
      \end{minipage} 
      \\
       \begin{minipage}[c]{0.45\textwidth}
            \includegraphics[scale=0.6]{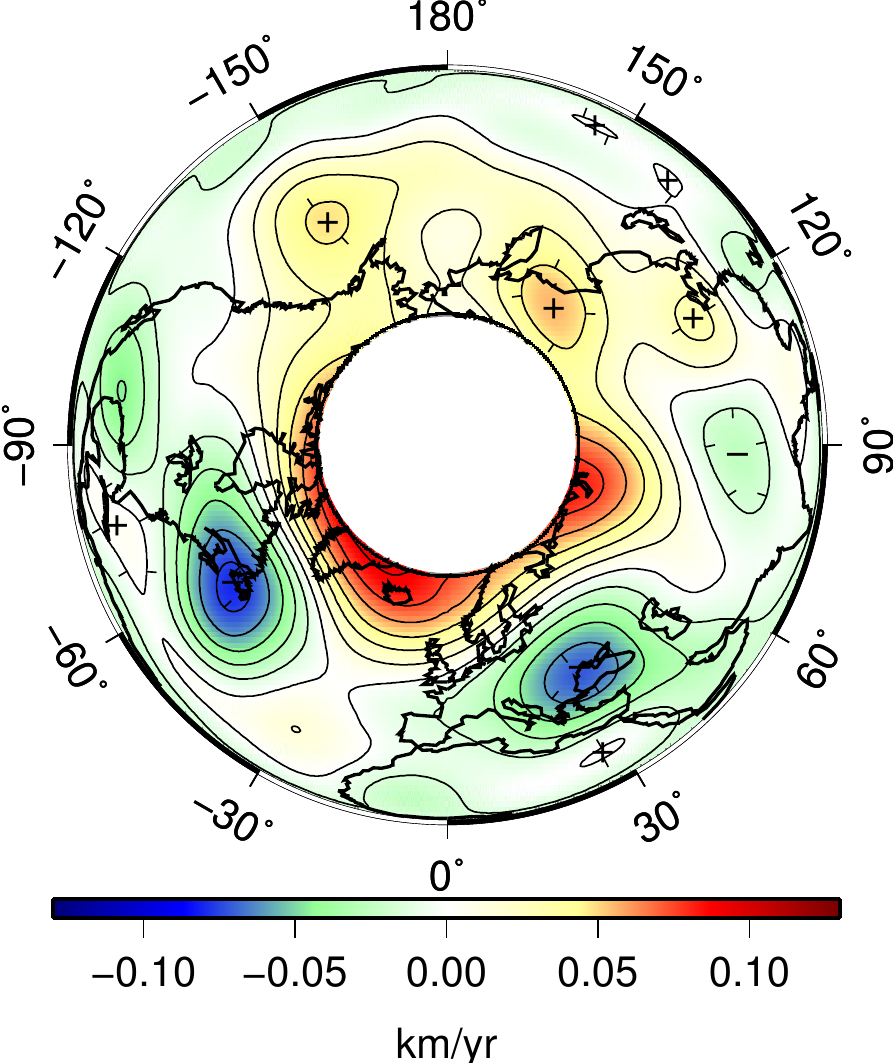}
      \end{minipage} 
      \begin{minipage}[c]{0.45\textwidth}
            \includegraphics[scale=0.6]{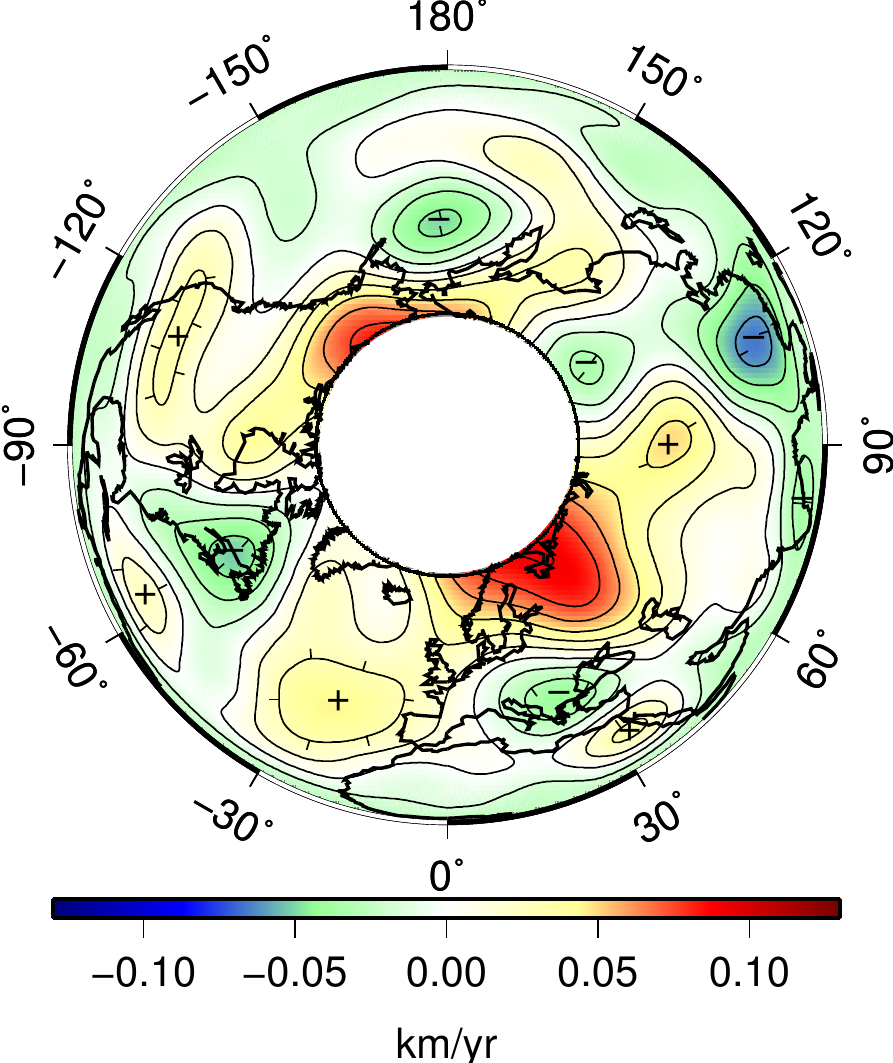}
      \end{minipage} 
      \\
       \begin{minipage}[c]{0.45\textwidth}
            \includegraphics[scale=0.6]{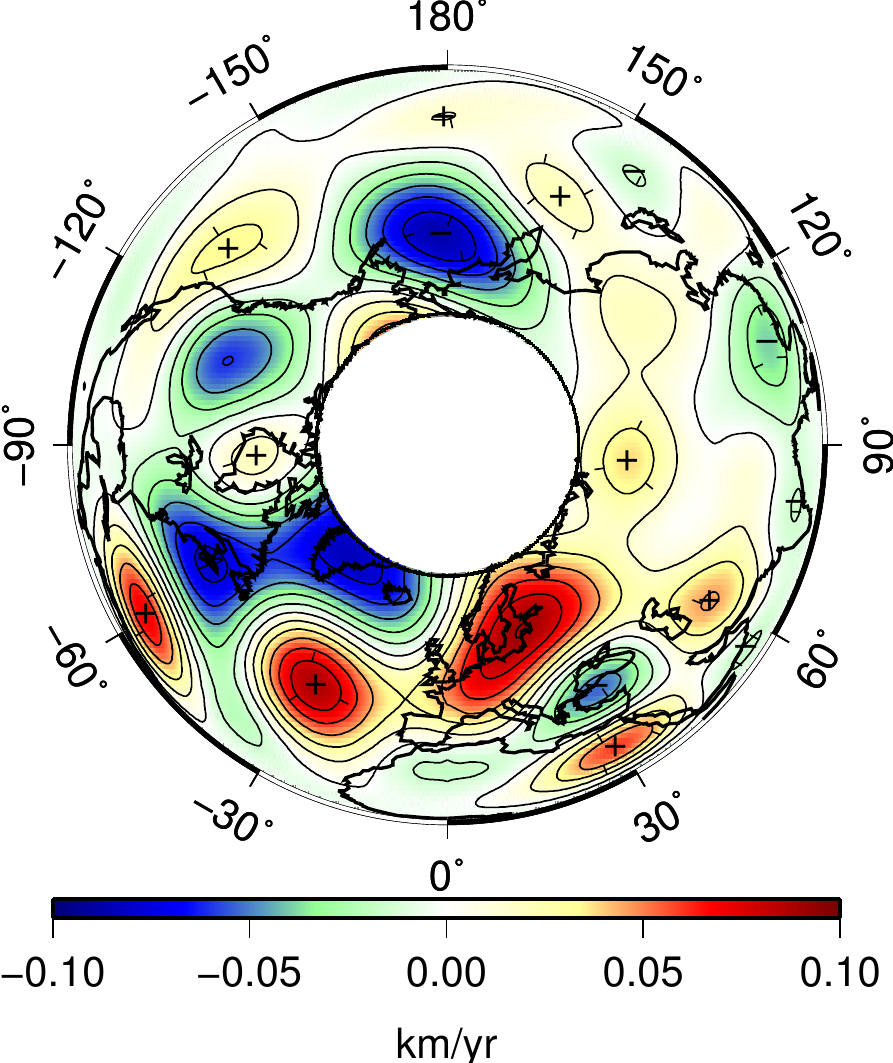}
      \end{minipage} 
      \begin{minipage}[c]{0.45\textwidth}
            \includegraphics[scale=0.6]{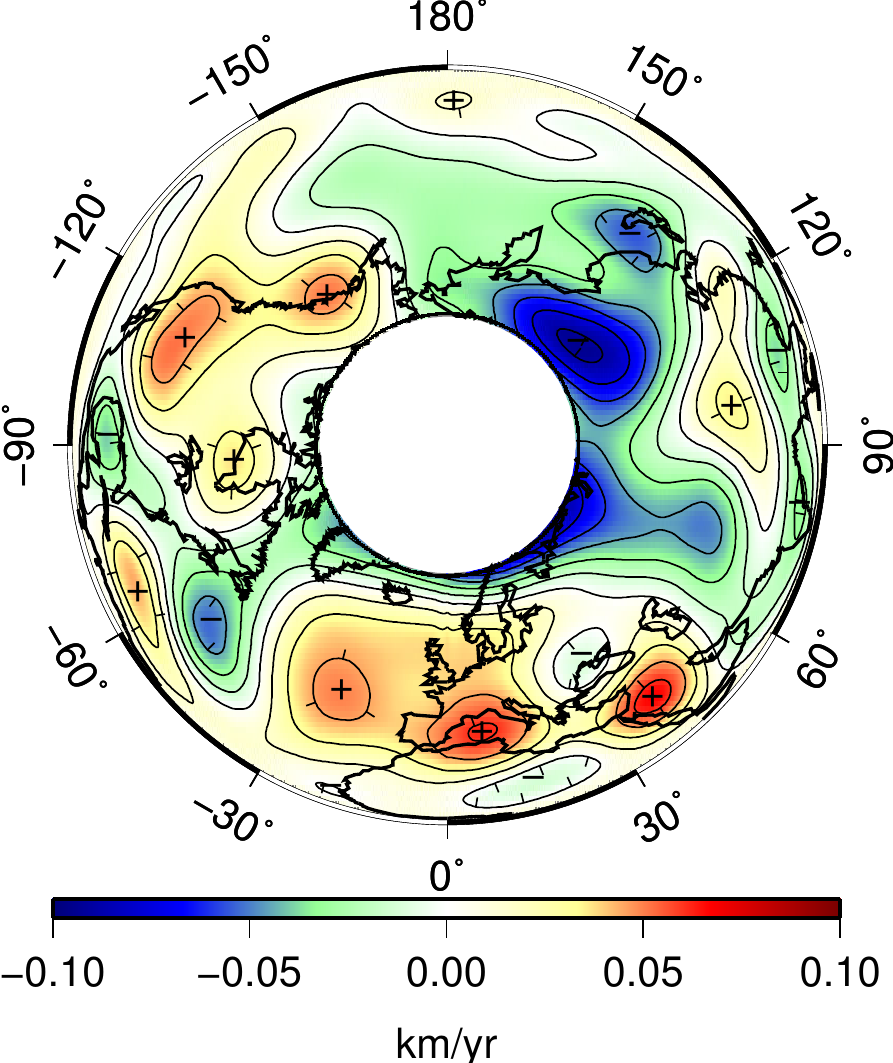}
      \end{minipage} 
 \caption{The three first EOFs resulting from PCA applied to {\it flow}$^{gufm1}$ (left) and {\it flow}$^{COV-OBS}$ in the 1840-1990 period (right): mode 1 (top), mode 2 (middle) and mode 3 (bottom). Charted values are directly those from columns of matrices $\mathbf{P}$ (see section \ref{secSEOF}) and are normalized such that $\mathbf{p}_i^T \mathbf{p}_j = \delta_{ij}$. Contours of $\xi / R_c$ streamfunction are used to visualize the flow, projected onto the equatorial plane, as seen from the North pole.}
\label{fig2PCA}
 \end{figure}

\begin{figure}
  \begin{center}
     \includegraphics[width=0.95\textwidth]{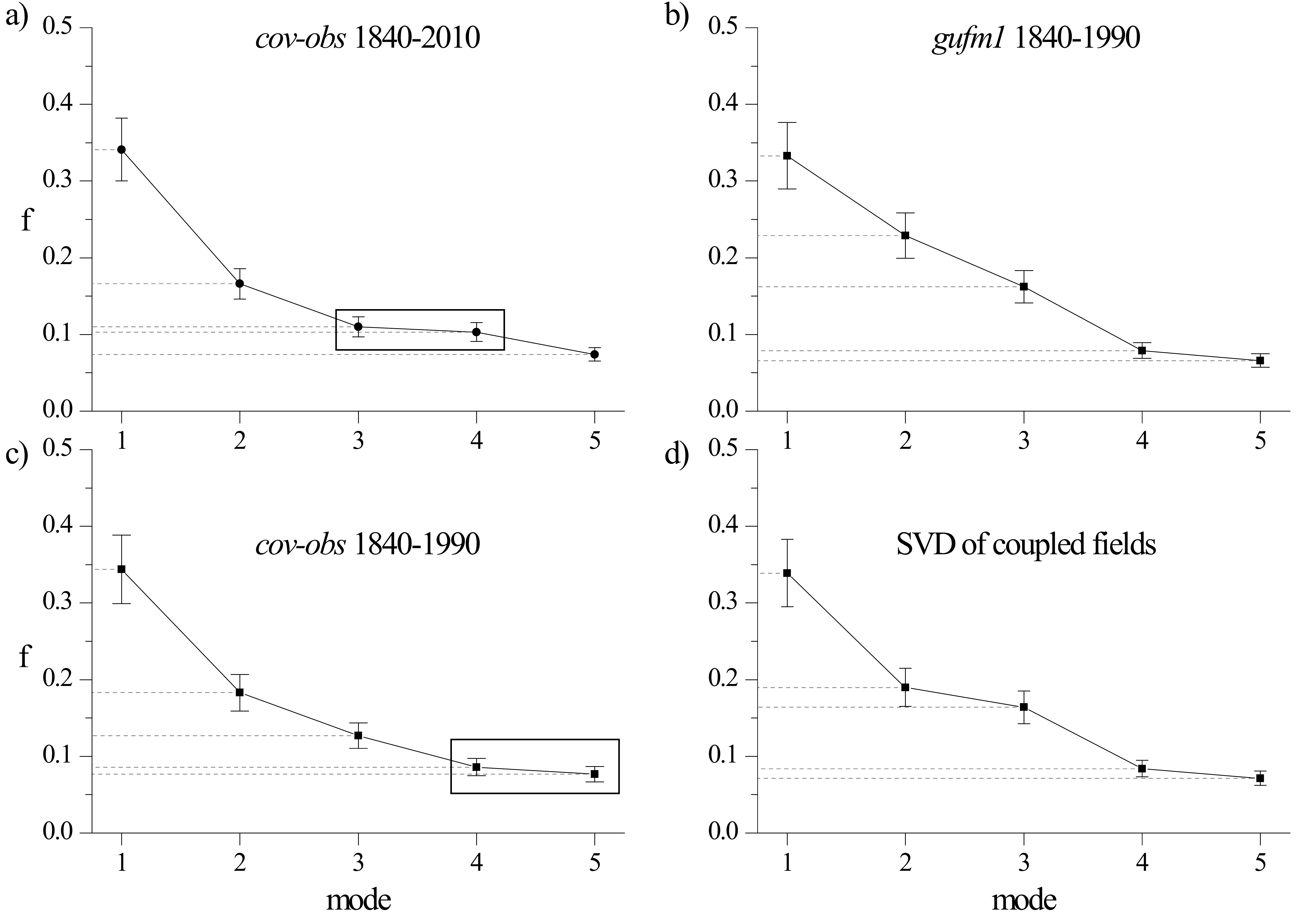}
  \end{center}
  \caption{Fraction of explained variablity $f_i$ for PCA applied to different datasets and for SVD of coupled ${\it flow}^{gufm1}$ and ${\it flow}^{COV-OBS}$ (on bottom, right), with error bars with total length 2$\delta f_i$, computed from eq. \ref{north_error}. Degenerate modes following North's criterion (eq. \ref{north}) are inside boxes}.
  \label{fig1-2}
\end{figure}

The first five PCA modes account for about 80\% of the flow total variability (see Table \ref{tab1}). The EOF patterns of the first three modes are shown in Figure \ref{fig2PCA}.
Results for {\it COV-OBS} considering the two time periods are similar. Note however that the first 3 modes represent a larger fraction of the signal for the shorter period. Besides, the 4th mode may not be completely separated from the 3rd when considering the longer time period. Comparing {\it COV-OBS} and {\it gufm1}, the  first mode explains very similar variances.
Using North's criterion (eq. \ref{north}) to detect mode degeneracy, modes 4 and 5 in {\it COV-OBS} (1840 - 1990) and modes 3 and 4 in {\it COV-OBS} (1840 - 2010) are not completely separated (see Figure \ref{fig1-2}), meaning that they may describe two aspects of a common structure. A simple illustration of  this effect is found in the example of a propagating wave that can be decomposed into two spatial patterns space-shifted by fourth of a wavelength, multiplied by two sinusoidal functions  time-shifted by fourth of a period. One unique structure (a propagating wave) would then appear in the PC analysis decomposed into two modes with exactly the same $f$ value.  In the case of the above-mentioned degenerate modes,  they should be considered together.

As a test for subdomain stability, the whole CMB domain was subdivided into two longitudinal hemispheres, and EOF/PC modes were recalculated for each of them. The chosen meridian for the separation goes through 70$^\circ$E and the two resulting hemispheres have longitudes 70$^\circ$E  to 250$^\circ$E (Pacific Hemisphere, PH)  and -110$^\circ$E to 70$^\circ$E (Atlantic Hemisphere, AH). Results are shown in Tables \ref{tab2} and \ref{tab2a}. The first five modes are still non-degenerate when computed independently for each hemisphere and according to North's criterion (eq. \ref{north}) (see Table \ref{tab2}). While the spatial and temporal descriptions of modes 4 and 5 can be different depending on if a global or an hemispherical grid of data values is used, the first three modes are recovered under AH with very close characteristics as in the global grid. The first three variability modes are not so well recovered using only data under the PH, especially for mode 2 (see Table \ref{tab2a}).

\begin{table}

                \begin{tabular}{ l r r r r r }
                  \hline \hline 
                  \, 	          &\multicolumn{5}{c}{Fraction of variance, $f_i$ [\%] ($\delta f_i$ [\%])}          \\
                  \cline{2-6}
                  & \textit{mode-1}  & \textit{mode-2}  & \textit{mode-3} & \textit{mode-4} & \textit{mode-5}  \\
                  \cline{1-6}       
                  {\it COV-OBS} (1840 - 1990), PH            & 29.7 (3.7)          & 19.3 (2.4)         & 16.2 (2.0)               & 9.5 (1.2)           & 7.7 (1.0)      \\

                  {\it COV-OBS} (1840 - 1990), AH            & 42.8 (5.5)          & 15.7 (2.0)        & 13.0 (1.7)               & 8.3 (1.1)           & 5.6 (0.7) \\

                  {\it gufm1} (1840 - 1990), PH              &  40.3 (5.0)          & 23.4 (3.0)         & 11.5 (1.5)               & 7.4 (0.9)          & 5.3 (0.7)   \\

                  {\it gufm1} (1840 - 1990), AH 		&  33.5 (4.3)            & 27.0 (3.4)       & 20.3 (2.6)             & 5.6 (0.7)          & 4.4 (0.6)   \\
                  \hline \hline
                \end{tabular}
                \caption{Subdomain stability of EOF/PCs for ${\it flow}^{gufm1}$ and ${\it flow}^{COV-OBS}$, when considering separately the Pacific and the Atlantic hemispheres. Percentage of variance explained by each mode and, in parenthesis, estimate of corresponding standard error, calculated as in Table  \ref{tab1}. }
                \label{tab2}
\end{table}

\begin{table}

                \begin{tabular}{ l r r r r r }
                  \hline \hline
                 \,          &\multicolumn{2}{c}{$|g(PH, global)|$ / $|r(PH, global)|$} &   \, & \multicolumn{2}{c}{$|g(AH, global)|$ / $|r(AH, global)|$} \\
                  \cline{2-6}    
                  & {\it gufm1} & {\it COV-OBS} & \, & {\it gufm1} & {\it COV-OBS}  \\            
                  \cline{1-6}         
                 \textit{mode-1}           & {\bf 0.99} / {\bf 0.95}         & {\bf 0.87}  / 0.77 & \,        	&  {\bf 0.94} / {\bf 0.95}  			& {\bf 0.98}  / {\bf 0.98}       \\

                  \textit{mode-2}          & {\bf 0.88} / 0.77         & 0.62 / 0.65  & \,             & {\bf 0.93} / {\bf 0.94} 	  		& {\bf 0.93} / {\bf 0.96}         \\

                  \textit{mode-3}          & {\bf 0.85} / {\bf 0.81}         & {\bf 0.96} / {\bf 0.90} & \,              & {\bf 0.96} / {\bf 0.97}    		& {\bf 0.92} / {\bf 0.90}         \\

                  \textit{mode-4}          & 0.25 / 0.30         & 0.68 / 0.79  & \,             & 0.72 / 0.70    		& 0.39 / 0.34         \\

                  \textit{mode-5}          & 0.25 / 0.32         & {\bf 0.87} / {\bf 0.85}  & \,             & 0.73 / {\bf 0.84}    		& 0.21 / 0.28         \\
                 \hline \hline
                \end{tabular}
               \caption{Subdomain stability of EOF/PCs for ${\it flow}^{gufm1}$ and ${\it flow}^{COV-OBS}$, when considering separately the Pacific and the Atlantic hemispheres. Absolute values of congruence coefficients between corresponding EOFs computed using the global grid and the grid under only the Pacific or the Atlantic hemispheres ($|g(PH, global)|$ and $|g(AH, global)|$, respectively). Absolute values of correlation coefficients between corresponding PCs computed using the global grid and the grid under Pacific or Atlantic  hemispheres ($|r(PH, global)|$ and $|r(AH, global)|$, respectively). Bold is used for $g$ and $r$ values above 0.8, indicating high similarity.}   
                \label{tab2a}               

\end{table}

Overall, when applying PCA to {\it gufm1} and {\it COV-OBS} separately, some significant differences come out: (i) mode 2 is more important to explain the variability of {\it flow}$^{gufm1}$ than of {\it flow}$^{COV-OBS}$, possibly due to a less clear expression of mode 2 under the PH; (ii) also, differences between recovered modes in PH and AH hemispheres are more important for {\it flow}$^{COV-OBS}$ than for {\it flow}$^{gufm1}$; (iii) most importantly, corresponding modes in {\it flow}$^{gufm1}$ and {\it flow}$^{COV-OBS}$ exhibit significant spatial and temporal differences in general, as shown by relatively low values of congruence and correlation coefficients in square brackets in Table \ref{tab3} (see also Fig \ref{fig2PCA}).

\begin{table}
         
                \begin{tabular}{l r r r }
                  \hline \hline
                 \,         &      $f_i$ [\%] ($\delta f_i$ [\%]) &  $| g(EOF_i^{\it gufm1},EOF_i^{\it COV-OBS}) | $  & $| r(PC_i^{\it gufm1},PC_i^{\it COV-OBS})| $  \\
                  \cline{1-4}       
                 \textit{mode-1}            & 33.9 (4.4)            & 0.80 [0.56]       & 0.96 [0.75]                  \\

                 \textit{mode-2}           & 19.0 (2.5)          & 0.74 [0.45]          & 0.93 [0.40]                \\

                  \textit{mode-3}               & 16.4 (2.1)             & 0.62 [0.24]         & 0.94 [0.38]                  \\

                  \textit{mode-4}               & 8.4 (1.1)             & 0.61 [0.63]          & 0.92 [0.75]                  \\

                  \textit{mode-5}               & 7.1 (0.9)             & 0.47 [0.31]         & 0.91 [0.70]                 \\
                \hline \hline
                \end{tabular}  
                \caption{SVD of coupled flows ${\it flow}^{gufm1}$ and ${\it flow}^{COV-OBS}$. In the first column, values for the percentage of cross-covariance explained by each coupled mode and, in parenthesis, estimates of corresponding standard error according to eq. \ref {north_error}. Second and third columns are for congruence (eq. \ref{coefg}) and correlation coefficients that compare corresponding EOFs and PCs, respectively, that reconstruct ${\it flow}^{gufm1}$ and ${\it flow}^{COV-OBS}$. In square brackets, values comparing corresponding EOFs and PCs from independent PCA applied to ${\it flow}^{gufm1}$ and ${\it flow}^{COV-OBS}$.}     
                \label{tab3}               

\end{table}

\begin{figure}
\centering
      \begin{minipage}[c]{0.45\textwidth}
            \includegraphics[scale=0.6]{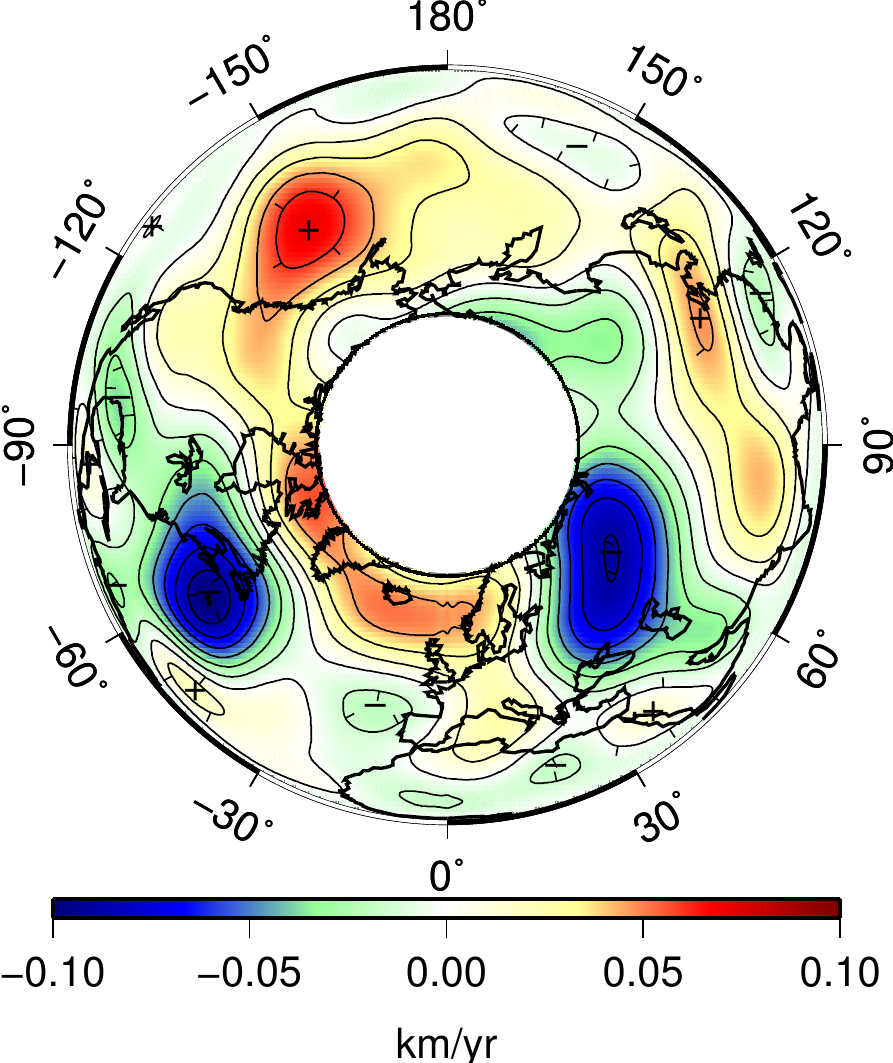}
      \end{minipage} 
      \begin{minipage}[c]{0.45\textwidth}
            \includegraphics[scale=0.6]{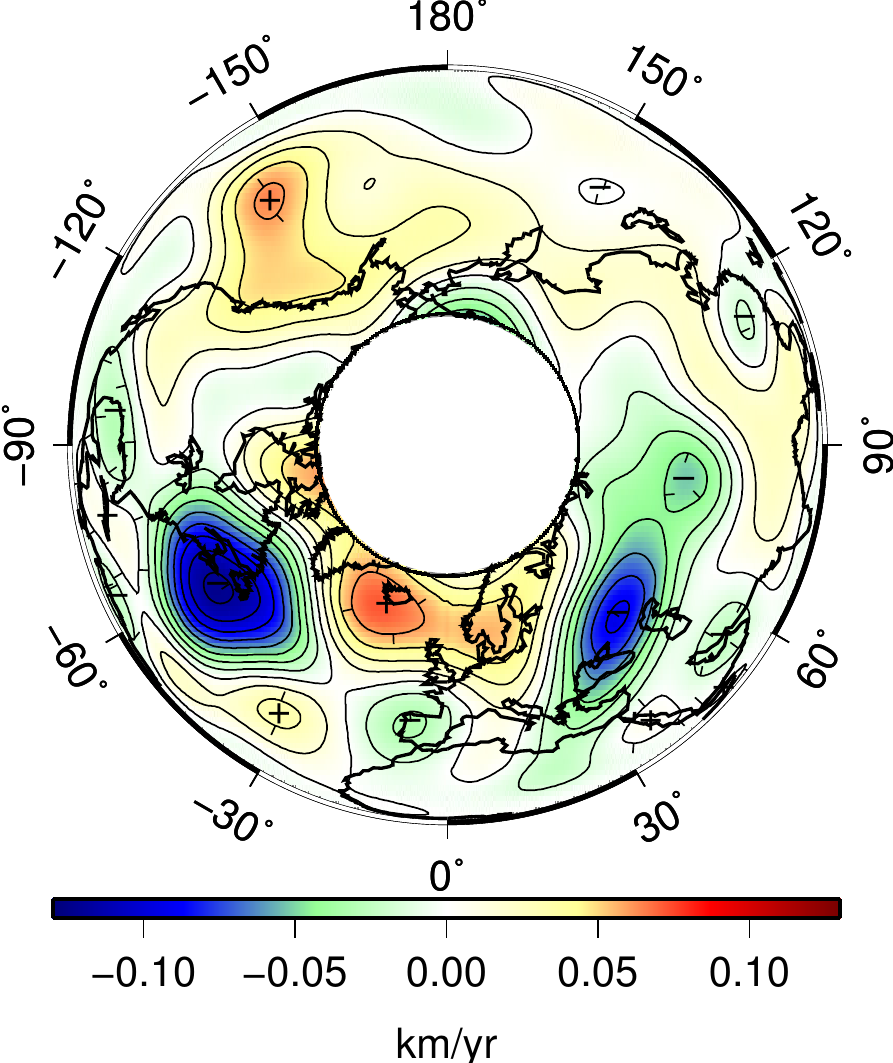}
      \end{minipage} 
      \\
       \begin{minipage}[c]{0.45\textwidth}
            \includegraphics[scale=0.6]{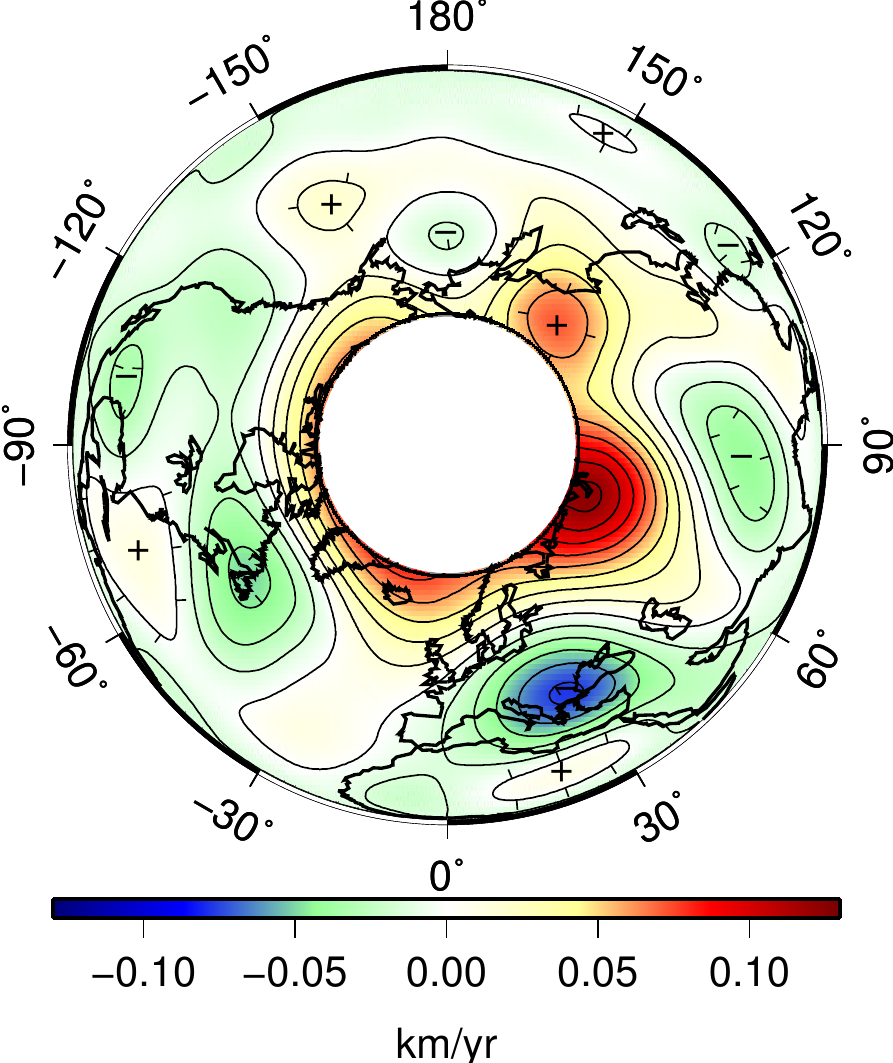}
      \end{minipage} 
      \begin{minipage}[c]{0.45\textwidth}
            \includegraphics[scale=0.6]{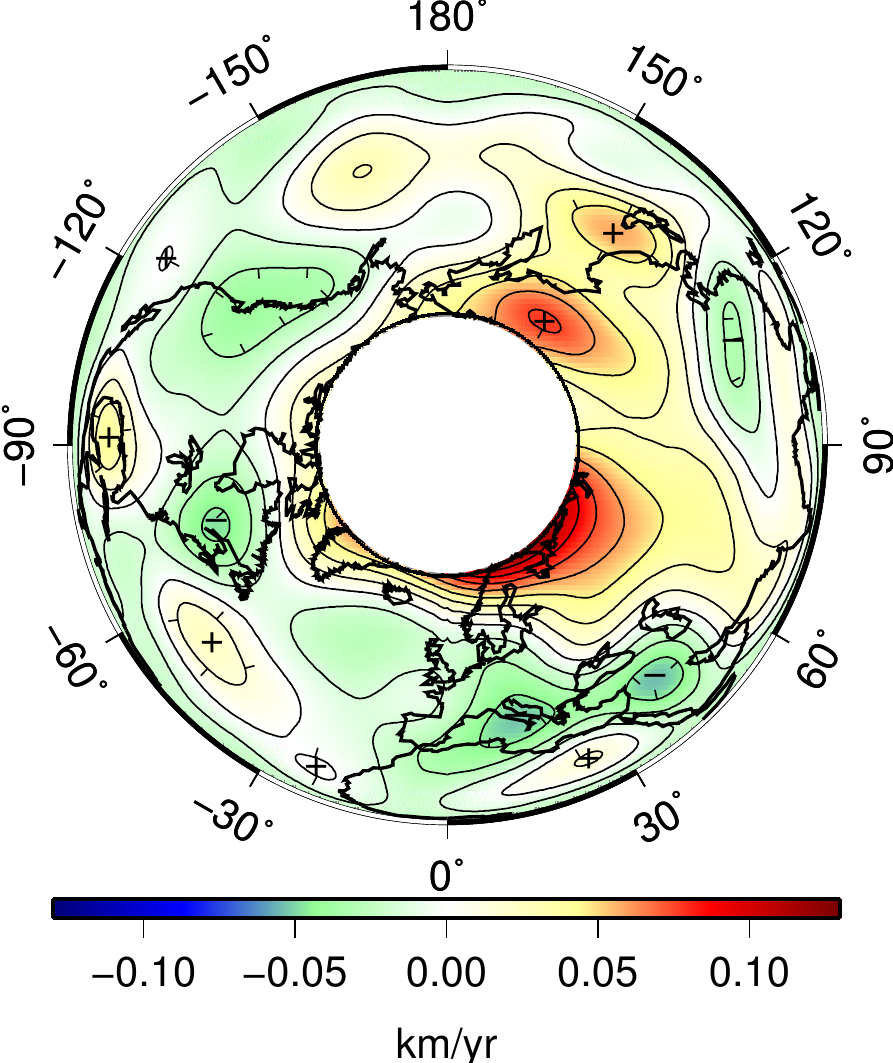}
      \end{minipage} 
      \\
       \begin{minipage}[c]{0.45\textwidth}
            \includegraphics[scale=0.6]{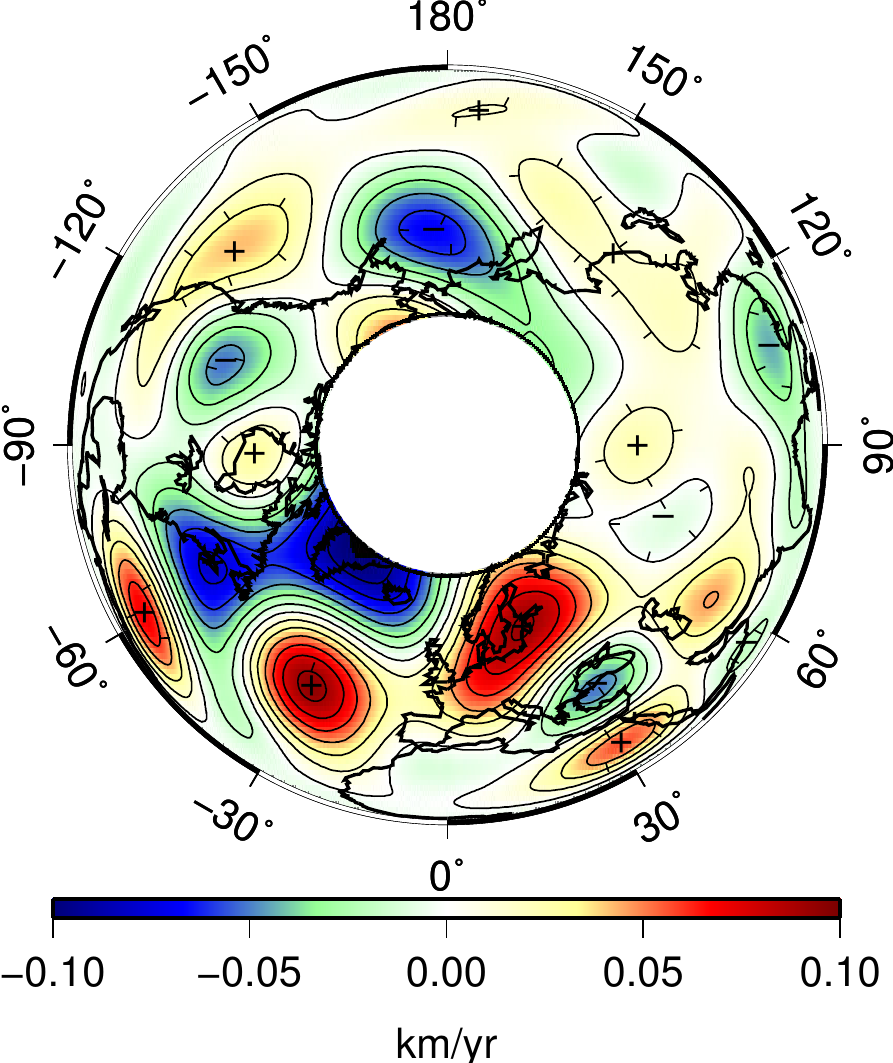}
      \end{minipage} 
      \begin{minipage}[c]{0.45\textwidth}
            \includegraphics[scale=0.6]{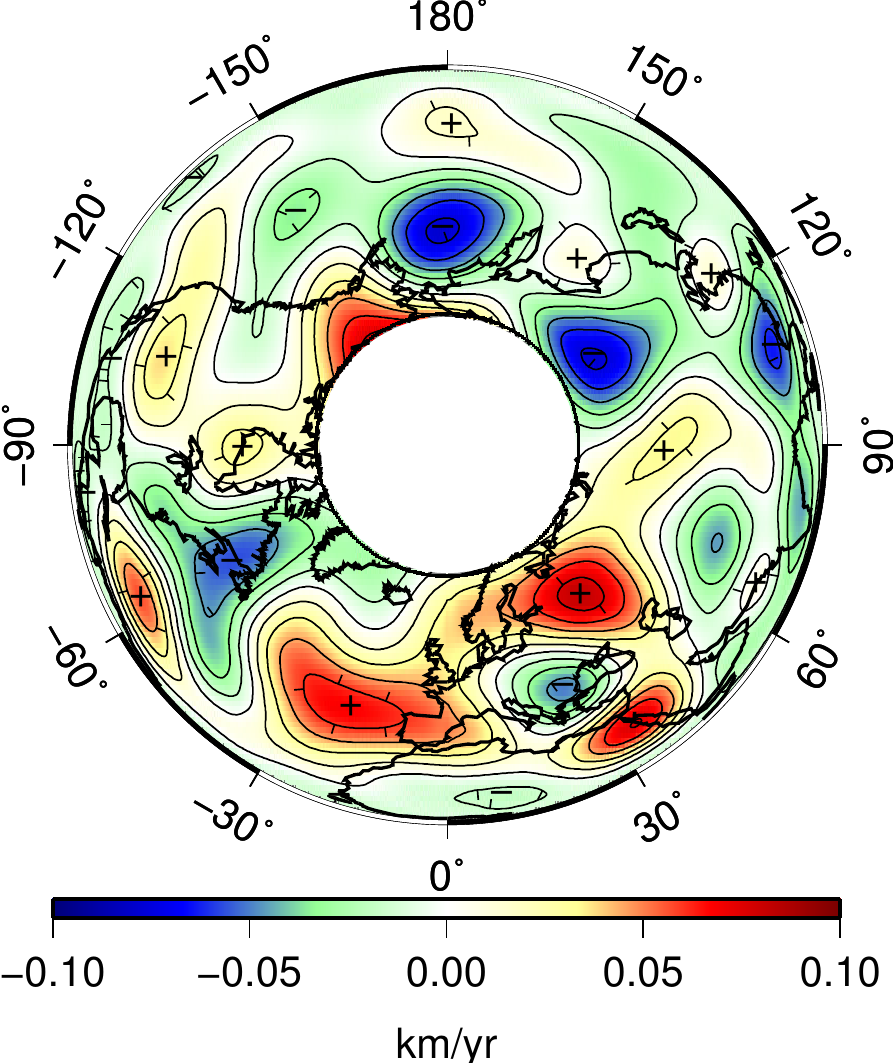}
      \end{minipage} 
 \caption{The three first pairs of EOFs resulting from SVD of coupled fields applied to {\it flow}$^{gufm1}$ and {\it flow}$^{COV-OBS}$. For mode 1 (top), mode 2 (middle) and mode 3 (bottom), the EOFs allowing to reconstruct {\it flow}$^{gufm1}$ (left) and {\it flow}$^{COV-OBS}$ (right). Values are directly those from columns of matrices $\mathbf{U}$ and $\mathbf{V}$ (see section \ref{secSVD}) and are normalized such that $\mathbf{u}_i^T \mathbf{u}_j = \mathbf{v}_i^T \mathbf{v}_j = \delta_{ij}$. Contours of $\xi / R_c$ streamfunction are used to visualize the flow, projected onto the equatorial plane, as seen from the North pole.}
\label{fig2}
 \end{figure}

\begin{figure}
  \begin{center}
     \includegraphics[width=0.95\textwidth]{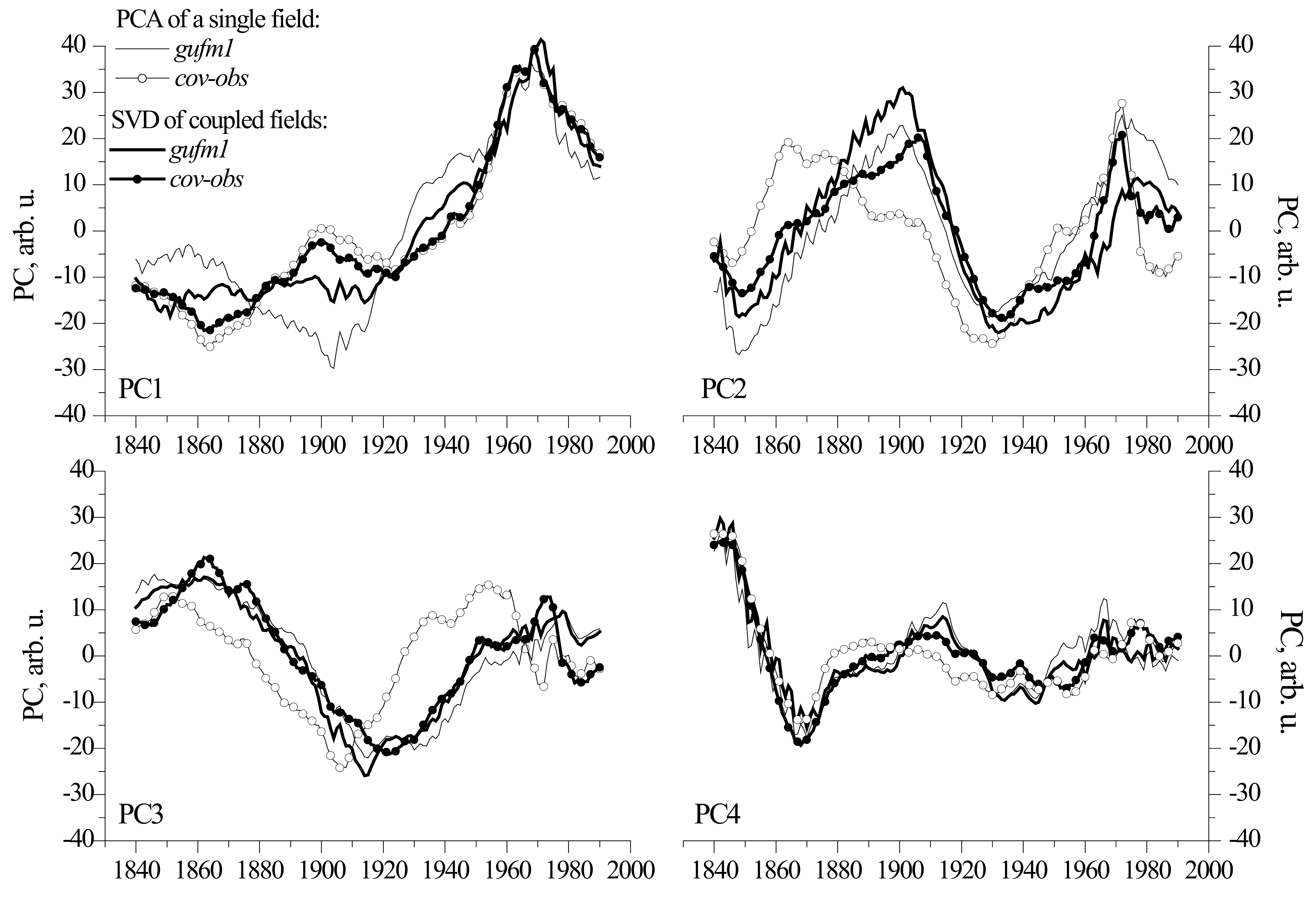}
  \end{center}
  \caption{Temporal functions representing: the expansion coefficients of PCA modes that reconstruct {\it flow}$^{gufm1}$ (thin solid line) and {\it flow}$^{COV-OBS}$ (thin line with empty circles); the expansion coefficients of SVD coupled modes that reconstruct {\it flow}$^{gufm1}$ (thick solid line) and {\it flow}$^{COV-OBS}$ (thick line with filled circles).}
  \label{fig4}
\end{figure}

\subsection{SVD of coupled {\it gufm1} and {\it COV-OBS} }
\label{secSVD2}

In order to identify structures correlated in both flows, we apply SVD of coupled fields to {\it flow}$^{gufm1}$ and {\it flow}$^{COV-OBS}$ for the common period 1840-1990 (see Table \ref{tab3}).

In Figure \ref{fig2} we represent the contours of streamfunction $\xi$ for the three first coupled modes.
As all modes show important spatial structures at medium-high latitudes, where visualization of the flow through $\xi$ contours is quite straightforward, the representation adopted uses these contours projected onto the equatorial plane as seen from the North pole. Each pattern EOF$_i$ contributes to the global flow with the flow circulation retrieved directly  from Fig. \ref{fig2} in epochs of positive PC$_i$ and opposite circulation in epochs of negative PC$_i$.
We easily recognize common main structures in the pair of EOFs for the first two modes, namely three large rolls at medium-high latitudes for mode 1, two anti-cyclonic under the AH and one cyclonic under the PH, and an important zonal component around the TC in mode 2.
Mode 2 also presents a strong zonal component at the equator, which is not apparent in Figure \ref{fig2} because of the chosen projection (see however Fig. \ref{fig7} and the Discussion in section \ref{discussion}).
Special care must be taken in interpreting the EOF$_2$ pattern in Figure \ref{fig2}. Indeed, the streamfunction $\xi$ has an important negative zonal component at the equatorial region which contributes to the flow through the first term in the RHS of eq. \ref{eq1}, but not through the second term where only nonzonal components of $\xi$ appear. Then as $\mathbf{\nabla}_H \xi$ points radially inwards, the corresponding flow circulation is eastward (in Fig. \ref{fig2}), the same direction as around the TC. The relative sign of the $\xi$ streamfunction of mode 2 near the TC and at the equator appears in Figure \ref{fig5} and the common direction of the zonal longitudinal flow appears in Figure \ref{fig7}.
The two coupled spatial functions of mode 3 differ more than for modes 1 and 2 (Table \ref{tab3}).
In particular, EOF$_3$ for {\it flow}$^{gufm1}$ (Fig. \ref{fig2} bottom left) is mainly localized under the Atlantic, a feature not seen in the EOF$_3$ that reconstructs {\it flow}$^{COV-OBS}$ (Fig. \ref{fig2} bottom right).
There are nonetheless some corresponding features in these EOF$_3$ charts: a cyclone under West of Iberia and another under East of the Caribbean Sea, an anticyclone under Eastern Canada and another under the Bering Sea.

As to the expansion coefficient time series (PCs), notice that for a given mode the {\it flow}$^{gufm1}$ and {\it flow}$^{COV-OBS}$ PCs for the SVD analysis are always closer to each other than the PC's of separated EOF/PC analysis of {\it flow}$^{gufm1}$ and {\it flow}$^{COV-OBS}$ (see Figure \ref{fig4}).
In addition, for mode 1 (respectively mode 2), the expansion coefficients obtained by SVD are closer to the one obtained by PCA of {\it flow}$^{COV-OBS}$ (respectively {\it flow}$^{gufm1}$).
This ability of SVD to bring the two flow descriptions closer together is also observed for spatial structures: the EOFs for the SVD analysis shown in Figure \ref{fig2} are closer together than the corresponding EOFs for PCA analysis of individual flows (see Figure \ref{fig2PCA}). This result is also apparent in Table \ref{tab3}, from comparison of values inside and outside square brackets for congruence (spatial similarity) and correlation (temporal resemblance) coefficients. 

A quasi-periodicity is seen in the temporal variation of mode 2. Given the importance of this mode in terms of data variability explained and significance of its spatial features close to the TC and at the equator, we further characterize this temporal behaviour.
To this end, all $PC_2$s obtained from PCA applied to {\it flow}$^{gufm1}$ and {\it flow}$^{COV-OBS}$ separately, both in a global grid and hemispherical subdomain grids, together with $PC_2$s obtained from SVD of coupled flows, in a total of 9 time series, were fitted with a single sinusoid where amplitude, frequency and initial phase were free parameters.
An average and a standard deviation of the estimated period $T$ for the ensemble PC$_2s$ were produced. When using the whole 1840-1990 dataset we found $T=90.5 \pm 4.6$ yr and, when using the more recent dataset for 1900-1990 for which errors in geomagnetic field models are smaller, we found $T=80.1 \pm 2.7$ yr.

\begin{figure}
\begin{center}
            \includegraphics[width=0.24\textwidth]{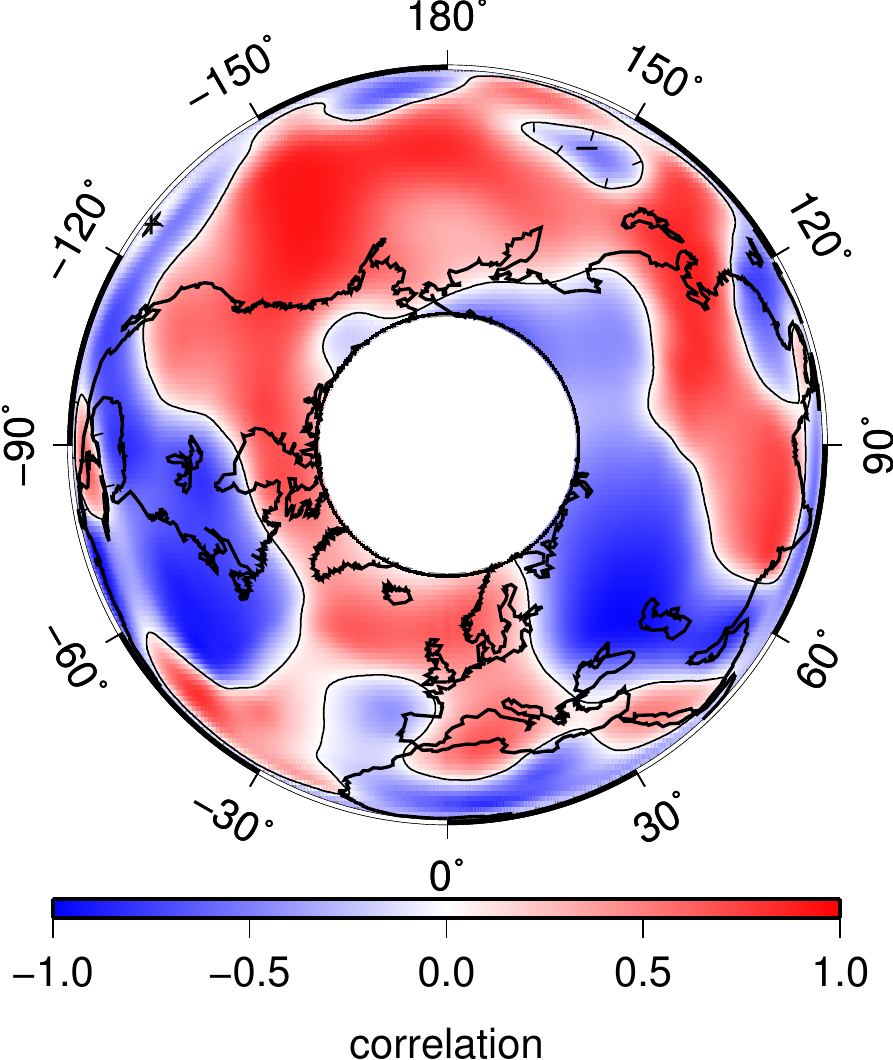}
            \includegraphics[width=0.24\textwidth]{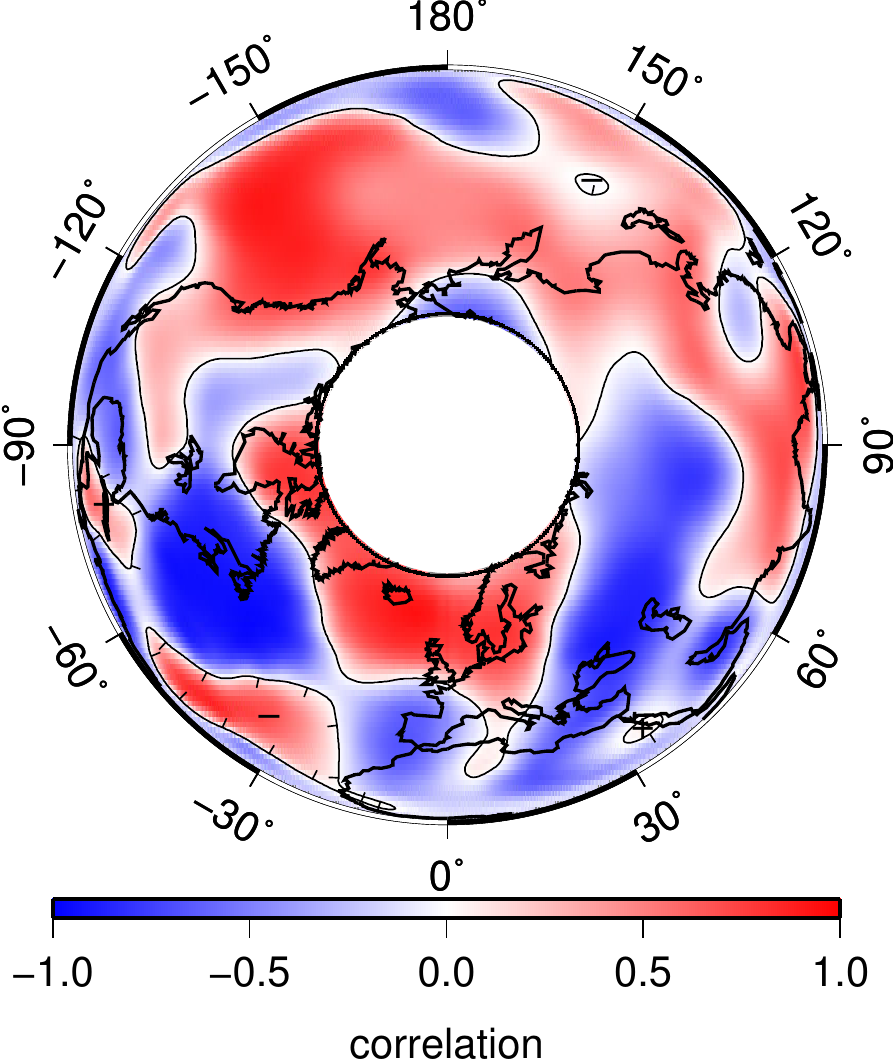}
            \includegraphics[width=0.24\textwidth]{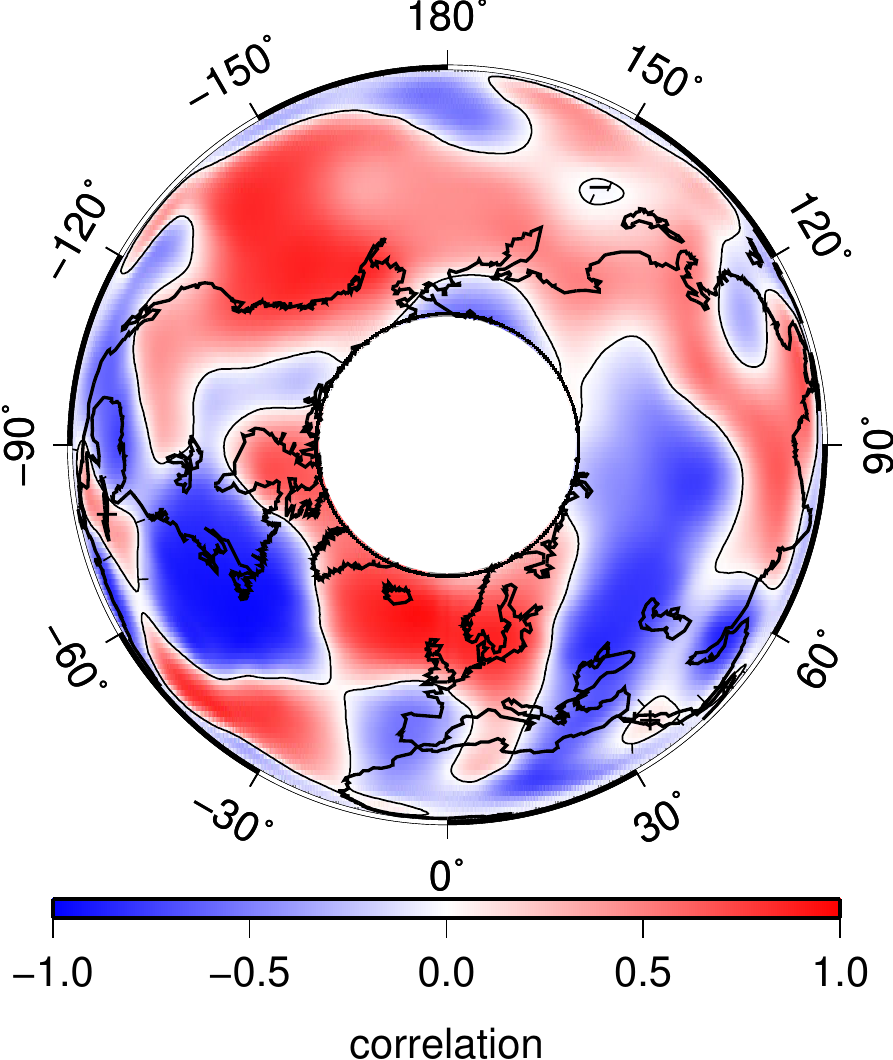}
            \includegraphics[width=0.24\textwidth]{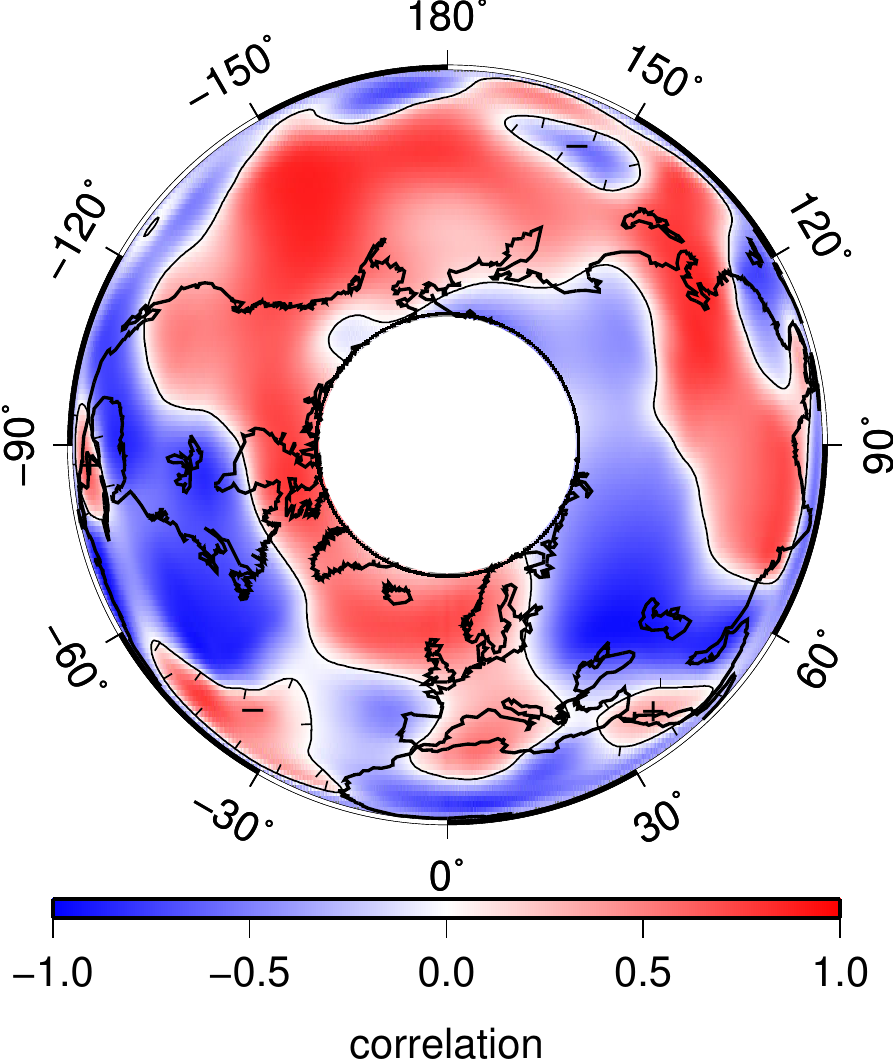}
      \\
            \includegraphics[width=0.24\textwidth]{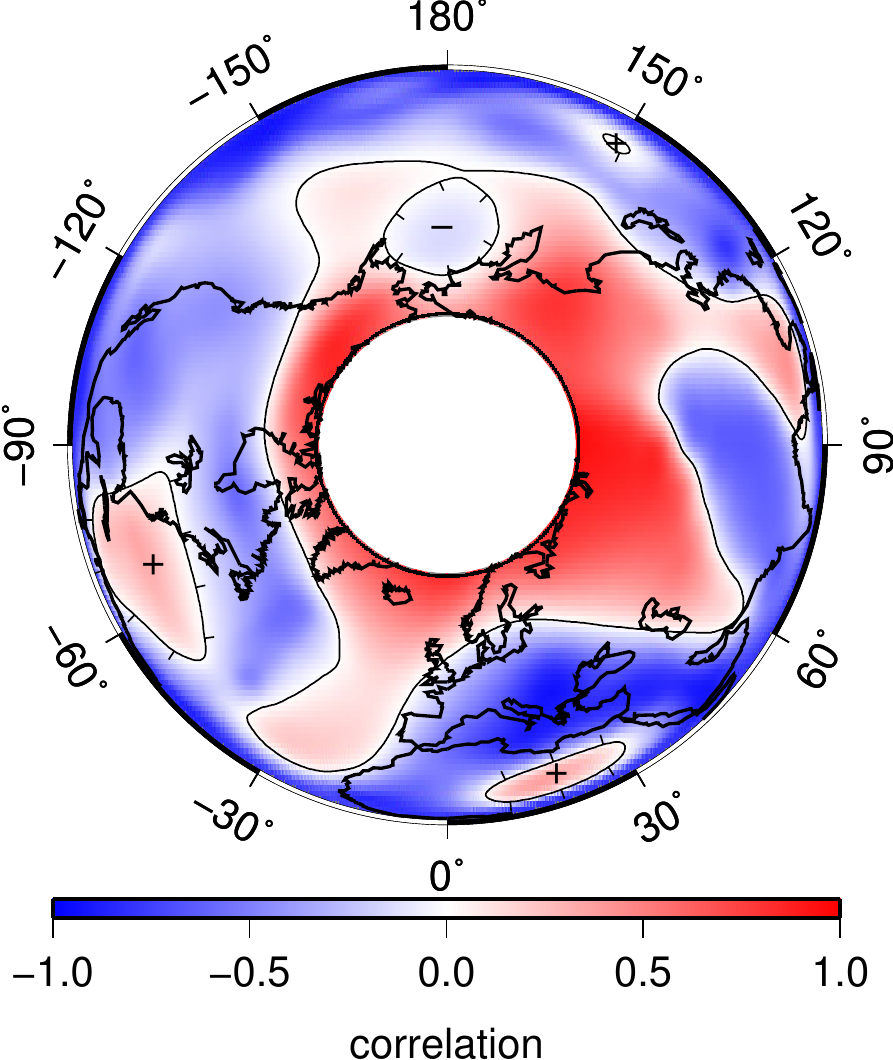}
            \includegraphics[width=0.24\textwidth]{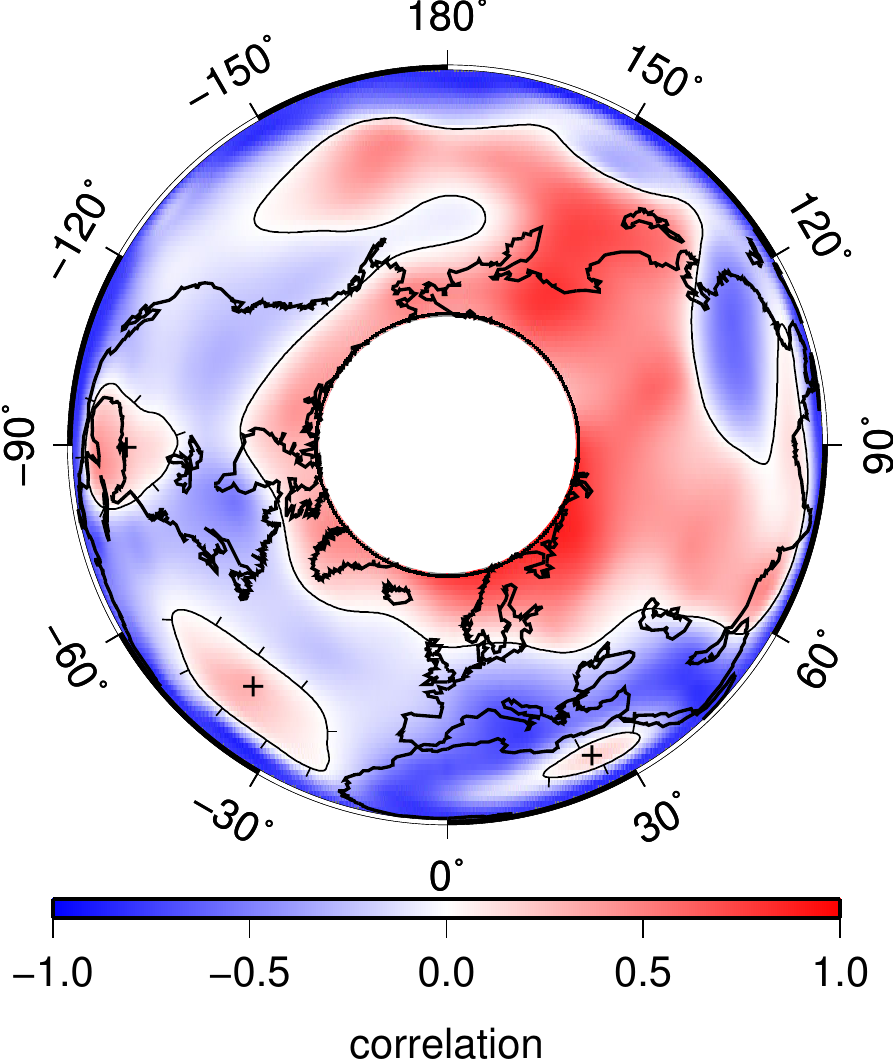}
            \includegraphics[width=0.24\textwidth]{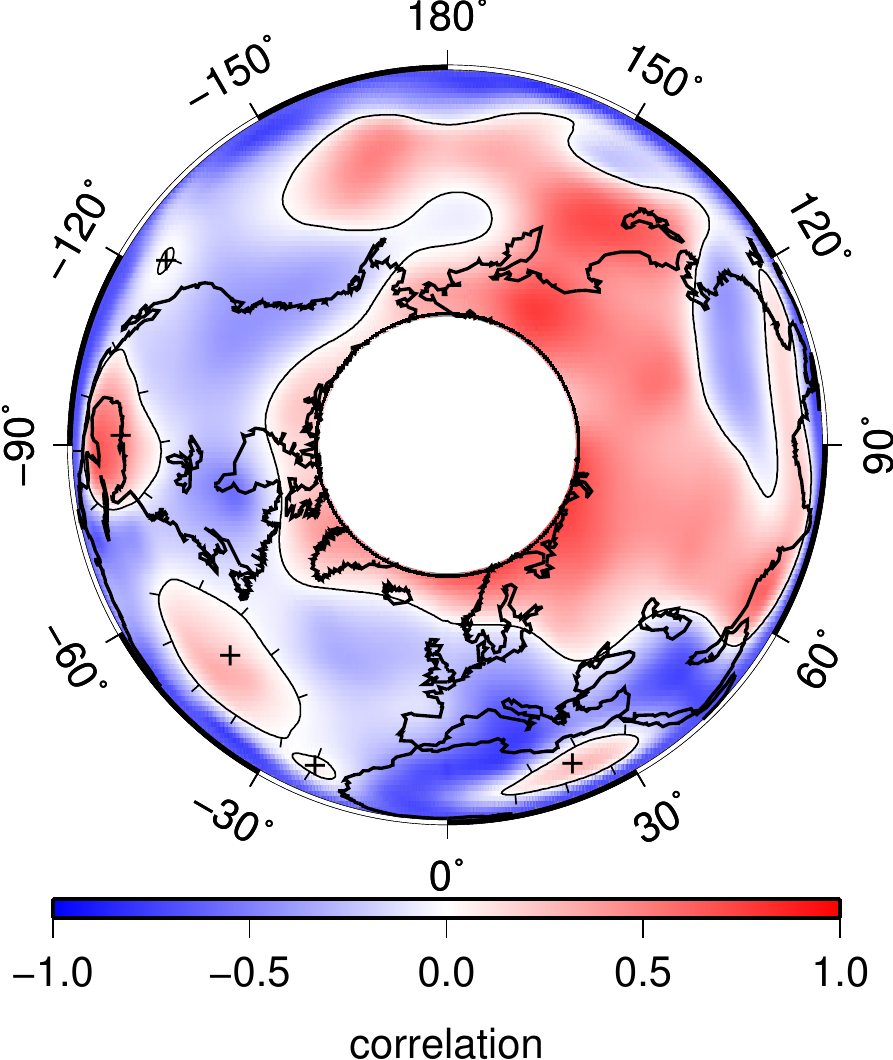}
            \includegraphics[width=0.24\textwidth]{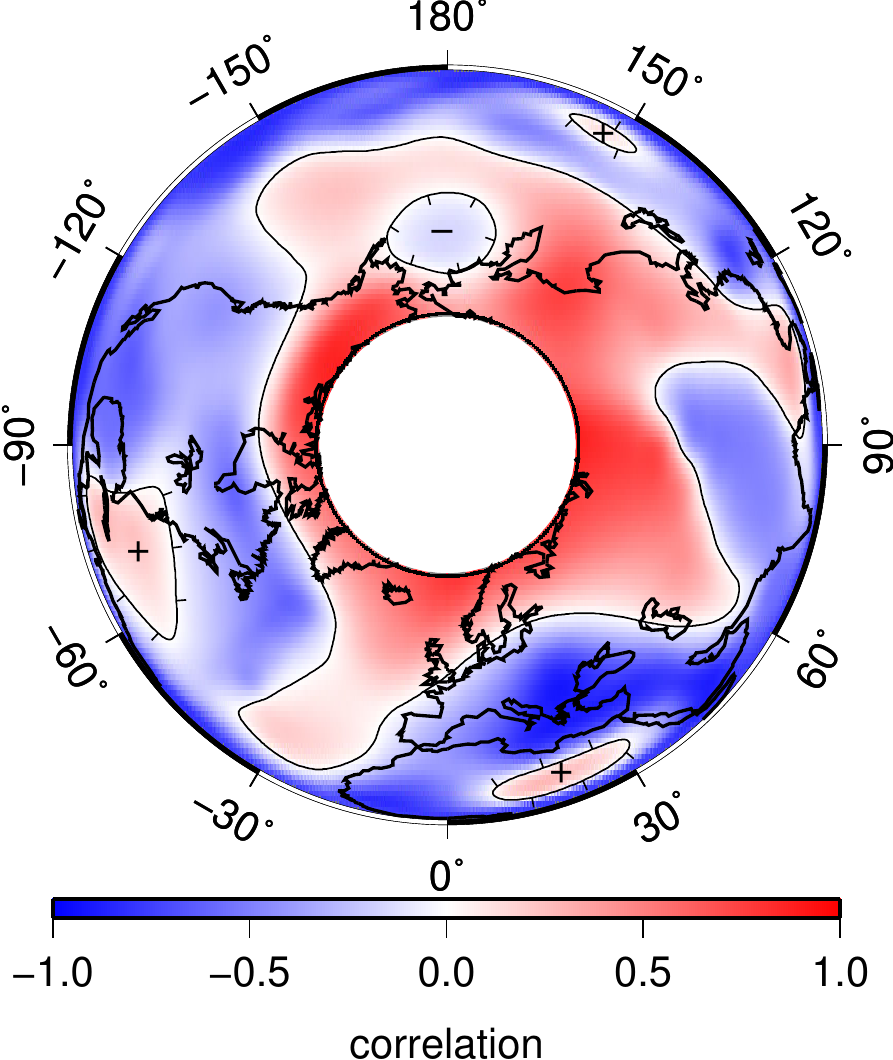}
 \end{center}
 \caption{Homogeneous (two left columns, first for PC$_i^{gufm1}$ correlated with {\it flow}$^{gufm1}$, second for PC$_i^{COV-OBS}$ correlated with {\it flow}$^{COV-OBS}$) and heterogeneous (two right columns, first for PC$_i^{gufm1}$ correlated with {\it flow}$^{COV-OBS}$, second for PC$_i^{COV-OBS}$ correlated with {\it flow}$^{gufm1}$) correlation maps for SVD modes 1 (top row) and 2 (bottom row). Results are projected onto the equatorial plane and seen from the North pole.}
\label{fig3}
 \end{figure}

\begin{figure}
\begin{center}
            \includegraphics[width=0.95\textwidth]{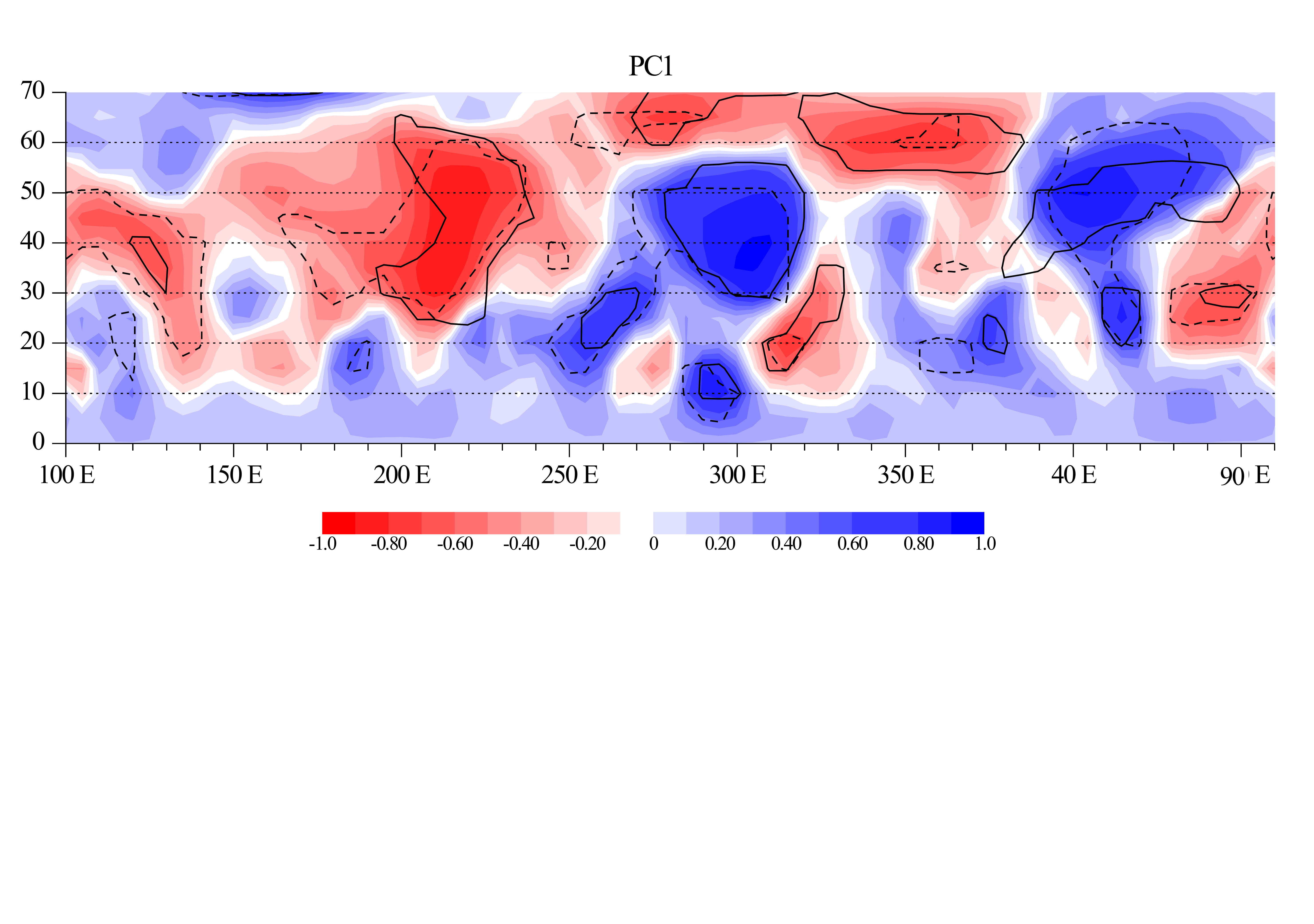}
            \includegraphics[width=0.95\textwidth]{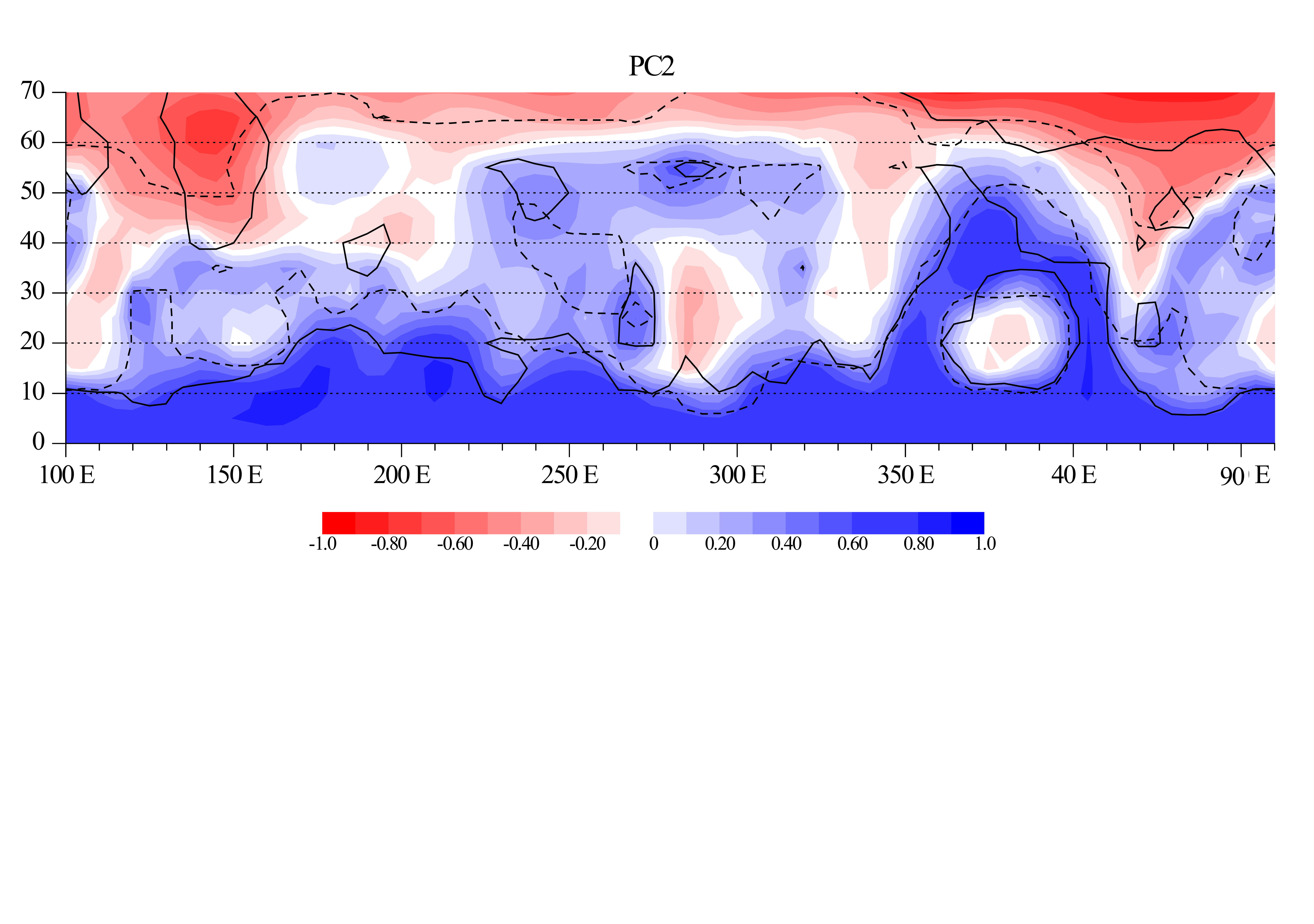}
            \includegraphics[width=0.95\textwidth]{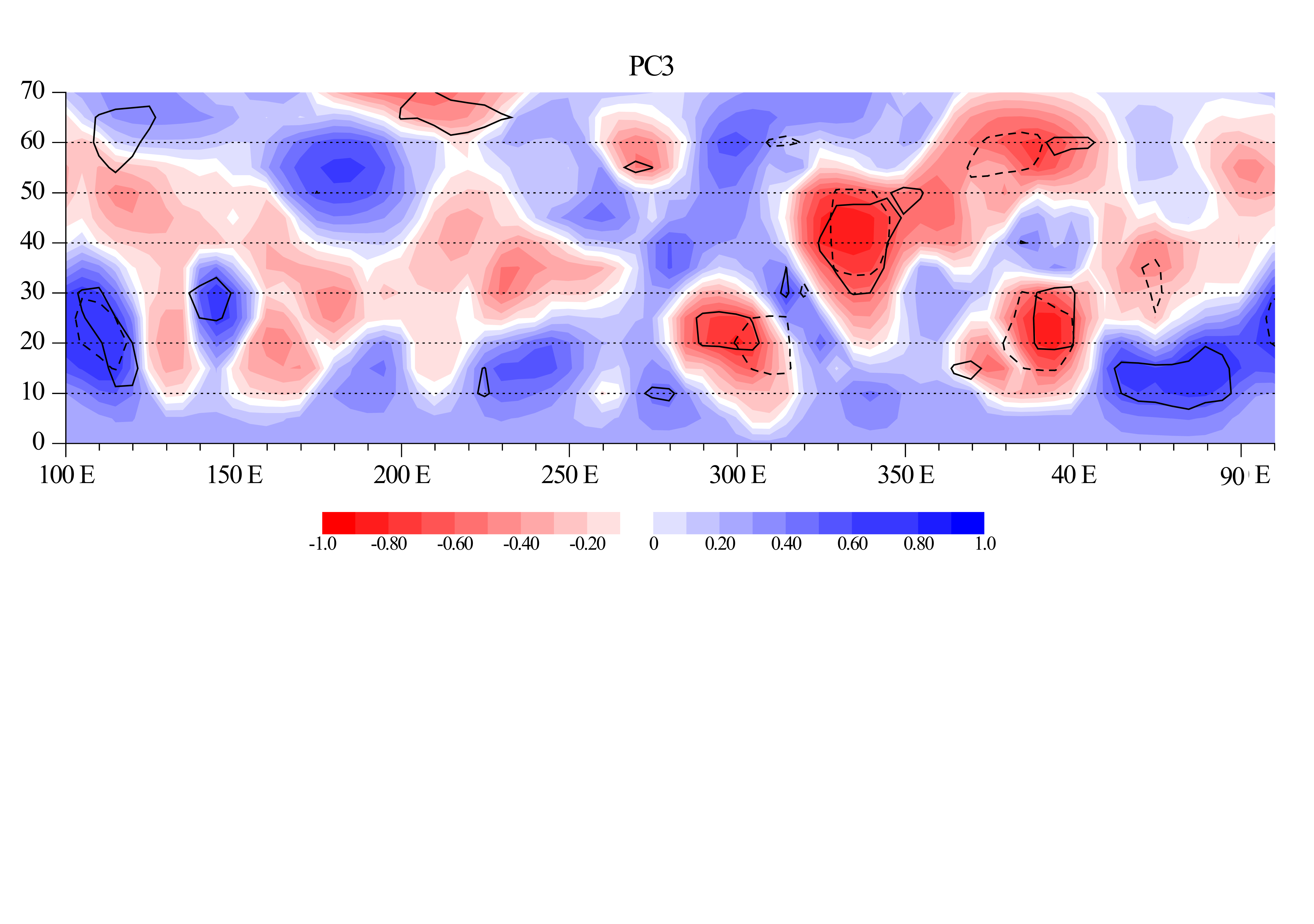}
 \end{center}
 \caption{ Maps for the parameters $(r_i^{gufm1}\, r_i^{COV-OBS})^{1/2} \, (r_i^{gufm1}/|r_i^{gufm1}|)$ (colored code), overlaid with p-level lines of 0.1 for the correlation coefficients  $r_i^{gufm1}$ (dashed) and $r_i^{COV-OBS}$ (solid).  Top: $i=1$, middle: $i=2$, bottom: $i=3$.}
\label{fig5}
 \end{figure}

Correlating PC$_i$ of {\it flow}$^{gufm1}$ (resp. {\it flow}$^{COV-OBS}$) with the data matrix corresponding to {\it flow}$^{gufm1}$ (resp. {\it flow}$^{COV-OBS}$) yields an homogeneous correlation map, whereas correlating it with the data matrix corresponding to  {\it flow}$^{COV-OBS}$ (resp. {\it flow}$^{gufm1}$) yields an heterogeneous correlation map (see Figure \ref{fig3}).
Homogeneous correlation maps provide a representation of EOF$_i$, while heterogeneous correlation maps are of special interest, since they reveal the correlated structures present in the two flows.
Strong blue and red patches show regions of {\it flow}$^{gufm1}$ or {\it flow}$^{COV-OBS}$ where the flow evolution is strongly correlated with a certain PC$_i$ (red for positive correlation, blue for anti-correlation).
Note that the black solid lines, which separate regions of positive and negative correlation with PC$_i$, are also the zero iso-$\xi$ lines of EOF$_i$.

The correlation maps of Figure \ref{fig3} highlight a spiraling structure in mode 1, better seen in {\it flow}$^{gufm1}$ than in {\it flow}$^{COV-OBS}$ and that was not obvious in Figure \ref{fig2} because of disparate flow amplitudes.
It is brought out by the normalization underlying the computation of correlation coefficients.

If we denote the mean flow column vector as $\overline{\mathbf{x}}$, with the same dimension as $\mathbf{u}_i$ and $\mathbf{v}_i$ vectors (the EOFs to retrieve {\it flow}$^{gufm1}$ and {\it flow}$^{COV-OBS}$, respectively), then we may write  the mean flows as a combination of orthogonal SVD modes $\overline{\mathbf{x}}^{\, gufm1,T} = \mathbf{\alpha}^T \mathbf{U}^T$ and $\overline{\mathbf{x}}^{\, COV-OBS,T} = \mathbf{\beta}^T \mathbf{V}^T$. The column vectors $\mathbf{\alpha}$ and $\mathbf{\beta}$ contain the expansion coefficients of the mean flows in the orthogonal basis of EOFs. Note that, because the mean flows $\overline{\mathbf{x}}^{\, gufm1}$ and  $\overline{\mathbf{x}}^{\, COV-OBS}$ have been removed prior to PCA or SVD analysis applied to matrices $\mathbf{X}$, there is no orthogonality imposed between them and the variability modes.

\begin{table}
           
                \begin{tabular}{ l r r l l rr}
                  \hline \hline
                  \,         &\multicolumn{2}{c}{$u_{rms}$ (km/yr)} &   \multicolumn{2}{c}{$\langle \mathbf{\dot{B}}_i \rangle / \langle \mathbf{\dot{B}} \rangle$} &   \multicolumn{2}{c}{mean flow projection}     \\
                  \cline{2-7}    
                  & {\it gufm1} & {\it COV-OBS} & {\it gufm1} & {\it COV-OBS} & {\it gufm1} & {\it COV-OBS} \\            
                  \cline{1-7}         
                 \textit{mean flow}            & 12.2            & 10.0         & 0.78 $\pm$ 0.16  & 0.75 $\pm$ 0.11   &  --   & --                    \\

                 \textit{mode-1}           & 14.0 $\pm$ 3.2          & 12.2 $\pm$ 3.6         & 0.88 $\pm$ 0.19  & 0.86 $\pm$ 0.15   &  0.17   & 0.06        \\

                  \textit{mode-2}               & 14.8 $\pm$ 3.1             & 13.1 $\pm$ 3.2          & 0.94 $\pm$ 0.20 &   0.91 $\pm$ 0.13   & 0.35  & 0.42   \\

                  \textit{mode-3}               & 15.9 $\pm$ 4.0             & 14.6 $\pm$ 2.9          & 0.98 $\pm$ 0.17  &  0.95 $\pm$ 0.08     & 0.22  & 0.32           \\

                  \textit{total flow}               & 17.1 $\pm$ 5.1             & 16.7 $\pm$ 3.1          & 1.00   &  1.00      & --  & --           \\
                 \hline \hline
                \end{tabular}
                \caption{First and second columns show cumulative values of $u_{rms, i}=\langle  \mathbf{u}_i \rangle = \left((1/4 \pi R_c^2) \int \mathbf{u}_i \cdot \mathbf{u}_i \, dS\right)^{1/2}$, starting from the mean flow over the 1840-1990 period; average and standard deviation (std) values are displayed. Third and fourth columns show the cumulative fraction of SV predicted, $\langle \mathbf{\dot{B}}_i \rangle / \langle \mathbf{\dot{B}} \rangle$ where $\langle \mathbf{\dot{B}}_i \rangle =  \left((1/4 \pi R_c^2) \int \mathbf{\dot{B}}_i \cdot \mathbf{\dot{B}}_i \, dS\right)^{1/2}$, when comparing to the SV signal induced by the total flows, $\langle \mathbf{\dot{B}} \rangle$;  average and std values are displayed. Fifth and sixth columns show the normalized contribution of each orthogonal mode in the mean flow expansion, i.e., $\left( \alpha_i^2/\sum_{k=1}^{N_p} \alpha_k^2 \right)^{1/2}$ and $\left( \beta_i^2/\sum_{k=1}^{N_p} \beta_k^2 \right)^{1/2}$, respectively (see text). The last row shows values of $u_{rms}$ mean and standard deviation for the whole flows ${\it flow}^{gufm1}$ and ${\it flow}^{COV-OBS}$ (first and second column). Note that they induce a SV signal that doesn't fit exactly the SV models, mainly because of the parameterization error (see appendix \ref{secA.1}).  The percentage of SV root mean square explained by both global flows is on average 87\%.}    
                \label{tab4}

\end{table}

The cumulative importance of the three first SVD modes both in terms of $u_{rms}$ and the energy in the secular variation that they account for is gathered in Table \ref{tab4}.
We can conclude that the compressed description of the flows using the mean and the three first variability modes, already accounts for about 90\% of the total $u_{rms}$. The percentage of SV explained by this simplified flow description is even greater, amounting to 95\% or more.
To see this, each mode added to the mean flow and previous variability modes, was made to interact with the corresponding geomagnetic field model for the whole 1840-1990 period and the relative SV energy was computed for each epoch. From Table \ref{tab4} we can conclude that the variability in {\it flow}$^{COV-OBS}$ is higher than in {\it flow}$^{gufm1}$, since larger fractions of the whole flow $u_{rms}$ and induced SV are explained by variability modes.
Finally, it is also shown in the last two columns the fraction of mean flow projected onto each one of the three modes.
Mode 2, in particular, contributes significantly to both mean flows.

Estimates of significance of the correlation coefficients were computed as explained in appendix \ref{sec_signif}.
To build one single map that could concentrate information from the SVD analysis, we computed for mode $i$, with $i=1,2,3$, the mean of the two PCs, PC$_i^{gufm1}$  and PC$_i^{COV-OBS}$ since they are very close (see Figure \ref{fig4}).
Then, the resulting time function  $\overline{PC}_i$ was correlated with {\it flow}$^{gufm1}$ and with {\it flow}$^{COV-OBS}$, giving two correlation maps $r_i^{gufm1}$ and $r_i^{COV-OBS}$, respectively.
To condense information in one single map, a new parameter $(r_i^{gufm1}\, r_i^{COV-OBS})^{1/2} \, (r_i^{gufm1}/|r_i^{gufm1}|)$ is introduced and its geographical distribution over the northern hemisphere is shown in Figure \ref{fig5} using a color code.
For each of the correlation coefficients $r_i^{gufm1}$ and $r_i^{COV-OBS}$, a corresponding $p$-value chart was computed (see sec. \ref{sec_signif}).
The $p=0.1$ iso-contours for correlation of $\overline{PC}_i(t)$ with {\it flow}$^{gufm1}$ (dashed lines) and for correlation of $\overline{PC}_i(t)$ with {\it flow}$^{COV-OBS}$ (solid lines) are plotted on Fig. \ref{fig5}.
These contours enclose the regions with correlation coefficients $r_i^{gufm1}$ or $r_i^{COV-OBS}$ statistically significant at level 90\% or higher (i.e, the probability that these correlations are obtained by chance is no more than 10\%), which means their variability is well explained by $\overline{PC}_i(t)$ in both flow models. In deep blue (or red) regions, both flows evolve in time strongly correlated (or anticorrelated) with the corresponding $\overline{PC}_i(t)$. Furthermore, if such patches are inside $p=0.1$ iso-contours of both  $r_i^{gufm1}$ and $r_i^{COV-OBS}$, we can conclude with a relatively high confidence (i.e., a 90\% probability of success) that they represent spatial features of variability modes common to both flows, their temporal variation being represented by  $\overline{PC}_i(t)$.
From Figure \ref{fig5} top, we conclude that not only the main 3 vortices are significant features of the first mode, but also several smaller ones.
Likewise, the zonal equatorial flow characterizing mode 2 is also a relevant feature in both flows, unlike the zonal flow component next to the tangent cylinder which is significant in {\it flow}$^{gufm1}$ but not in {\it flow}$^{COV-OBS}$. As explained above, the longitudinal flow component is in the same direction near the TC and in the equatorial region, in spite of different signs of $\xi$ there (resulting in different signs of the correlation with $\overline{PC}_2(t)$).
As for mode 3, the presence of correlated rolls under AH is significant, cyclonic around 1860 and 1970, anticyclonic around 1920 (see also Fig. \ref{fig4}).
Their localization is however fuzzy as we can conclude from a relatively weak superposition of corresponding p-level curves from correlation with {\it flow}$^{gufm1}$ and {\it flow}$^{COV-OBS}$.
Under the PH, this third mode shows more variability in {\it flow}$^{COV-OBS}$ than in {\it flow}$^{gufm1}$.

\begin{figure}
\begin{center}
\includegraphics[width=0.32\linewidth]{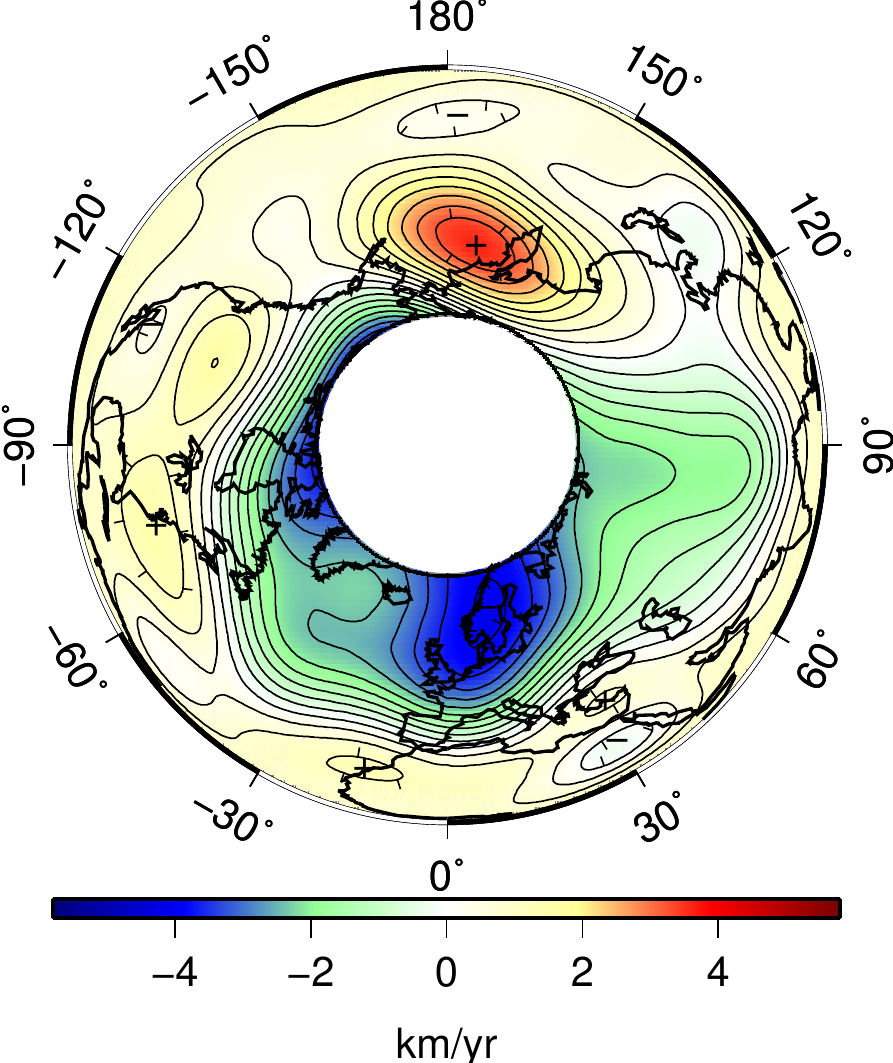}
\includegraphics[width=0.32\linewidth]{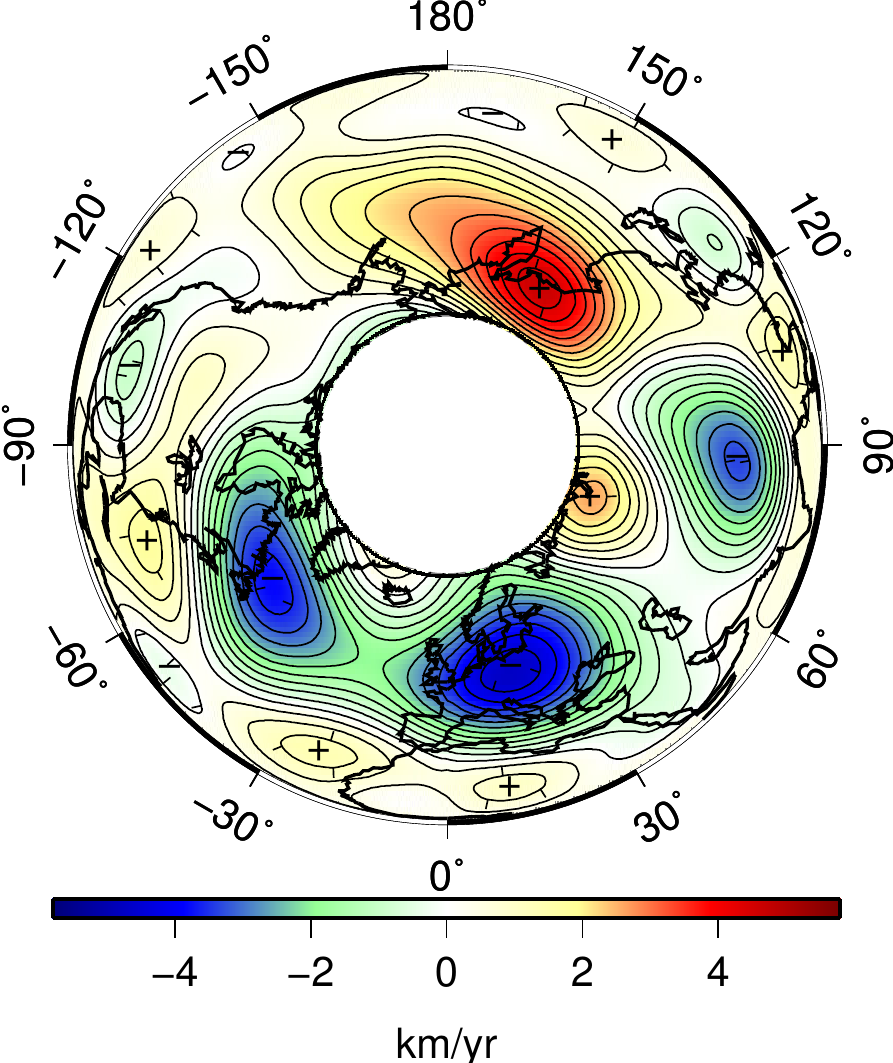}
\includegraphics[width=0.32\linewidth]{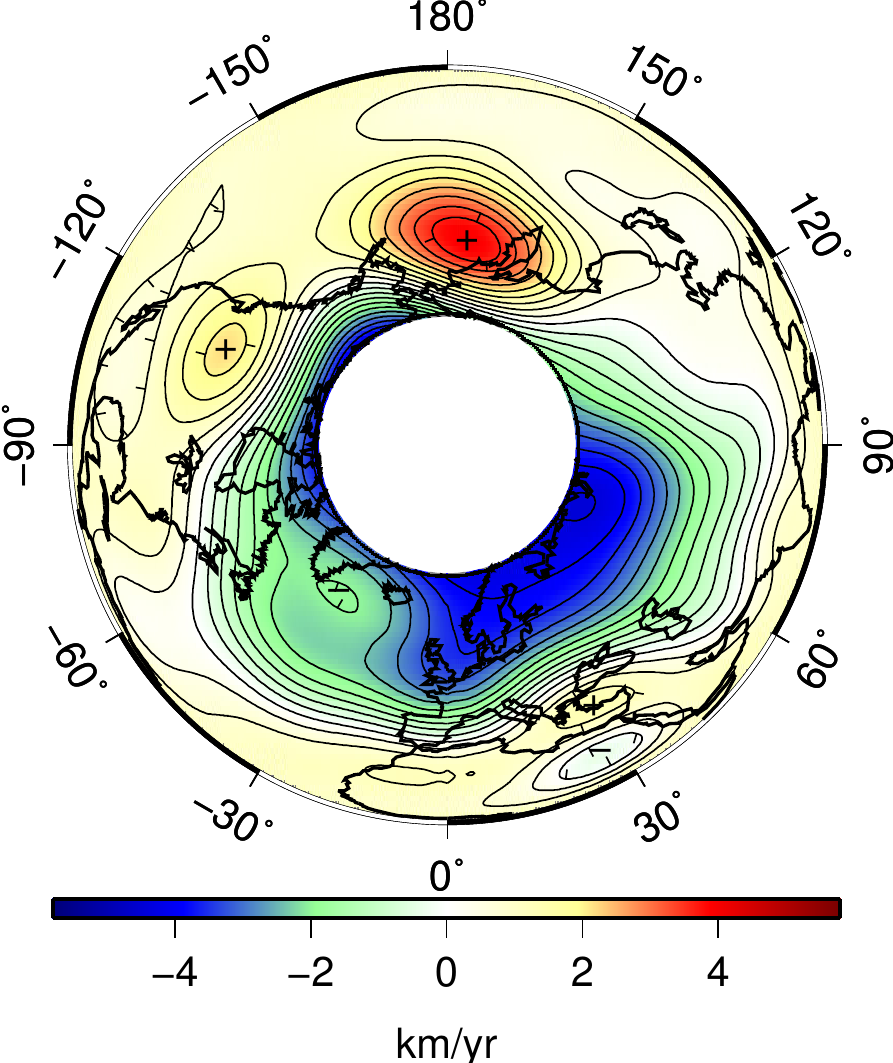}

\includegraphics[width=0.32\linewidth]{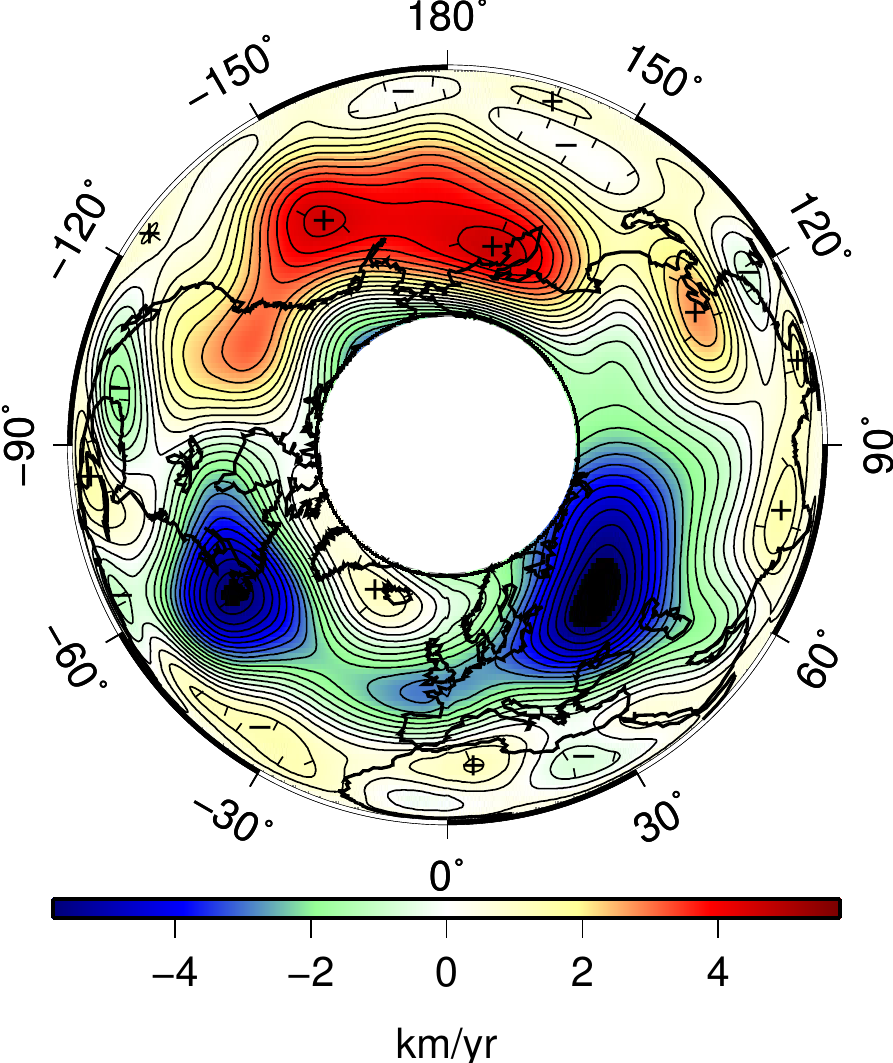}
\includegraphics[width=0.32\linewidth]{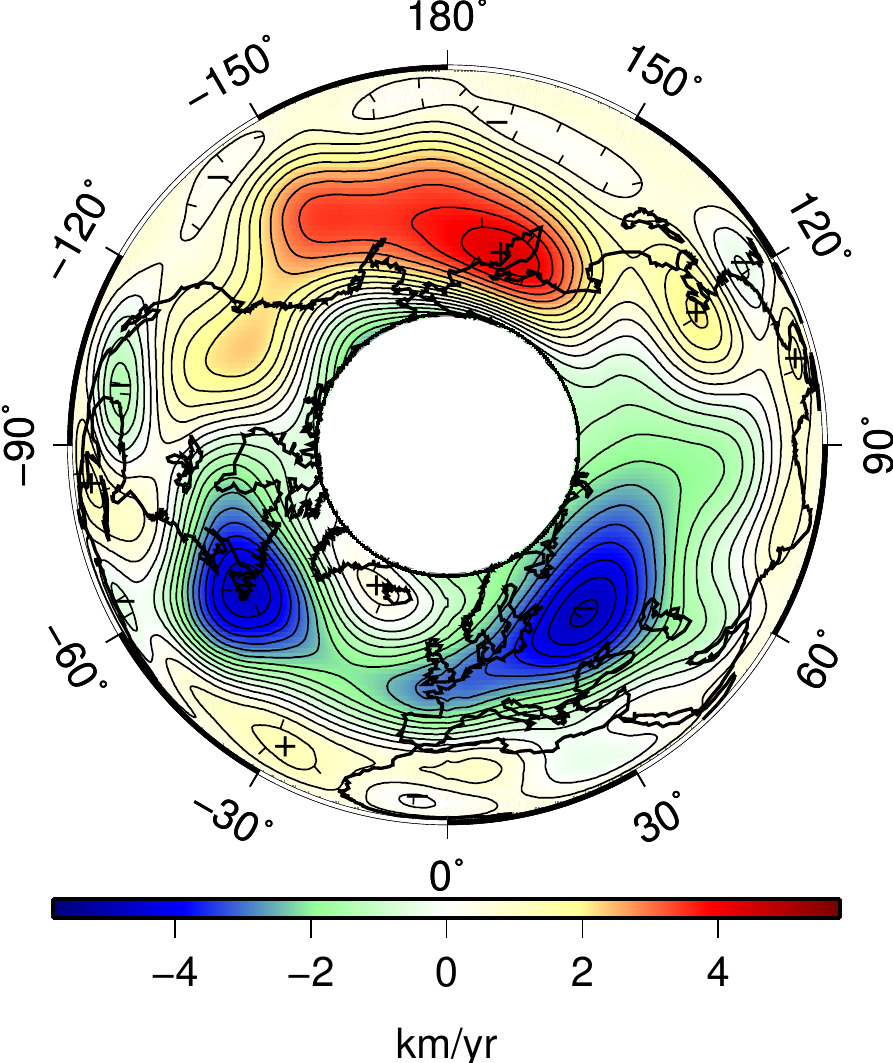}
\end{center}
\caption{Contours of pseudo streamfunction $\xi / R_c$ in units of km/yr, for a reconstruction of {\it flow}$^{gufm1}$ using the mean flow and the two first SVD modes.
Shown epochs are, from top left to bottom right: 1860 (both $PC_1$ and $PC_2$ are weak, negative), 1900 ($PC_2$ strong, positive), 1930 ($PC_2$ strong, negative), 1970 ($PC_1$ very strong, positive) and 1980 ($PC_1$ and $PC_2$ strong, positive).
In epochs of high positive values of PC$_1$ and/or PC$_2$, the flow features in these two modes act to destroy the large eccentric jet, their amplitude being of the same order of magnitude as the mean flow. See also the movies \citep{pais14movie}. }
\label{fig6}
\end{figure}

In order to visualize how and when these significant flow structures can be seen in the flow, Figure \ref{fig6} shows the flow reconstructed from the mean and the two first SVD modes for {\it flow}$^{gufm1}$, at epochs 1860, 1900, 1930, 1970 and 1980.
Referring only to the most significant features simultaneously spotted in both flows (see Fig. \ref{fig5}), the jet breakup can occur either because of an intensification of the cyclone centered at $\sim -150^{\circ}$E longitude (mode 1), either because of an intensification of the high latitude cyclones cutting the TC under Asian continent, the most important centered at $\sim 60^{\circ}$E longitude (mode 2).
Negative values of PC$_1$ and/or PC$_2$ tend to reinforce the jet, as can be seen in 1930.

\section{Discussion}\label{discussion}
 
\subsection{Mean flow}

The mean flows obtained from our inversions of {\it gufm1} and {\it COV-OBS}, shown on Fig. \ref{fig1} are not a surprise, both displaying an eccentric gyre already described by \cite{pais08}.
There are however several differences between the two: {\it flow}$^{COV-OBS}$  needs a much weaker Pacific recirculation, and a lower overall rms value for the mean flow.
As can be seen, the centennial timescale flow representing the mean shows strong latitudinal jets as main features, an anisotropy which is seen on different natural rotating flow systems, with or without magnetic fields.
This fact was used by \cite{schaeffer11} to propose an anisotropic regularization favoring zonation and leading to a mean flow not much different from that in Figure \ref{fig1}.
The dominance of $m=1$ azimuthal wavenumber in the mean flow suggests that an $m=1$ forcing strongly influences the core flow.
With direct numerical simulations of the core, \cite{aubert13} advocate for a differential inner-core growth \citep{monnereau10,alboussiere10} to explain the jet eccentricity. In their model, the westward circulation is an effect of uppermost liquid core lagging behind the mantle which, due to gravitational coupling with the inner core is indirectly pushed eastward by thermochemical winds inside the tangent cylinder. However, the pattern observed here is strikingly close to the one found by \cite{hori13} (compare their Fig 4c with our Fig. \ref{fig1}) with a geodynamo simulation with internal sources and an heterogeneous ($m=1$) heat flux imposed by the mantle.
Thus the mean flows inverted from both {\it COV-OBS} and {\it gufm1} could suggest a control of the convection by heterogeneous heat extraction from the mantle. They might indicate that more heat is extracted from the core in an hemisphere centered on the Western Atlantic, between $-60^{\circ}$ and $-90^{\circ}$ longitude, if we follow the findings of \cite{hori13}.

\subsection{First mode}

The principal component analysis of both flow models show similar empirical orthogonal modes. The singular value decomposition of the coupled flows further pinpoints common features. The method is particularly convenient in  making corresponding time functions (PCs) and spatial patterns (EOFs) that reconstruct {\it flow}$^{gufm1}$ and {\it flow}$^{COV-OBS}$ more similar (Table \ref{tab3}).
The empirical mode that carries the largest variability of the flow consists mainly of three big vortices (see Fig. \ref{fig2}): two anti-cyclones located at mid-latitudes around $-60^{\circ}$ and $+45^{\circ}$ longitude respectively;
a cyclone located under the Eastern Pacific ocean around $-150^{\circ}$ longitude, but which appears weaker on the flow inverted from {\it COV-OBS}.
The significance analysis carried out on this mode further indicates that all the main three rolls (as well as a few weaker ones) are meaningful, the centers of the most significant regions being located at medium latitudes between $40^{\circ}$ and $50^{\circ}$ (see Fig. \ref{fig5}).
Note also the cyclonic features close to the tangent cylinder under the North Atlantic ocean.
The main effect of this empirical mode when added to the mean flow is to slightly reinforce the mean eccentric gyre before 1940, and later break it by making the three vortices largely dominant after 1960, and especially around 1970 (see Fig. \ref{fig6}).

\subsection{Second mode}

\begin{figure}
\begin{center}
      \includegraphics[width=0.48\textwidth]{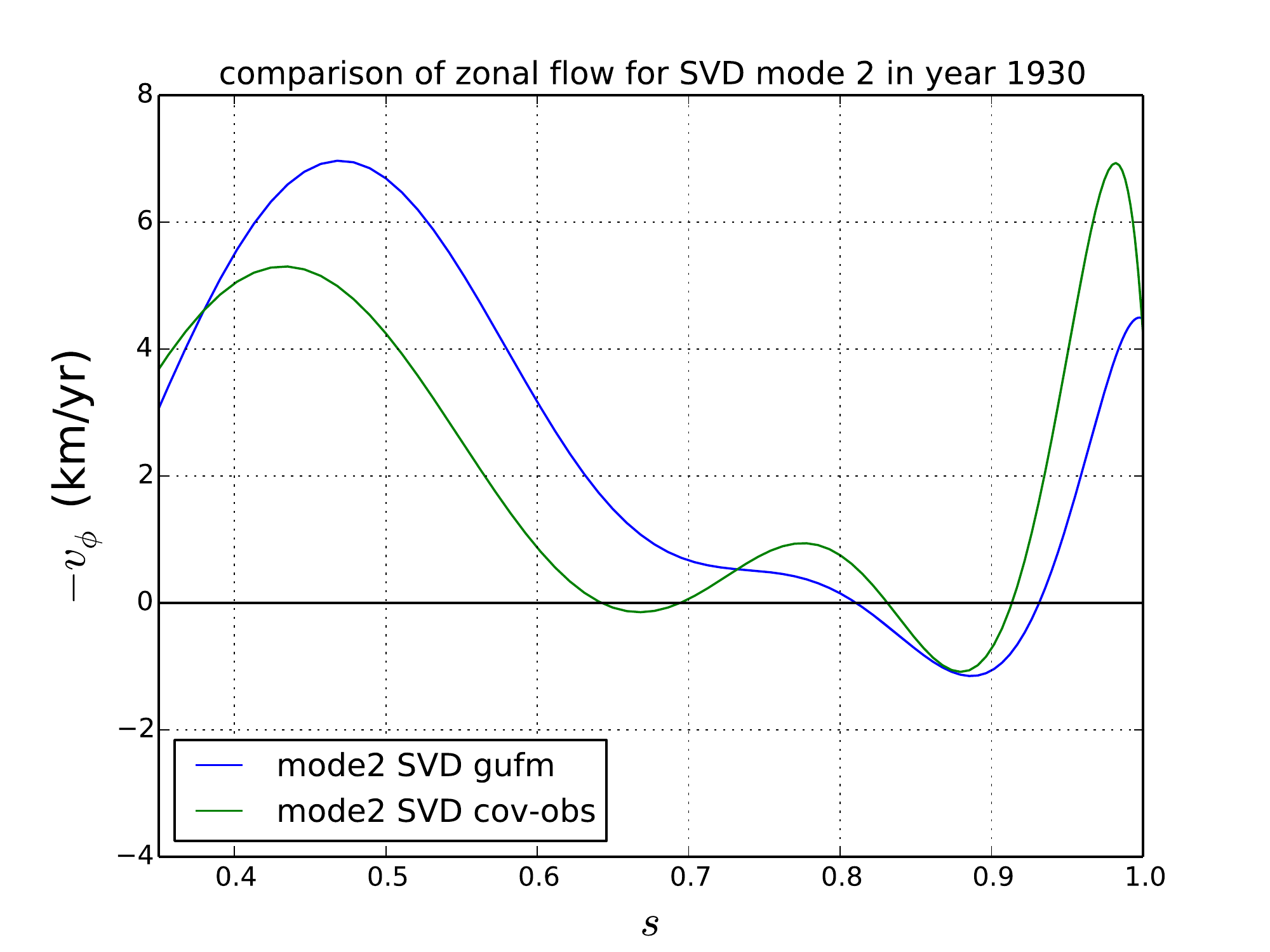}
      \includegraphics[width=0.48\textwidth]{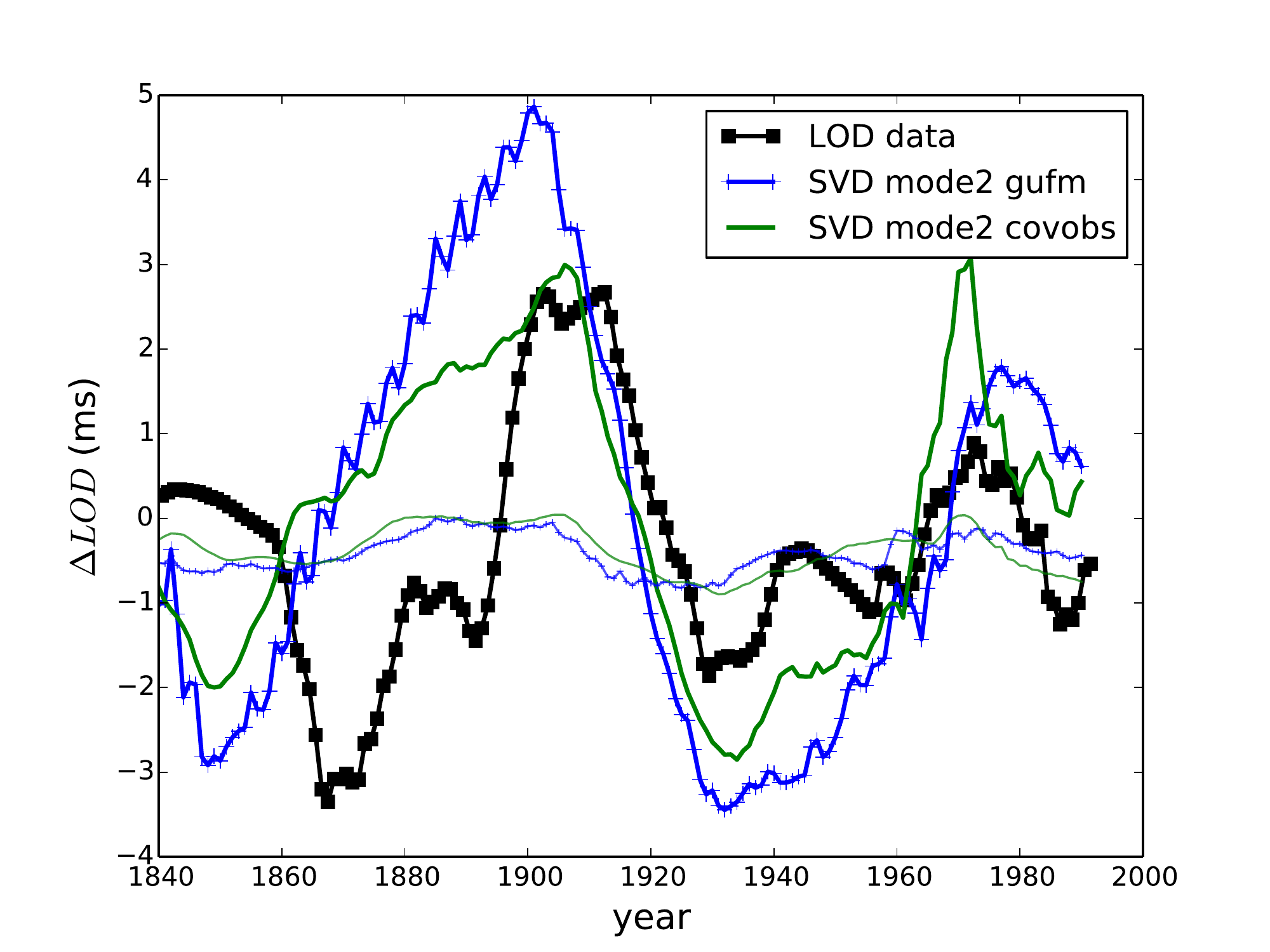}
\end{center}
\caption{On the left, the zonal component of $u_{\phi}$ for SVD mode 2, in km/yr, as a function of normalized distance to the rotation axis. On the right, the observed $LOD$ variation (thick black line with squares), estimates from our SVD mode 2 assuming a columnar flow \citep[thick blue and green lines, see][for the formula]{jault14} or assuming the flow is restrained to a stratified layer at the top of the core of thickness 140 km (thin lines). }
\label{fig7}
\end{figure}

The second mode coming out of our analysis has the interesting property of carrying most of the angular momentum of our core flows, as shown in Fig. \ref{fig7} (right) where the estimations for length of day (LOD) variations $\delta T$ have been computed for our core flows and compared to LOD observations.
The flow contributing to these variations is concentrated near the equator and the tangent cylinder (see Fig. \ref{fig7}, left).
In addition to the zonal part, mode 2 also shows several vortices and especially two co-rotating ones located close to the tangent cylinder under Eastern and Western Russia. There are also vortices with opposite circulation, the most significant one being located under north of the Mediterranean basin, at mid latitudes (see Figure \ref{fig5}, region with 10$^{\circ}$E $<$ long. $<$ 20$^{\circ}$E and 30$^{\circ}$ $<$ lat. $<$ 50$^{\circ}$N).
During the period 1840-1990, this mode shows two main oscillations basically reinforcing the eccentric gyre around 1850 and 1930, and weakening it especially around 1900 with vortices and an opposite circulation around the tangent cylinder.
The fact that most of the variation in zonal motion is correlated to these vortices might indicate either an excitation of a (zonal) eigenmode by the convecting flow, or the presence of slave secondary vortices excited by the interaction of a zonal eigenmode with the background magnetic field.

In this study, we show that these oscillations, have a mean period between 80 and 90 years (see Fig. \ref{fig4}).
These rather long periods cannot be explained by torsional oscillations which are thought to have much shorter periods of about 6 years \citep{gillet10}, but could, in principle, be a match for (a combination of) MAC oscillations \citep{braginsky93}, as shown recently by \cite{buffett14}.
However, these MAC oscillations are restricted to a thin stratified layer at the top of the core.
Fig. \ref{fig7} (right) shows that the simple restriction of our QG core flow to a thin layer largely underestimates the LOD variations. Hence, with the hypothesis of MAC oscillations, the core flow cannot explain the LOD variations directly (at least not for the rms values of 10-20 km/yr constrained by the SV), while a QG model extending in the whole core nicely explain these \citep[e.g. ][]{gillet10, schaeffer11}.
It might be worth mentioning that the solar activity exhibits a period of 80 to 100 years, the Gleissberg cycle \citep[e.g.][]{hathaway10}, which might provide an excitation for our mode 2.
Periods of 96 yrs and 76 yrs associated to damped  waves with parameters fitted to core surface flows have been reported by \cite{zatman97,zatman98}.
A core flow mode with a similar period has also been identified by \cite{dickey09} (85 yrs) and \cite{buffett09} (86.3 yrs) using different methods.
However, all those studies focused on the zonal longitudinal flow component inverted from {\it gufm1} or a prior version, {\it ufm1} \citep{bloxham92}.
In some cases, a shorter time span (1900-1990) was analysed \citep[e.g.][]{zatman97,zatman98}.
The standing oscillation reported in our study also involves nonzonal features, mainly close to the TC, and is detectable during the whole 1840 -1990 period. Furthermore, it combines information retrieved from two geomagnetic field models, {\it gufm1} and {\it COV-OBS}.
It must be emphasized that the time-dependence of mode 2 is not a nice sine wave, and that several frequencies are needed to reconstruct the signal.
We obtain a dominant period of 80 to 90 years for the time-span 1840-1990 as well as for SVD applied to the shorter interval 1900-1990. However, if we restrict the SVD analysis from 1920 to 1990 the period decreases to $\sim$ 70 years and a 60 years sinusoid fitted to mode 2 gives an almost equally high R-squared value, in agreement with other studies \citep{roberts07,buffett14}.

\subsection{Flow reconstruction and other modes}

The flow reconstructed from the first two modes and the mean flow is shown in Figure \ref{fig6} and basically exhibits a very strong eccentric gyre around 1860 and 1930, which is destroyed by the appearance of several vortices around 1900 (especially two counter rotation vortices under Asia) and by three very strong and large vortices around 1970, basically splitting the unique global gyre into two circulation systems, a cyclonic one located under the Pacific ocean, and an anticyclonic one under Europe and the Atlantic ocean.

The third mode is less important in terms of the pseudo-streamfunction $\xi$ variability explained, still it accounts for more than 10\% either using PCA or SVD (Fig. \ref{fig1-2}).
It's spatial pattern is also more complex showing a distribution of smaller-scale vortices, with often no clear correspondence between {\it flow}$^{gufm1}$ and {\it flow}$^{COV-OBS}$.
The correlated vortices clustered under the Atlantic region show nonetheless some correspondence in both flows, significant only in reduced regions of Fig. \ref{fig5}.

We also checked higher order modes, which concentrate smaller fractions of the data variability.
We looked in particular for a 6 year periodicity which has been reported by \cite{gillet10}.
Such periodicities can  be glimpsed at modes 8 and 10 of PCA applied to {\it COV-OBS} for the time period 1960-2010.
However, they never stand out above other relatively small periods that are also present, as is the case of a period of 9 years.
Indeed, to isolate this component, \cite{gillet10} need to extract the zonal flow of the ensemble average which is then band-pass filtered around 6 years, resulting in a signal several order of magnitudes smaller than the whole flow.
Other tools should be employed to look at small signals of specific periods, as the EOF/PC analysis is not the adequate tool to isolate such a relatively low amplitude variability.

\section{Conclusions}

We use Principal Component Analysis tools applied to `data' consisting of the pseudo-streamfunction $\xi$ evaluated on a regular spatial grid at the core surface outside the tangent cylinder.
Our main goal is to identify some underlying structure in the flow and reconstruct the data from a linear combination of a small number of spatial patterns multiplied by time-varying coefficients.
In this attempt for a simplification of observations, we first extracted the mean flow over the observatory era period (1840 to present):
a large eccentric jet, flowing at lower latitudes under the Atlantic Hemisphere (AH)  and at higher latitudes under the Pacific  Hemisphere (PH)  and  a large cyclone under the PH, centered at medium-latitudes.
The clear dichotomy revealed in this mean flow can be due to heterogeneous thermal forcing at the core-mantle boundary or at the solid core boundary \citep{aubert07,aubert13,hori13}.

We further identified three main circulation modes that account together for about 70$\%$ of the observed variability, 90\% of the total flow $u_{rms}$ and 95\% of the SV predicted by the total flow.
Mode 1, consists of three large QG vortices indicative of large scale convection. It has an aperiodic variability during the inspected period, with a main boost around the 1970 epoch.
At this time, the amplitude of mode 1 largely dominates the mean flow $u_{rms}$ and the three-roll flow breaks the large eccentric jet.
The same dichotomy as in the mean flow  reveals the dominance of anticyclonic circulation under the AH and cyclonic one under the PH.

A second mode was identified, which concentrates variations of core angular momentum, with a quasi-periodicity of 80 to 90 yrs.
Periods close to 80 or 90 yrs have been found in magnetic field observatory data \citep{jackson10}, geomagnetic field models \citep{kang08} and inverted flows \citep{zatman97, zatman98, dickey09, buffett09}.
Never before, however, have they been put forward without focusing the analysis on the zonal component of the flow and resorting to flows inverted from different geomagnetic field models.
Here, this period emerges associated with an empirical orthogonal mode of the whole system, naturally associated to a large zonal flow but also, interestingly, to some significant small scale circulations mainly next to the tangent cylinder.
These vortices evolve correlated with the equatorial flow and can also contribute to destroy or reinforce the large jet.
The spatial structure of this mode has an important projection on the mean flow suggesting, if we believe on the presence of this flow component inside the core, that it can give some insight into the morphology of the sources responsible for accelerating the large eccentric jet. The physical existence of this mode needs further testing, however. The presence of a mode of 80-90 yrs period both in historical observatory data series and field models spanning centennial time periods seems confirmed. However, its absence from decade LOD historical data where a period of $\sim$ 60 yr is seen instead can sound suspicious, if we accept that decade LOD variations have their origin in angular momentum exchanges between the mantle and the core. A possible explanation would be that this flow mode results from a leakage of an external magnetic field signal (associated to the Gleissberg Sun-cycle) into internal geomagnetic field models and inverted core flows. Unfortunately, we are not in a position to deny (or confirm) this hypothesis based only on results from this study.

Mode 3 shows the presence of a few correlated rolls under AH but their localization is fuzzy due to the fact that they are retrieved differently in {\it flow}$^{gufm1}$ and {\it flow}$^{COV-OBS}$. This precludes the emergence of significant spatial features.

Significance tests were carried out with results that support the spatial stability of modes 1, 2 and 3 in general, and of  specific spatial features present in these modes in particular.
The PCA tools used in this study look for the  directions  of  maximum  variance  in data space. The modes corresponding to projection along these directions will be of special interest in the case that higher variability modes correspond to interesting dynamics and lower ones   correspond to noise. We can not guarantee from our results that this condition is fulfilled. However, we point out the similarities we found between certain identified modes (modes 1 and 2), significant both in terms of spatial features described and amount of variance explained, and other time/spatial features identified in previous studies of core dynamics.

This study gives important insight into the variability of the core flows on centennial time-scales. We have shown that the mean flow and three empirical modes can account for most of the secular variation of the geomagnetic field.
Direct physical interpretation of the observed variability modes is not straightforward, although we have been able to link the first one to
a dominant vortex pattern, and the second one to observed length-of-day variations.
These modes may also guide future analytical and numerical studies on QG modes \citep[e.g.][]{canet14} and maybe help to constrain geometry of the internal magnetic field.

Finally, decomposition of quasi geostrophic flows inverted from geomagnetic field models into a few number of spatial structures and time dependent coefficients is of practical interest to serve as input for dynamical studies of the Earth's core interior where conditions closer to geomagnetic data are sought.
An interesting question that this decomposition in empirical modes could help to answer is whether and how the core flow that we can reconstruct from magnetic models contributes to the generation of the Earth's magnetic field.
We plan to address this question in a future study.

\section*{Acknowledgments}
M. A. Pais is supported by FCT (PTDC/CTE-GIX/119967/2010) through the project COMPETE (FCOMP-01-0124-FEDER-019978). A. L. Morozova is supported by a postdoc FCT scholarship (SFRH/BPD/74812/2010). M. A. Pais is grateful to the Grenoble University for funding her stay as an invited professor for two months. ISTerre is part of Labex OSUG@2020 (ANR10 LABX56).
We thank two anonymous reviewers for helping us to improve this paper.

\bibliography{coimbra}


\appendix

\section{Iterative inversion of {\it gufm1} and {\it COV-OBS}}
\label{secA.1}

In this appendix we outline the procedure to invert the geomagnetic field models {\it gufm1} in the time-period 1840-1990 and {\it COV-OBS} in 1840-2010, for a QG core flow with surface core expression $\mathbf{u}$. The method is regularized weighted least squares inversion, whereby the objective function $\Phi(\mathbf{m})$ that expresses a linear combination of the discrepancy between observables and theoretical predictions and a term with quadratic forms on the flow model $\mathbf{u}$ is minimized \citep[e.g][]{gubbins83}:
\begin{equation}
\Phi(\mathbf{m})= \left(\mathbf{A}(\mathbf{b}) \, \mathbf{m} - \mathbf{\dot{b}} \right)^T \mathbf{C}_{\mathbf{e}}^{-1} \left(\mathbf{A}(\mathbf{b})\,  \mathbf{m} - \mathbf{\dot{b}} \right) + \lambda \, \mathbf{m}^T \mathbf{C}_{\mathbf{m}}^{-1}  \mathbf{m} \label{obj_func} \quad.  
\end{equation}
$\mathbf{m}$ is the vector of poloidal ($s_{\ell}^{m(c,s)}$) and toroidal ($t_{\ell}^{m(c,s)}$) coefficients of the spherical harmonic (SH) expansion of the scalars  $\mathcal{S}$ and $\mathcal{T}$ that determine $\mathbf{u}= R_c \mathbf{\nabla}_H \mathcal{S} -R_c \hat{r} \wedge \mathbf{\nabla}_H \mathcal{T}$; $\mathbf{\nabla}_H=\mathbf{\nabla} - \hat{r} \partial/\partial r$ is the horizontal nabla operator; $\mathbf{b}$ is the vector of SH time-dependent coefficients of the scalar potential $V$ that determines the internal component of the geomagnetic field at the Earth's surface $\mathbf{B}= - \mathbf{\nabla} V$ and $\mathbf{\dot{b}}$ is the vector of the corresponding first time derivatives; $\mathbf{A}$ is the interaction matrix or matrix of equations of condition, with elements that depend on the geomagnetic field model and on the Elsasser and Adams-Gaunt integrals \citep[e.g.][]{whaler86}:
\begin{equation}
\mathbf{\dot{b}} = \mathbf{A}(\mathbf{b})\, \mathbf{m} + \mathbf{e}  \label{eqA-2} \quad.
\end{equation}

The solution to this minimization problem is of the form
\begin{equation}
\mathbf{\hat{m}}= \left(\mathbf{A}^T  \mathbf{C}^{-1}_{\mathbf{e}} \mathbf{A} + \lambda  \mathbf{C}^{-1}_{\mathbf{m}} \right)  \mathbf{A}^T \mathbf{C}_{\mathbf{e}}^{-1} \, \mathbf{\dot{b}}  \label{eqA-2a} \quad.
\end{equation}

The formalism above is standard and can be found in different studies \citep[see e.g.][]{pais00}. We now concentrate on specific features of our inversions.

\subsection{The error covariance matrix, $\mathbf{C}_{\mathbf{e}}$}
\label{secA.1b}

In the inversions for the present study, coefficients up to degree and order $\ell_\mathbf{b}= \ell_\mathbf{\dot{b}}=13$ from model {\it mod} are used, for the deterministic part of the main field and its SV, where {\it mod} stands for {\it gufm1} or {\it COV-OBS}. As to the dimension of vector $\mathbf{m}$,  it is that required to solve for flow coefficients up to degree and order $\ell_\mathbf{m}= 26$, the maximum degree that can be constrained (even if only slightly) by the first 13 spherical harmonic  degrees of the MF and of the SV.

The error $\mathbf{e}$ is treated as a Gaussian random variable with mean zero and covariance matrix $\mathbf{C_{\mathbf{e}}}$, which can be condensed using the notation $\mathbf{e} \sim \mathcal{N}(0, \mathbf{C_{\mathbf{e}}})$.
We consider two contributions for $\mathbf{C_{\mathbf{e}}}$:
(i) $\mathbf{C}^{mod}_e$ is the {\it a posteriori} covariance matrix for the SV coefficients coming from the geomagnetic field model; 
(ii) $\mathbf{C}^r_e$ is the covariance matrix characterizing the spatial resolution or modeling errors. This second term is dependent on the flow solution $\mathbf{m}$:
\begin{equation}
\mathbf{C}^{\,}_{\mathbf{e}} = \mathbf{C}^{\it mod}_{\mathbf{e}}+ \mathbf{C}^{r}_{\mathbf{e}}(\mathbf{m}) \label{eqA-2b} \quad.
\end{equation}
For $\mathbf{C}^{\it mod}_{\mathbf{e}}$, we will be using the diagonal $\mathbf{C}^{gufm}_{\mathbf{e}}$ matrix from \cite{jackson97} and the dense  $\mathbf{C}^{COV-OBS}_{\mathbf{e}}$ matrix  provided by \cite{gillet13}. Previous studies have identified the modeling errors arising from an incomplete knowledge of the main field ($\mathbf{C}^{r}_{\mathbf{e}}(\mathbf{m})$) as having a higher contribution to the large scales of the SV than do observational errors \citep[e.g.][]{eymin05,pais08,gillet09}. Inversions reported in recent papers do already take into account these errors, though different approaches have been followed, in particular the iterative approach in \cite{pais08} and the stochastic approach in \cite{gillet09}.
More recently another method has been proposed, called inverse geodynamo modeling, whereby the estimation of the effects of underparametrization is based on a statistical study of a numerical dynamo used as a prior model \citep{aubert13b,aubert14}.
 
A law is required to prolongate the spectrum of the geomagnetic field to length scales that are unperceivable at the Earth's surface and above. We  used the form $R(\ell, R_c)=W/(2 \ell +1)$ for the Lowes-Mauersberger spectrum at the CMB, from \cite{mcleod96}.
This function was fitted to the geomagnetic spectrum computed from gufm-sat-E3 \citep{finlay12} for the 2005 epoch and using only $ 3 \le \ell \le 13$. The value $W=1.11 \times 10^{11}$nT$^2$/yr$^2$ was found. The small-scale magnetic field ($14 \le \ell \le 40$) was then treated as a Gaussian random variable of zero mean and variance-covariance matrix $\mathbf{C}_{\mathbf{b}^S}$, i.e. 
\begin{equation}
\mathbf{b}^S \sim \mathcal{N}(0, \mathbf{C}_{\mathbf{b}^S}) \label{eqA-4} \quad,
\end{equation}
where $\mathbf{C}_{\mathbf{b}^S}$ is a diagonal matrix with elements given by
\begin{equation}
\sigma^{2}_{\mathbf{b}^S} (\ell) = \dfrac{W}{(\ell + 1)(2 \ell +1)^2} \left(\dfrac{R_c}{R_E}\right)^{2\ell +4} \label{eqA-2c} \quad,
\end{equation}
$R_E$ being the mean Earth radius.

To compute $\mathbf{C}^{\,}_{\mathbf{e}}$ for a certain flow solution $\mathbf{m}$, an ensemble of $K$ modeling error vectors $\mathbf{e}^{r,k}(\mathbf{m})=\mathbf{A}(\mathbf{b}^{S,k})\, \mathbf{m}$ are calculated, from an ensemble of $K$ small-scale magnetic fields $\mathbf{b}^{S,k}$ with statistical properties given by (\ref{eqA-4}) and (\ref{eqA-2c}). From this sampling, estimates of the mean and covariance matrix for the modeling (or spatial resolution) error are computed according to standard formulas, the (non-diagonal) covariance matrix being denoted $\mathbf{C}^{r}_{\mathbf{e}}(\mathbf{m})$. The updated error covariance matrix is then computed from eq. \ref{eqA-2b}. In all our computations we used $K=80$.

\subsection{The flow regularization}
\label{secA.1c}

The second term in the RHS of (\ref{obj_func}), also referred to as model regularization, incorporates the assumptions on the flow; its relative weight in the linear combination  is controlled by the magnitude of the regularization parameter $\lambda$. Two important assumptions are considered in this study: (a) that most of the observed SV is due to a large scale flow interacting with the magnetic field and (b) that this flow is the surface expression of incompressible columnar convection inside the core. To impose condition (a) a $\ell^3$ norm that minimizes flow gradients was used \citep{gillet09}.
As to condition (b), it requires the flow at the CMB to be
equatorially symmetric and verifying $\mathbf{\nabla}_H \cdot (\mathbf{u} \cos^2 \theta)=0$ \citep[e.g.][]{pais08,schaeffer11,amit13}.
This second condition is imposed using a very high $\lambda$ multiplier, such that deviations are zero to the machine precision.
The regularizing parameter multiplying the $\ell^3$ norm, denoted $\lambda_R$, is allowed to change from one epoch to the other, though, in order for the normalized misfit to be  nearly $1$ at each epoch. For each inversion the following condition is used to select an acceptable $\lambda_R$ \citep{pais04}:
\begin{equation}
0.8 < \sqrt{\dfrac{\chi^2}{N}} < 1.2   \label{eqA-3}
\end{equation}
where 
\begin{equation}
\chi^2=\left(\mathbf{A}(\mathbf{b})\,  \mathbf{m} - \mathbf{\dot{b}} \right)^T \mathbf{C}_{\mathbf{e}}^{-1} \left(\mathbf{A} (\mathbf{b})\, \mathbf{m} - \mathbf{\dot{b}} \right) \label{eqA-3b}
\end{equation}
is the normalized misfit and $N$ is the total number of SV coefficients.

\subsection{The inversion algorithm}
\label{secA.1d}

In this study, the iterative approach of \cite{pais08} was used, whereby the effects of the small-scale magnetic field are computed for each step $i$ of the flow calculation and the result is assimilated as a modeling error into the covariance matrix $ \mathbf{C}^i_{\mathbf{e}}$. The procedure is repeated iteratively until a convergence is met, when either the modeling error or the flow solution no longer suffer significant changes.
At this point, the found solution is consistent with the error covariance matrix $\mathbf{C}_{\mathbf{e}}$ used to invert it.
Another iterative search is made with iteration index $j$, within each step $i$, in order to find the regularization parameter $\lambda_R$ that provides a flow solution $\mathbf{m}$ satisfying (\ref{eqA-3}).

The whole procedure can be described in the following way, for each epoch.
Each step $i$ of the cycle involves three intermediate flows $\mathbf{m}^{i,0}$, $\mathbf{m}^{i,1}$ and $\mathbf{m}^{i,2}$ in four intermediate steps within step $i$, denoted $A)$ to $D)$:
\begin{enumerate}
\item[$A$)] $\mathbf{m}^{i,0}$ is used to compute a first estimate $\mathbf{C}^{i,0}_{\mathbf{e}}(\mathbf{m}^{i,0})$ of the error covariance matrix, as explained in section \ref{secA.1b};
\item[$B$)] $\mathbf{C}^{i,0}_{\mathbf{e}}$ is used in eq. \ref{eqA-2a} to invert for $\mathbf{m}^{i,1}$ and from this a new error covariance matrix $\mathbf{C}^{i,1}_{\mathbf{e}}(\mathbf{m}^{i,1})$ is computed as in previous item;
\item[$C$)] $\mathbf{C}^{i,1}_{\mathbf{e}}$ is used to invert for $\mathbf{m}^{i,2}$ and  $\mathbf{m}^{i,2}$ is used to compute $\mathbf{C}^{i,2}_{\mathbf{e}}(\mathbf{m}^{i,2})$;
\item[$D$)] The flow solution resulting from step $i$ is  $\mathbf{m}^{i}=(\mathbf{m}^{i,1}+\mathbf{m}^{i,2})/2$, which is used as $\mathbf{m}^{i+1,0}$ in the next step.
\end{enumerate}
To monitor the convergence of flow solutions $\mathbf{m}^{i,1}$ and $\mathbf{m}^{i,2}$ and of covariance matrices $\mathbf{C}^{i,1}_{\mathbf{e}}(\mathbf{m}^{i,1})$ and $\mathbf{C}^{i,2}_{\mathbf{e}}(\mathbf{m}^{i,2})$ as $i$ increases, they are compared using as diagnostic parameters the L$^2$ norm of the vector difference $\mathbf{m}^{i,2}-\mathbf{m}^{i,1}$ and the Frobenius norm of the matrix difference $\mathbf{C}^{i,2}_{\mathbf{e}}(\mathbf{m}^{i,2})-\mathbf{C}^{i,1}_{\mathbf{e}}(\mathbf{m}^{i,1})$.

Each of the two flow inversions made at step $i$ is done according to a second (internal) iterative cycle with index $j$, whereby the found solution must verify condition (\ref{eqA-3}):
\begin{enumerate}
\item[$j=0,1$:] Two trial solutions are first computed from (\ref{eqA-2a}), using $\lambda_{R,0}$ and $\lambda_{R,1}=\lambda_{R,0} + \Delta \lambda^{\,}_0$. Computing $\chi^2$ for each of these two inversions, makes possible to estimate the derivative of $\chi^2$ with respect to $\lambda$. Then a Newton-Raphson type algorithm is applied to choose $\Delta \lambda^{\,}_1$ such that $\lambda_{R,2} = \lambda_{R,1}+ \Delta \lambda^{\,}_1$ makes $\chi^2$ approach the due condition (\ref{eqA-3}).
\item[$1<j<J$:] For each new value $\lambda_{R,j} = \lambda_{R,j-1}+ \Delta \lambda^{\,}_{j-1}$ a new inversion using (\ref{eqA-2a}) is made and the corresponding $\chi^2$ computed. A Newton-Raphson type algorithm is applied to choose $\Delta \lambda^{\,}_j$.
\item[$j=J$:] The cycle ends at iteration $J$, when $\lambda_{R,J}$ verifies condition (\ref{eqA-3}). This final estimate gives $\lambda_{R}$ of the inversion.
\end{enumerate}
In the first of all $j$-cycles, we used in our calculations  $\lambda_{R,0}= 1.0 \times 10^6$ and $\Delta \lambda_0=\lambda_{R,0}/4$, from previous tests. In all other cycles, the converged $\lambda_{R}$ from the previous cycle is used.

Finally, the two particular $i$-cycle steps are:
\begin{enumerate}
\item[$i=0$:] The initial flow $\mathbf{m}^{0,0}$ is computed from (\ref{eqA-2a}), using as error covariance matrix $\mathbf{C}^{\,}_{\mathbf{e}}=\mathbf{C}^{\it mod}_{\mathbf{e}} + \mathbf{C}^{r}_{\mathbf{e}}$, where only diagonal elements are considered for $\mathbf{C}^{r}_{\mathbf{e}}$, given by $ \sigma^r_{\mathbf{e}}(\ell)^2 = 36 \exp(-\ell)$ as in e.g. \cite{schaeffer11}. 
\item[$i=I$:] Convergence is met for $i=I$ if $\lambda^{I,1}_R=\lambda^{I,2}_R$ AND $\chi^{I,1}_{\,}(\mathbf{m}^{I,1})=\chi^{I,2}_{\,}(\mathbf{m}^{I,2})$ (see eq. \ref{eqA-3b}), within a certain precision imposed by \ref{eqA-3}. These two conditions ensure that the inversion is well-converged, as confirmed by very small values of root squares of the L$^2$ norm of flow differences and of the Frobenius norm of error covariance matrices differences (10$^{-7}$ to 10$^{-8}$ and 10$^{-3}$ to 10$^{-5}$, respectively.)
\end{enumerate}

\begin{figure}
\begin{center}
            \includegraphics[scale=0.55]{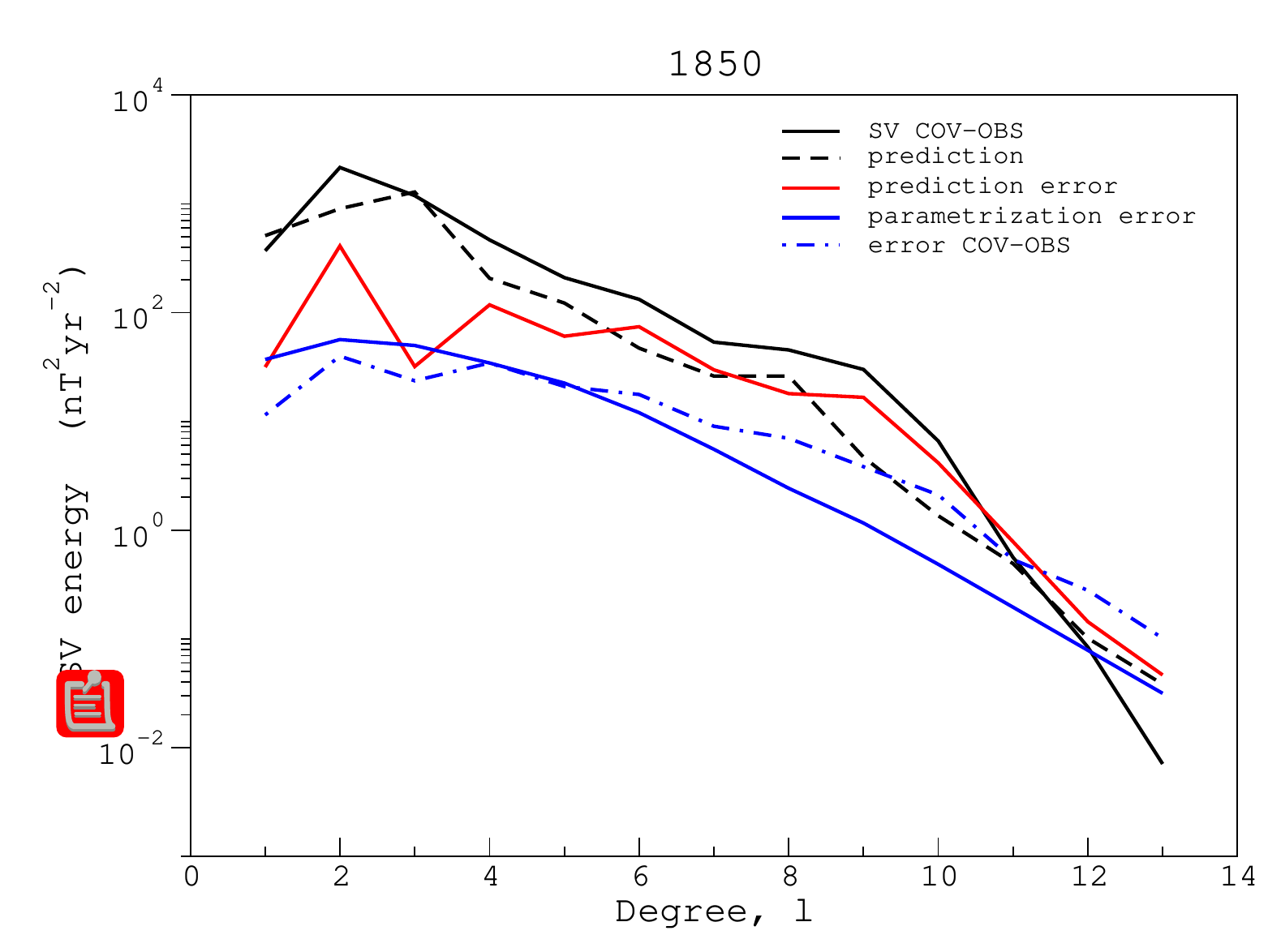}      
            \includegraphics[scale=0.55]{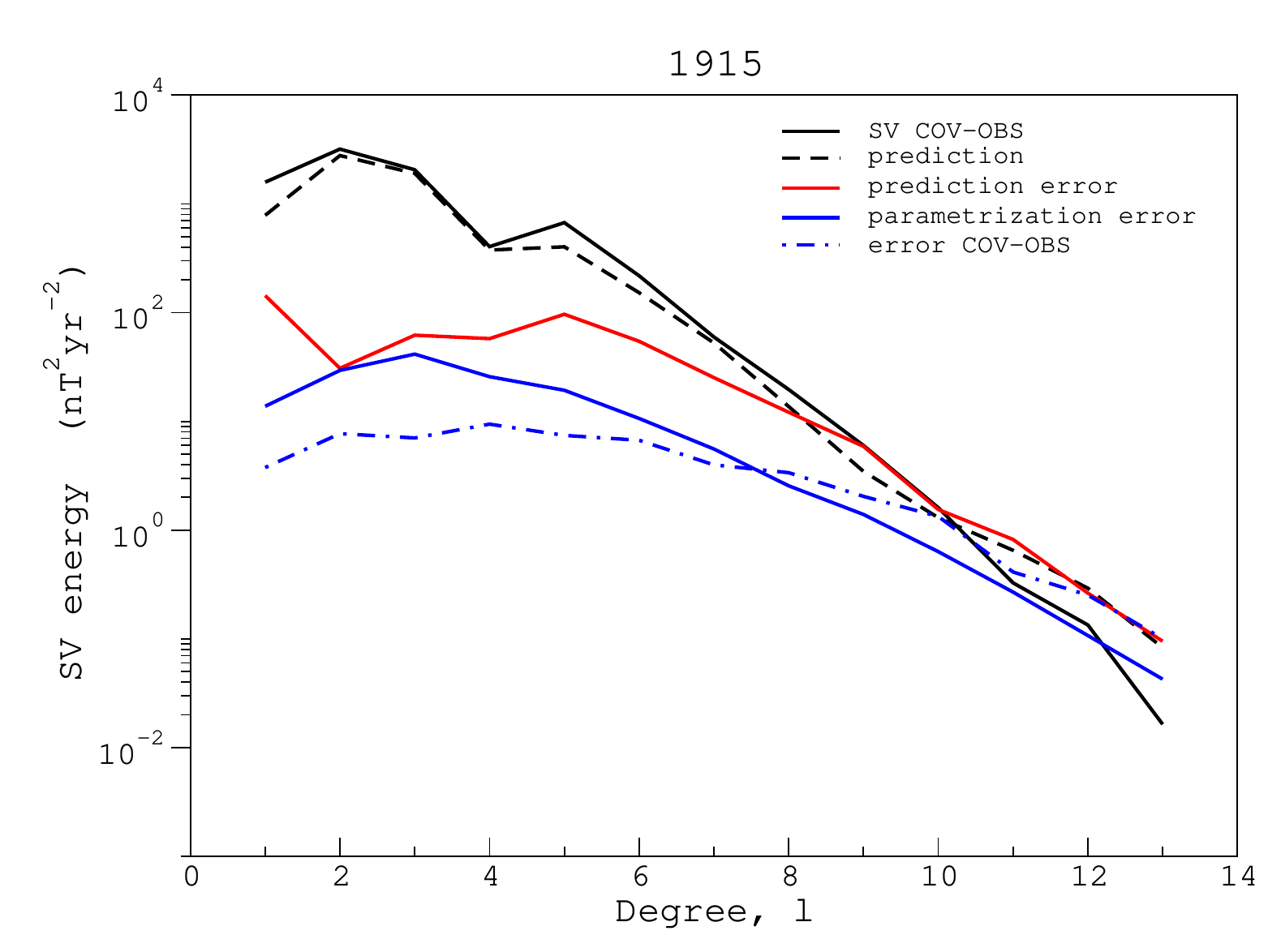}
            \includegraphics[scale=0.55]{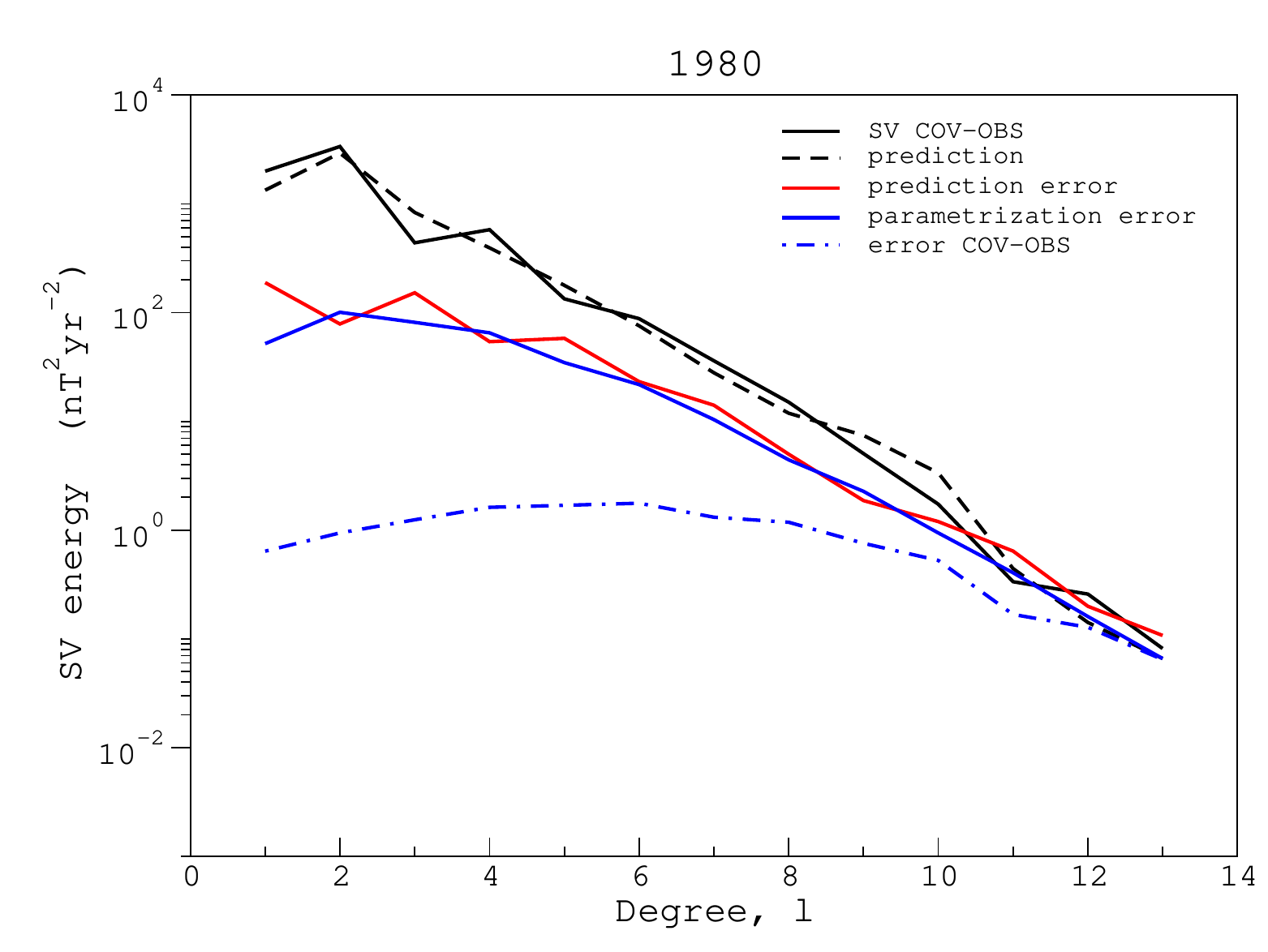}
      
 \end{center}
 \caption{The Lowes-Mauersberger power spectrum at the Earth's surface of: the SV of model {\it COV-OBS} (black solid line), flow predictions (black dashed line), misfit errors (red line), SV prior data errors from $\mathbf{C}^{COV-OBS}_{\mathbf{e}}$ (blue dashed line) and modeling errors from $\mathbf{C}^{r}_{\mathbf{e}}(\mathbf{m})$ (blue solid line). }
\label{fig8}
\end{figure}

Fig. \ref{fig8} shows the Lowes-Mauersberger power spectrum at the Earth's surface of the SV of model {\it COV-OBS}, together with predictions from our inverted flows and different errors referred above: misfit errors, SV prior data errors from $\mathbf{C}^{COV-OBS}_{\mathbf{e}}$ and modeling errors from $\mathbf{C}^{r}_{\mathbf{e}}(\mathbf{m})$.
For the two last cases, although the covariance matrices are dense, only diagonal elements are shown. Besides, rotationally invariant errors are calculated by averaging the diagonal elements (variances) within each degree $\ell$. Along the time interval from 1840 to 1990 the prior data errors decrease as the quality and amount of data improve, but the modeling errors are always of the same order of magnitude with very similar spectra. It is worth noting that the prediction error for epoch 1850 is  very similar to the corresponding error obtained by \cite[][see his Figure 1]{aubert14}, in spite of a more involved approach whereby the error covariance matrix to use is obtained from an ensemble of direct numerical simulations of the geodynamo.

When dealing with the two different geomagnetic field models {\it gufm1} and {\it COV-OBS}, the main differences come from the fact that we used the diagonal $\mathbf{C}^{gufm}_{\mathbf{e}}$ matrix from \cite{jackson97} and the dense  $\mathbf{C}^{COV-OBS}_{\mathbf{e}}$ matrix  provided by \cite{gillet13}. The total matrices $\mathbf{C}^{\,}_{\mathbf{e}}$ are always non-diagonal, contrary to most previous inversions, thus allowing to take into account the fact that errors in the SV coefficients are correlated.

For {\it COV-OBS}, this correlation comes in part from the geomagnetic field model computation, but the most important contribution comes from the parameterization errors 
iteratively computed during the inversion as explained in section \ref{secA.1b}.

\section{Estimation of significances for identified modes}
\label{sec_signif}

A test on the covariance matrix spectrum allows to assess the uncertainty of each eigenvalue of the covariance matrix $\mathbf{C}_X$ (section \ref{secSEOF}) or each singular value of $\mathbf{C}_{X_1 X_2}$ (section \ref{secSVD}), that reflects the fact that the sampling EOF's and PC's differ from the asymptotic ones (when $N_e \rightarrow \infty$).
Following \cite{north82}, the sampling error in the eigenvalue $\lambda_i$ of  $\mathbf{C}_X$ is, to first order, 
\begin{equation}
\delta \lambda_i \sim \lambda_i \sqrt{\dfrac{2}{N}}    \label{north_error}
\end{equation}
where $N$ is the number of independent realizations or the number of degrees of freedom, which we take equal to $N_e$.
Then, two consecutive eigenvalues $\lambda_i$ and $\lambda_j$ are considered degenerate if their difference $\Delta \lambda=\lambda_i - \lambda_j$ is such that
\begin{equation}
\Delta \lambda \leq \lambda_i \sqrt{\dfrac{2}{N_e}}  \label{north}  \quad.
\end{equation}
This is known as North's rule of thumb. When two (or more) eigenvalues of $\mathbf{C}_X$ are degenerate it means they cannot be clearly distinguished and the corresponding EOFs patterns cannot be considered independent \citep{hannachi07}. The same rule of thumb can be used to decide on the separation of consecutive singular values of the cross-covariance matrix when applying SVD \citep[e.g.][]{venegas97}.

Temporal functions are compared using the correlation coefficient $r$ \citep[e.g.][]{vonstorch95}, which measures the extent to which there is a linear relationship between the two functions. It always takes values in the interval $[-1,1]$, the two extreme values meaning there exists a perfect linear relationship between the two compared functions. For 
$\mathbf{t}_1$ a column vector representing a certain time series and $\mathbf{t}_2$ a column vector representing another time series, with the same dimension,
\begin{equation}
r(\mathbf{t}_1, \mathbf{t}_2) =\dfrac{\left(\mathbf{t}_1-\overline{\mathbf{t}_1}\right)^T \left(\mathbf{t}_2-\overline{\mathbf{t}_2}\right)}{\sqrt{\left(\mathbf{t}_1-\overline{\mathbf{t}_1}\right)^T\left(\mathbf{t}_1-\overline{\mathbf{t}_1}\right)} \sqrt{\left(\mathbf{t}_2-\overline{\mathbf{t}_2}\right)^T\left(\mathbf{t}_2-\overline{\mathbf{t}_2}\right)}} \label{coefr}
\end{equation}
where $\overline{\mathbf{t}_1}$ and $\overline{\mathbf{t}_2}$ are the average values of the two series over the time period considered.

To estimate the level of similarity between two different spatial structures, we use congruence coefficients, $g$ \citep[e.g.][]{harman76}. If $\mathbf{s}_1$ is  a column vector representing a given spatial structure and $\mathbf{s}_2$ a column vector for some other spatial structure, both with the same dimension, then
\begin{equation}
g(\mathbf{s}_1, \mathbf{s}_2) = \dfrac{ \mathbf{s}^T_1 \mathbf{s}^{\,}_2 }{ \sqrt{\mathbf{s}^T_1 \mathbf{s}^{\,}_1} \sqrt{ \mathbf{s}^T_2 \mathbf{s}^{\,}_2} } \label{coefg}
\end{equation}
Congruence coefficients have the same range as correlation coefficients, i.e.  from -1 (negative similarity) to +1 (positive similarity). However, as noticed by \cite{richman85} who uses this parameter in the context of PCA analysis, unlike the correlation coefficient the congruence coefficient does not subtract the mean value of the spatial patterns to compare. It therefore measures not only the pattern similarity  but the magnitude similarity as well.

Finally, a Monte Carlo (MC) approach is used to estimate the significance of the correlation coefficients $r$ for each grid point shown in correlation maps (see Figure \ref{fig5}). To this end, the two original time series to be correlated are shuffled, one at a time, using a technique called bootstrapping with moving blocks.
The theoretical discussion of the method can be found in \cite{kunsch89} and \cite{liu92}. Here, we use it in a similar context as in several studies on coupled atmosphere-ocean variability \cite[e.g.][]{wallace92,peng96,venegas97}. The method is used for resampling and is applicable to weakly dependent stationary data that is, data which is nearly independent if  far apart in time. There is nonetheless a correlation time to consider and that should be kept in resampled as in original data, i.e. the autocorrelation structure of the original series should be preserved. The reason is that shuffling the data in time randomly would increase the temporal degrees of freedom, giving in the end lower p-values for the statistical tests and improving artificially the significance of estimated parameters. So, in order to avoid this effect, the shuffling in time is applied not to every single data point randomly, but to a number of blocks containing the original data.
The length of the blocks ($b$) depends on the autocorrelation structure of the original data. 
For a data set $x_i$  of  total time-length $N_e$, a number of blocks $N_e-b+1$ are created so that the data from $x_1$ to $x_b$ will be in block 1, the data from $x_2$ to $x_{b+1}$ will be in block 2, etc. After creating all the blocks, $N_e/b$ of these blocks are randomly selected to create a new pseudo-time series $x'_i$, from which a certain quantity to test may be recalculated (if $N_e$ is not an integer multiple of $b$, $x'_i$ is truncated to have the same length as $x_i$). The length $b$ of the blocks was chosen after inspection of the flow data autocorrelation function in each point of the spatial grid. Having verified that the time lag at which the autocorrelation function becomes zero or very closely zero does not exceed 55 yrs, this time interval was therefore used as the time-length of the moving blocks. 
For the significance tests on correlation maps produced from the SVD of coupled {\it flow}$^{gufm1}$ and {\it flow}$^{COV-OBS}$, each of the Monte-Carlo runs included two sub-runs.
During the first sub-run the first of the series to correlate, (e.g. the $\overline{PC}_i$  time series for SVD mode $i$, see section \ref{secSVD}) was shuffled and the resulting artificial series was correlated in each grid point with the second series which is the observed time series from {\it flow}$^{gufm1}$  or from  {\it flow}$^{COV-OBS}$ .
During the second sub-run the shuffling in time procedure was applied to the flow data matrix ({\it flow}$^{gufm1}$ or {\it flow}$^{COV-OBS}$, respectively), and the resulting resampled data series for each grid point were then correlated with $\overline{PC}_i$. The main goal is always to break the chronological order of one field relative to the other.
For each MC run, the mean of the two artificial correlation coefficients for each grid point obtained from the two sub-runs was calculated. After a large number of MC runs ($2 \times 1000$), the number of the correlation coefficients of magnitude higher than the original $r$ divided by the total number of the runs is the p-value of this specific $r$. This p-value gives the probability to obtain this specific $r$ just by chance.
For instance, a p-value = 0.1 associated to a certain correlation coefficient, means that the probability to obtain such correlation coefficient for this specific pair of correlated time series just by chance is only 10\%.

\end{document}